\documentclass[journal,12pt,onecolumn,draftclsnofoot]{IEEEtran}
\usepackage{cite}
\usepackage{caption}
\usepackage{subcaption}
\usepackage{graphics}
\usepackage{booktabs}
\usepackage{algorithm,algorithmic}
\usepackage[pdftex]{graphicx}
\usepackage{graphicx}
\usepackage{epstopdf}
\usepackage[section]{placeins}
\usepackage{color}
\usepackage{pgfplots}
\pgfplotsset{compat=1.3}
\usepackage[cmex10]{amsmath}
\usepackage{amsfonts}
\usepackage{amssymb}
\usepackage{multirow}

\usepackage{array}
\usepackage{footnote}
\usepackage{fancyhdr}
\usepackage[caption=false,font=normalsize,labelfont=sf,textfont=sf]{subfig}

\pgfplotscreateplotcyclelist{bler_list_1}{
	{red,thick,solid, mark = triangle},
	{blue,thick, solid, mark = triangle},
	{red,thick,solid, mark = square},
	{blue,thick, solid, mark = square},
	{red,thick,solid, mark = o},
	{blue,thick, solid, mark = o},
	{red,thick,solid, mark = diamond},
	{blue,thick, solid, mark = diamond},
	{red,thick,dashed, mark = triangle, mark options=solid},
	{blue,thick, dashed, mark = triangle, mark options=solid},
	{red,thick,dashed, mark = square, mark options=solid},
	{blue,thick, dashed, mark = square, mark options=solid},
	{red,thick,dashed, mark = o, mark options=solid},
	{blue,thick, dashed, mark = o, mark options=solid},
	{red,thick,dashed, mark = diamond, mark options=solid},
	{blue,thick, dashed, mark = diamond, mark options=solid},
}
\pgfplotscreateplotcyclelist{bler_list_sync_2}{
	{red,thick,solid, mark = triangle},
	{blue,thick, solid, mark = triangle},
	{red,thick,solid, mark = square},
	{blue,thick, solid, mark = square},
	{red,thick,solid, mark = o},
	{blue,thick, solid, mark = o},
	{red,thick,solid, mark = diamond},
	{blue,thick, solid, mark = diamond},
}
\pgfplotscreateplotcyclelist{bler_list_async_2}{
	{red,thick,dashed, mark = triangle, mark options=solid},
	{blue,thick, dashed, mark = triangle, mark options=solid},
	{red,thick,dashed, mark = square, mark options=solid},
	{blue,thick, dashed, mark = square, mark options=solid},
	{red,thick,dashed, mark = o, mark options=solid},
	{blue,thick, dashed, mark = o, mark options=solid},
	{red,thick,dashed, mark = diamond, mark options=solid},
	{blue,thick, dashed, mark = diamond, mark options=solid},
}
\pgfplotscreateplotcyclelist{mse_list_1}{
	{red,thick,solid, mark = o},
	{blue,thick, solid, mark = o},
	{red,thick,dashed, mark = o, mark options=solid},
	{blue,thick, dashed, mark = o, mark options=solid},
}
\pgfplotscreateplotcyclelist{vtvchantf_list_1}{
	{red, only marks,mark=asterisk},
	{blue, only marks,mark=asterisk},
	{black, only marks,mark=asterisk},
	{magenta, only marks,mark=asterisk},
	{green, only marks,mark=asterisk},
	{red, only marks,mark=triangle},
	{blue, only marks,mark=triangle},
	{black, only marks,mark=triangle},
	{magenta, only marks,mark=triangle},
	{green, only marks,mark=triangle},
	{red, only marks,mark=square},
	{blue, only marks,mark=square},
	{black, only marks,mark=square},
	{magenta, only marks,mark=square},
	{red,thick,dashed, },
	{blue,thick, dashed},
	{black,thick, dashed},
}

\pgfplotscreateplotcyclelist{mi_list_1}{
	{black,thick,solid},
	{red,thick,solid, mark = triangle},
	{blue,thick, solid, mark = triangle},
	{red,thick,solid, mark = square},
	{blue,thick, solid, mark = square},
	{red,thick,solid, mark = o},
	{blue,thick, solid, mark = o},
	{red,thick,dashed, mark = triangle, mark options=solid},
	{blue,thick, dashed, mark = triangle, mark options=solid},
	{red,thick,dashed, mark = square, mark options=solid},
	{blue,thick, dashed, mark = square, mark options=solid},
	{red,thick,dashed, mark = o, mark options=solid},
	{blue,thick, dashed, mark = o, mark options=solid},}

\pgfplotscreateplotcyclelist{bler_list_2}{
	{red,thick,solid, mark = square},
	{blue,thick, solid, mark = square},
	{red,thick,solid, mark = o},
	{blue,thick, solid, mark = o},
	{red,thick,solid, mark = diamond},
	{blue,thick, solid, mark = diamond},
	{red,thick,dashed, mark = square, mark options=solid},
	{blue,thick, dashed, mark = square, mark options=solid},
	{red,thick,dashed, mark = o, mark options=solid},
	{blue,thick, dashed, mark = o, mark options=solid},
	{red,thick,dashed, mark = diamond, mark options=solid},
	{blue,thick, dashed, mark = diamond, mark options=solid},
}
\begin{document}
\title{\textcolor{black}{Pulse Shaped OFDM for 5G Systems}}

\vspace{3cm}
\author{\IEEEauthorblockN{Zhao Zhao \footnote{Corresponding author: Zhao Zhao, 
erc.zhaozhao@huawei.com.}, 
      Malte Schellmann, Xitao Gong, Qi Wang, \\ Ronald B\"{o}hnke, and Yan Guo} \\
\IEEEauthorblockA{ \normalsize{German Research Center, 
Huawei Technologies Duesseldorf GmbH\\
Riesstr. 25, 80992 Munich, Germany\\
Email: \{erc.zhaozhao, malte.schellmann, xitao.gong, qi.wang1, ronald.boehnke\}@huawei.com}}}
  
\maketitle
\section*{Abstract} \addcontentsline{toc}{section}{Abstract}
OFDM-based waveforms with filtering or \textcolor{black}{windowing} functionalities are considered key enablers for a flexible
air-interface design for multi-service support in future \textcolor{black}{5G} systems. One candidate from this category of waveforms is pulse shaped OFDM, which follows the idea of subcarrier filtering and aims at fully maintaining the advantages of standard OFDM systems while addressing their drawbacks. In this paper, we elaborate on several pulse shaping design methods, and show how pulse shapes can be exploited as an additional degree of freedom to provide 
better frequency localization and more efficient spectrum utilization under a pre-defined spectrum mask.
The performance analysis and evaluation results show that, for a practical mobile communication systems, the application of pulse shaping is a simple and effective means to achieve a lower out-of-band leakage for OFDM systems at virtually no costs. In fact, the complexity increase amounts to $1-2\%$ compared to CP-OFDM only. In addition, the system's robustness against both time and frequency distortions is shown to be substantially improved by 
a proper pulse shape design. By allowing for the flexible
configuration of physical layer parameters per sub-band according to the diverse requirements of future 5G services, pulse shaped OFDM systems can efficiently facilitate asynchronous transmissions and fragmented spectrum access, rendering it beneficial for various
mobile-broadband and Internet-of-Things applications.

\begin{IEEEkeywords}
OFDM, pulse shaping, multi-carrier, Filter Bank.
\end{IEEEkeywords}

\IEEEpeerreviewmaketitle


%
\section{Introduction}
\textcolor{black}{Conventional cyclic prefix (CP)-OFDM is the widely-used multicarrier waveform in the current communication systems, such as 4G LTE/LTE-A. Despite its favourable features of enabling low-complexity equalization in fading channels and its simple implementation, it exhibits several drawbacks, such as the poor localization of the signal in frequency domain, yielding high out-of-band leakage (OOB) caused by the large sidelobes of the sinc-pulses used per subcarrier, and the loss in spectral and energy efficiency due to the CP overhead. 
Specifically, CP-OFDM is robust to the time dispersion only up to CP duration, but vulnerable to frequency dispersion of the channels, such as Doppler spread, carrier frequency offset (CFO) and phase noise. Based on the above properties, CP-OFDM signals are subject to distortions in the popular scenarios considered for 5G cellular systems, for instance, machine-type and vehicular to anything (V2X) communications.}

Since new scenarios and use cases impose new requirements to the 5G system, the waveform topic has been shifted into the research focus again. More specifically, enhanced mobile broadband services (eMBB), together with mission critical communication (MCC), and massive machine communication (MMC) have diverse requirements on the spectral efficiency, spectral containment, processing latency and synchronization \cite{METIS2013, 5GNOW2013}, posing new challenges to the waveform design.
Due to the flexibility in multiple access, MIMO compatibility as well as efficient implementations,
it is believed that OFDM based waveforms are the most promising waveform proposals for 5G, building on the success in 4G LTE and many other systems. 

This paper is dedicated to exploiting an additional degree of freedom of OFDM systems, namely the pulse shaping design, showing that it is possible to \textcolor{black}{improve frequency localization and spectral coexistence of OFDM systems. Furthermore, time-frequency robustness can be adjusted according to different scenario requirements. 
If properly designed, OFDM with pulse shaping could lead to both link and system benefits for wireless communication systems.}

The paper is organized as follows: Section \ref{sec:ofdmsystem} will introduce the system model with pulse shaped OFDM and state-of-the art OFDM systems. Section \ref{sec:pulseshapedesign} gives the principles for OFDM pulse shape design and some practical methods. Section \ref{sec:pulseshapeexample} evaluates the pulse shape examples designed in Section \ref{sec:pulseshapedesign} and Section \ref{sec:implementation} addresses the practical implementation and system impacts. Some application examples are illustrated in Section \ref{sec:appexamples}. Finally, section \ref{sec:conclusion} draws the conclusion for the paper.

\section{OFDM System and Pulse Shaping}
\label{sec:ofdmsystem}
In this section, a generic OFDM system model with pulse shaping is introduced.
The design methodology and the impact on the OFDM numerologies are briefly discussed.
\subsection{System Model}
\label{sec:systemmodel}
The transmit signal $s\left(t\right)$ of an OFDM-based multicarrier system can be generally represented as follows
\cite{MatzTWC2007}:
\begin{equation}
\label{eq:txmodel}
s\left(t\right)=\sum\limits_{n=-\infty}^{\infty}\sum_{m=0}^{M_A-1}a_{m,n}g_{m,n}\left(t\right)
\end{equation}
where 
$a_{m,n}$ is the information bearing symbol on the $m$th subcarrier of the $n$th symbol. $M_A$ is the number of active subcarriers. 
The transmit filter bank $g_{m,n}\left(t\right)$ is a time-frequency shifted version of the transmit pulse shape (also known as prototype filter)\footnote{Pulse shape and prototype filter are used interchangeably throughout the paper.} $g\left(t\right)$, i.e., 
\begin{equation}
\label{eq:gmn}
g_{m,n}\left(t\right)=g\left(t-nT\right)e^{j2\pi mF\left(t-nT\right)}.
\end{equation}
Note that subband-based filtering can be used on top of $s\left(t\right)$ with a band-pass filter, in order to further suppress the out-of-band (OOB) leakage. 

At the receiver side, the demodulated symbol $\tilde{a}_{m,n}$ is obtained by correlating the received signal $r\left(t\right)$ with the receive filter $\gamma_{m,n}\left(t\right)$:
\begin{equation}
\label{eq:rxmodel}
\tilde{a}_{m,n}=\left\langle r,\gamma_{m,n}\right\rangle=\int_{-\infty}^{\infty}r\left(t\right)\gamma_{m,n}^*\left(t\right)dt.
\end{equation}
where $\gamma_{m,n}\left(t\right)$ is a time-frequency shifted version of the receive pulse $\gamma\left(t\right)$\footnote{In this paper, we assume the power of both transceiver pulses $g(t)$ and $\gamma(t)$ are normalized to one.}
\begin{equation}
\label{eq:gammamn}
\gamma_{m,n}\left(t\right)=\gamma\left(t-nT\right)e^{j2\pi mF\left(t-nT\right)}
\end{equation}
In short, a generic OFDM-based system with pulse shaping can be presented by the following steps: the transmit signal is first synthesized using (\ref{eq:txmodel}), passes through propagation channels, and then analyzed at the receiver through (\ref{eq:rxmodel}). 

If the pulses employed at the transmitter and the receiver are the same, i.e., $g(t)=\gamma(t)$, the approach is \textit{matched filtering} \cite{SahinSurvey2014}. Alternatively, different pulses can be used at the transmitter and the receiver, i.e., $g(t) \neq \gamma(t)$, yielding the \textit{mis-matched filtering}. Generally, matched filtering aims at maximizing the SNR for AWGN channel, while mis-matched filtering may provide more flexibility to combat the inter-symbol interference (ISI) and inter-carrier interference (ICI) in doubly dispersive channels at the expense of suffering from noise enhancement.

Different from conventional CP-OFDM where pulse shapes are fixed to the rectangular pulses, pulse shaped OFDM (P-OFDM) follows the idea to fully maintain the signal structure of CP-OFDM, but allowing for the use of flexible pulse shapes other than the rectangular pulse to balance the localization of the signal power in time and frequency domain. Prototype filters $g(t)$ and $\gamma(t)$, together with the numerology $T$ and $F$, are the central design parameters for our pulse shaped OFDM system. 

\begin{figure}[!t]
	\centering
	\includegraphics[scale=0.5]{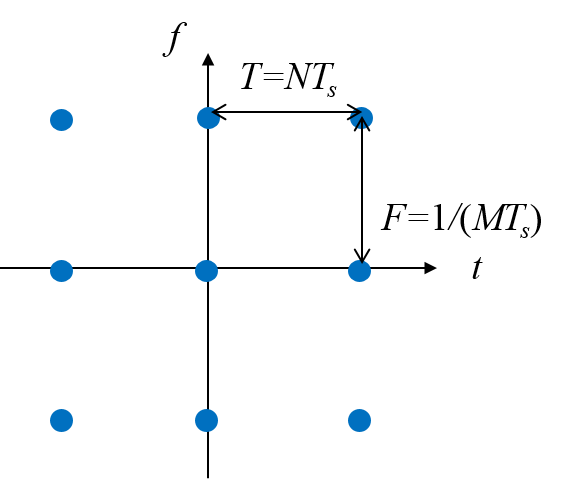}
	\caption{Rectangular lattice for QAM systems.} 
	\label{fig:lattQAM}
	\vspace{-0.8cm}
\end{figure}

A useful representation of numerology design is a lattice that contains the coordinates in the time-frequency plane. Assuming the symbol period is $T = N T_s$ and subcarrier spacing is set to $F = {1}/{M T_s}$, where $T_s$ is the sampling time and $M, N\in\mathbb{N}$ denotes the FFT size and the number of samples in one symbol period, respectively, Fig. \ref{fig:lattQAM} depicts the rectangular lattice representation for pulse shaped OFDM. The metric $1/TF$ can be considered as the data symbol density in rectangular sampling lattice and it is proportional to the spectral efficiency.
In this paper, we choose the numerology $T$ and $F$ such that $TF={N}/{M}>1$ holds \cite{MatzTWC2007}. Under this condition, orthogonality can be guaranteed for the signal space, yielding the full compatibility to current techniques developed for OFDM.

Pulse shaped OFDM allows the pulse shape to extend over the symbol period $T$, rendering successively transmitted symbols to overlap or partially overlap. 
The overlap is characterized by the overlapping factor $K$, which is defined as the ratio of filter length $L_g$ and the symbol period, i.e. $K=L_g/N$. The factor $K$ can be set to any rational number in pulse shaped OFDM. 

\subsection{Transceiver of Pulse Shaped OFDM}
\begin{figure}[!htbp]
\vspace{-0.5cm}
	\centering
	\includegraphics[scale=0.62]{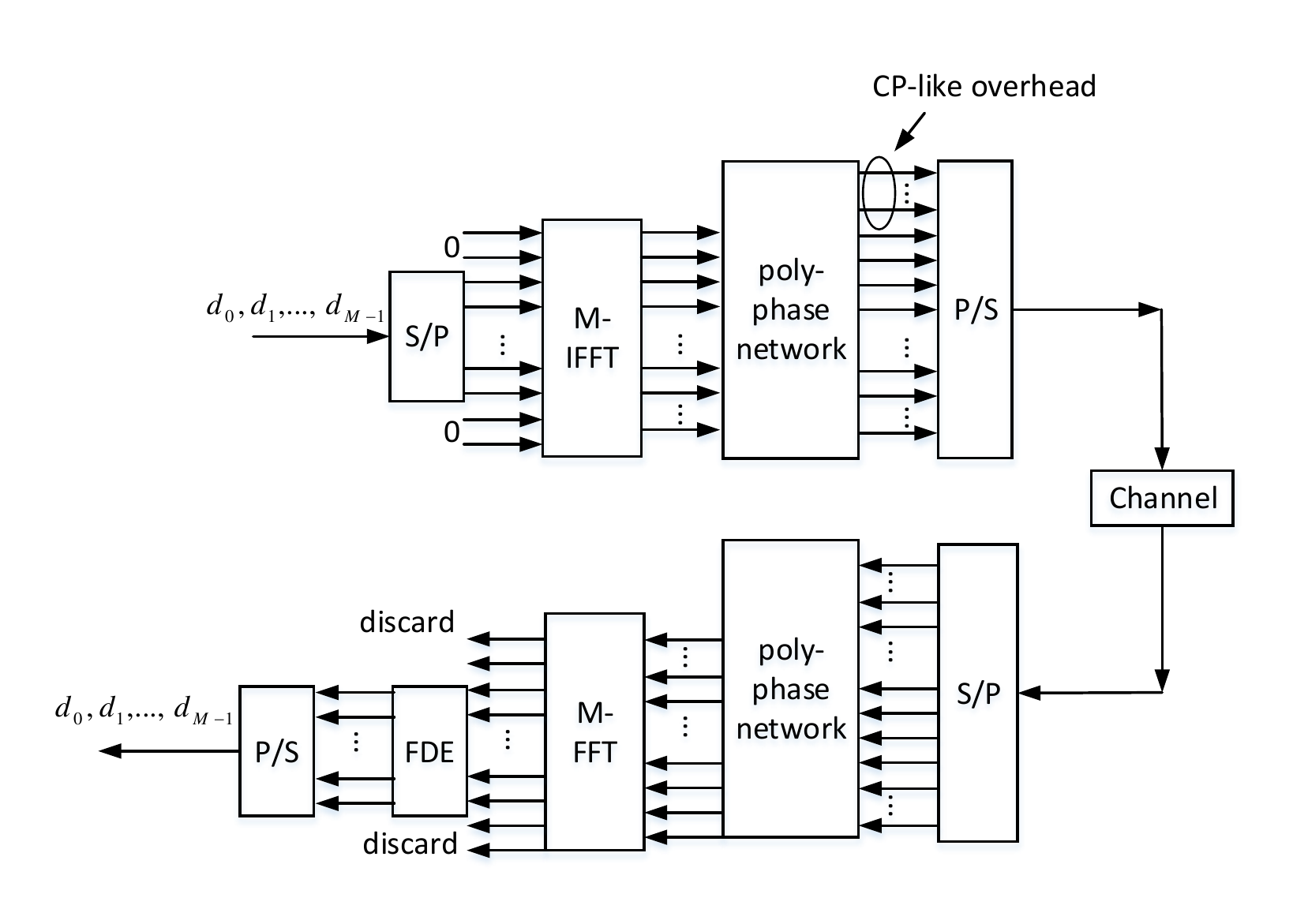}
	\caption{Pulse shaped OFDM transceiver structure with efficient implementation of pulse shaping by a polyphase network.} 
	\label{fig:transceiver}
\vspace{-0.8cm}
\end{figure}

As a typical uniform Filter Bank system, the overall transceiver structure of pulse shaped OFDM system is given in Fig. \ref{fig:transceiver}. The pulse shaping can be efficiently realized by polyphase network (PPN) \cite{VaidyanathanProcIEEE1990}
for arbitrary overlapping factor $K$. For short pulse shapes where $K \approx 1$, the PPN structure can be simplified to the "CP addition", "CP removing", "zero-padding" or "windowing" operations, etc. \textcolor{black}{For $K>1$, PPN implementation can be considered as a realization of overlap add.}

\subsection{State-of-the-art OFDM Systems and Numerology Design}
\label{sec:numerdesign}
Numerology design for multicarrier systems, including the determination of symbol period $T$ and subcarrier spacing $F$, is an essential part in system design. Its design needs a comprehensive consideration of many aspects, e.g., spectrum efficiency or propagation channel characteristics. In this section, we will introduce the numerology design of state-of-the-art OFDM based systems, where their
overview is provided in Appendix \ref{sec:relationmc}. Note that these systems can be all considered as special cases of the pulse shaped OFDM framework.
\subsubsection{CP-OFDM}
\label{sec:numerdesigncpofdm}
The derivation of OFDM numerology w.r.t. $(T, F)$ can be carried out by the following steps:
\begin{itemize}
\item Set the CP-length $T_{\text{cp}}$ in seconds according to channel characteristics, i.e., at least longer than the maximum channel excess delay $\tau_\text{max}$.
\begin{equation}
\label{eq:sotaofdm1}
T_{\text{cp}} = T-\frac{1}{F} \geq \tau_\text{max}.
\end{equation}
\item Determine the minimal subcarrier spacing $F$ in Hertz, in the way that the signal-to-interference ratio (SIR) in the case of maximum Doppler frequency ($\nu_{\text{max}}$) is higher than the minimal SIR requirement ($\text{SIR}_\text{min}$) supporting highest modulation requirement in the system. 
\item Determine the approximate values of $T$ and $F$ based on the above two steps.
\item Quantize $T$ and $F$ according to the sampling rate and sub-frame numerology.
\end{itemize}
The above steps are based on the speculation that CP-OFDM can support  the reliable transmission with no introduction of ISI nor ICI
when the maximum excess delay of the channel is smaller than the CP-length. It  is a pragmatic approach, since CP-OFDM provides more pronounced robustness in time domain than in the frequency domain. 
\subsubsection{W-OFDM}
Windowed-OFDM (W-OFDM) is originally introduced as an enhancement to CP-OFDM for reducing the OOB emission. Currently W-OFDM is often implemented in practice by the vendors to enable such favorable features, as long as the error vector magnitude (EVM) distortion (Appendix \ref{sec:appendixevm}) is within the system tolerance. 

Essentially, W-OFDM is a pulse shaped OFDM system, with a temporal-smoothing window instead of a rectangular one (used in CP-OFDM) is introduced to effectively reduce the side lobe. 
For link performance perspective, W-OFDM is aiming to trade off its robustness in the time domain (due to the relaxation of CP) for a better robustness in the frequency domain. We will show later that within general mobile system operational range, 
a proper designed W-OFDM can outperform CP-OFDM in overall time-frequency operational range and better fulfill  time-frequency (TF) localization requirements.

Normally, a windowing operation has a marginal impact on the system's robustness against time-frequency dispersions. Therefore, the numerology design for CP-OFDM can be similarly applied to W-OFDM. Windowing effect can additionally be compensated by slightly adapting $(T, F)$ according to the specific system requirement. 
\subsubsection{TF Localized OFDM}
Given the same spectral efficiency overhead, it has been shown that the link level performance can be improved over conventional CP-OFDM and its pragmatic $(T,F)$ numerology design
\cite{MatzTWC2007,StrohmerTCOM2003}. One solution is the TF localized OFDM aiming at minimizing the distortion resulting from time-frequency dispersive channels \cite{StrohmerTCOM2003}. The numerology design of this waveform can be comprised of following steps:
\begin{itemize}
\item Determine the ratio of $T$ and $F$: In order to reduce ISI and ICI, it has been suggested that the numerology $T$ and $F$ of TF-localized OFDM should be chosen such that they are adapted to the doubly dispersive channel property. Specifically, for given maximal time delay $\tau_\mathrm{max}$ and maximal Doppler spread $\nu_\mathrm{max}$, the choice of $T$ and $F$ should satisfy \cite{StrohmerTCOM2003}
\begin{equation}
\frac{T}{F}=\frac{\tau_\mathrm{max}}{\nu_\mathrm{max}}.
\end{equation}
\item If aiming at a fair comparison with CP-OFDM, identify the product of $TF$ according to the relative CP-overhead or spectral efficiency loss. Combined with the ratio of $T$ and $F$ in the first step, parameters $T$ and $F$ are obtained.
Otherwise, if $TF$ is not specified, the numerology needs to be defined as follows: first, initialize $TF$ with some pre-defined number, then calculate SIR using TF-localized pulses in case of maximum excess delay and Doppler frequency; finally, adapt the numerology $T$ and $F$ to guarantee the resulting SIR can support the highest modulation transmission.    
\end{itemize}
\section{OFDM Pulse Shape Design and Proposed Methods}
\label{sec:pulseshapedesign}
Future 5G mobile communication systems are envisioned to support
the coexistence of multi-service with diverse requirements. PHY setting including pulse shape configuration is thus anticipated to be adapted to different requirements for each service. For example, MCC requires low latency, yielding comparably short pulse is favorable. Narrowband internet-of-things (NB-IOT) service targets at good coverage extension and allows long pulse design. Machine-type communication (MTC) with mobility may require pulse design to be robust to asynchronicity and Doppler. In this section, we discuss several pulse shape design approaches and outline their features and applications.
\subsection{Pulse Shape Categorization}
In OFDM systems with pulse shaping, the ISI and ICI are determined by the transmit pulse $g(t)$ and the receive pulse $\gamma(t)$. In this paper, we use the pulse shape categorization according to the correlation property \cite{SahinSurvey2014}:
\begin{itemize}
\item \textit{Orthogonal pulse design} is the pulse shaping scheme when perfect reconstruction condition is fulfilled (details given in Sec. \ref{sec:pr}), and matched filtering is employed.
\item \textit{Bi-orthogonal pulse design} is the pulse shaping scheme when perfect reconstruction condition is fulfilled and mis-matched filtering is employed.
\item \textit{Nonorthogonal pulse design} is the pulse shaping scheme when perfect reconstruction condition is not fulfilled.
\end{itemize}
\subsection{Design Criteria}
Depending on the specific criteria for pulse shaped OFDM systems, pulse shapes are constructed to satisfy diverse requirements. Herein we review several commonly applied conditions that the pulse shape design needs to fulfill.
\subsubsection{Length Constraint}
Length constraint is the primary design criterion for the OFDM pulses shapes, as in many of the uses cases such as eMBB and MCC, large processing latency is not acceptable for the system as stringent timing is required for framed transmission. In such cases, waveform should not introduce large latency to the system and the pulse shapes of OFDM in this case should be constrained to the length comparable to one OFDM symbol duration (i.e. $K \simeq 1$). In other scenarios such as MMC or NB-IOT, this latency constraint may be relaxed to the length of several symbols ($K \geq 2$) if benefits can be shown in other aspects. 
\subsubsection{Orthogonal and Bi-orthogonal Condition}
\label{sec:pr}
Assuming an ideal channel where $r\left(t\right)=s\left(t\right)$,
perfect reconstruction condition holds if
$\left\langle g_{m^\prime,n^\prime},\gamma_{m,n}\right\rangle=\delta_{m^\prime m}\delta_{n^\prime n}$,
where $g_{m,n}$ and $\gamma_{m,n}$ follow the definition in (\ref{eq:gmn}) and (\ref{eq:gammamn}), respectively. Due to the practical issue encountered in wireless communications that certain level of self-interference is tolerated for the reliable transmission of modulated signals, we slightly relax the conventional PR condition in this paper to allow minor cross-correlation, i.e. 
\begin{align}
\label{eq:pr}
\left\langle g_{m^\prime,n^\prime},\gamma_{m,n}\right\rangle 
\begin{cases}
 = 1  \quad \quad  {m^\prime =  m} \text{ and } n^\prime = n \\
 \leq \epsilon \hspace{0.9cm} {m^\prime \neq  m} \text{ or } n^\prime \neq n
\end{cases}
\end{align}
where $\epsilon$ is determined according to the
EVM and SINR requirement in Appendix \ref{sec:appendixevm}. (\ref{eq:pr}) is also named as bi-orthogonality condition if assuming $g(t) \neq \gamma(t)$, as it is a prerequisite in bi-orthogonal division multiplexing (BFDM) system for perfectly reconstructing $\tilde{a}_{m,n}$ from $r\left(t\right)$ \cite{Farhang-Boroujeny2011}. Under the condition that matched filtering is employed, i.e. $g(t) =  \gamma(t)$, (\ref{eq:pr}) reduces to the orthogonality condition.
Opposing to the orthogonal transceiver pulse that SNR for AWGN channel is maximized, BFDM has a potential to further reduce ISI and ICI for dispersive channels at the cost of a noise enhancement. 
It has been shown in \cite{MatzTWC2007,StrohmerTCOM2003} that a necessary condition to achieve perfect reconstruction (either orthogonal or bi-orthogonal) is $TF \geq 1$. Larger values of $TF$ leads to larger spectral efficiency loss but more degree of freedom for pulse design.
\subsubsection{Time Frequency Localization}
\textcolor{black}{In general, the ISI and ICI can be reduced if both the pulse shapes at the transmitter and receiver are well TF localized \cite{MatzTWC2007}, thus yielding higher overall SINR.} The classical way to measure time-frequency localization (TFL)
of a filter involves the Heisenberg uncertainty parameter \cite{StrohmerTCOM2003, SahinSurvey2014}. Filters with good TFL properties have a Heisenberg parameter $\xi$ closer to 1. 
Assuming the center of gravity of $g\left(t\right)$ is at $\left(0,0\right)$, the ``width" of $g$ in the time and frequency domain is often measured using the second-order moments defined as
\begin{equation*}
\sigma_{t}=\left(\int_{-\infty}^{+\infty}t^2\left|g\left(t\right)\right|^2dt\right)^{1/2},\quad \sigma_f=\left(\int_{-\infty}^{+\infty}f^2\left|G\left(f\right)\right|^2df\right)^{1/2},
\end{equation*}
where $G\left(f\right)$ is the Fourier transform of $g\left(t\right)$. Then, the Heisenberg uncertainty parameter $\xi$ is given by 
\begin{align}
\label{eq:heisenpara}
\xi = \dfrac{\|g(t)\|^2}{4\pi\sigma_{t}\sigma_{f}} \leq 1
\end{align}
where equality holds if and only if $g(t)$ is a Gaussian function, rendering such filter leads to optimal TFL\cite{StrohmerTCOM2003}.

Note that the joint TFL of transceiver pulses considering channel dispersion is related to the TF concentration properties of both transmit and receive pulses. The work in \cite{MatzTWC2007} has asserted that excellent TFL characteristics can be simultaneously achieved by pulse pairs.
\subsubsection{SIR/SINR Optimization}
In wireless communication systems, the essential goal is to transmit signals reliably in practical channels. Hence, the above criteria can be slightly relaxed to increase the design degree of freedom, as long as the link performance with pulse shaping is optimized relative to certain dispersive channels of interest. 

One common criterion is SIR or signal-to-interference-plus-noise ratio (SINR) optimization, namely, the transceiver pulses are chosen to optimize the SIR/SINR under certain dispersive channels. The resulting pulse shapes may not be exactly satisfying perfect reconstruction condition, but offer better ISI/ICI robustness in dispersive channels compared to orthogonal/bi-orthogonal design.
In the following, we detail this optimization problem for both continuous and discrete channel models.
\begin{itemize}
\item Continuous Model:
Assuming a doubly dispersive fading channel satisfying
wide-sense stationary uncorrelated scattering (WSSUS) property, its scattering function (channel statistics) can be described as \cite{MatzTWC2007}
\begin{equation}
C_{\mathbb H}(\tau, \nu) = \int_{\Delta t} E [{h(t, \tau) h^*(t+\Delta t, \tau)}] e^{-j2\pi \nu \Delta t} d\Delta t,
\end{equation}
where $h(t, \tau)$ is the time-varying impulse response at time instance $t$ and delay $\tau$, $\nu$ is the Doppler frequency, and $\mathbb H$ indicates the random linear time-varying channel. We call the WSSUS channel $h\left(t,\tau\right)$ underspread, if the support of $C_{\mathbb H}(\tau, \nu)$ is constrained in a rectangular region of $\{-\tau_\text{max}\le\tau\le\tau_\text{max}, -\nu_\text{max}\le\nu\le\nu_\text{max}\}$ with
$\tau_\text{max} \nu_\text{max} \ll 1$.

The SINR involving pulse shaping is represented by 
\begin{equation}\label{eq:SINR_continuous}
\text{SINR}\mathtt{C}_{g,\gamma} (\mathbb H) = \frac{\int_\tau \int_\nu C_{\mathbb H}(\tau, \nu) |A_{g,\gamma}(\tau, \nu) |^2 d\tau d\nu}{\int_\tau \int_\nu Q_{\mathbb H}^{(0)}(\tau, \nu) |A_{g,\gamma}(\tau, \nu) |^2 d\tau d\nu+\sigma_\mathrm{n}^2},
\end{equation}
where $\sigma_\mathrm{n}^2$ is the noise variance, $A_{g,\gamma}(\tau, \nu) = \int_t \gamma(t) g^{*}(t-
\tau)e^{-j2\pi \nu tdt}$ is the cross-ambiguity function of transceiver pulse pair $\langle g\left(t\right),\gamma\left(t\right)\rangle$ \cite{MatzTWC2007}, $Q_{\mathbb H}^{(0)} (\tau, \nu)$ is defined as 
\begin{equation*}
Q_{\mathbb H}^{(0)} (\tau, \nu) \triangleq \sum_{(m, n)\neq (0, 0)} C_{\mathbb H}(\tau-n T, \nu - mF).
\end{equation*}
If the noise variance part $\sigma_\mathrm{n}^2$ is omitted in (\ref{eq:SINR_continuous}), it reduces to SIR metric. 
\item Discrete Model: 
Assuming the dispersive channel has $P$ paths, and the $p$th path is characterized by $\tau_p$, $\nu_p$, and $\eta_p$ representing the corresponding delay, Doppler frequency shift, and complex gain, respectively. 
Let $\boldsymbol{\eta}=\left[\eta_1,\dots,\eta_P\right]^T$ denote the channel impulse response with $P$ paths, where $\eta_p$ is the complex gain of the $p$th path,  $p = 1 \dots P$. Under the assumption of WSSUS property, the channel scattering function $C_{\mathbb{H}}$ can be further described by a diagonal matrix
$\mathbf{C}_\mathbb{H}=\mathbb{E}\{\boldsymbol{\eta}\boldsymbol{\eta}^H\}.
$ 
For a doubly dispersive channel, the scattering function satisfies $\mathbf{C}_\mathbb{H}\in\mathbb{R}^{P\times P}$. 
Assuming the transmit and receive filters are discretized as well as their power are normalized to one, it yields $\mathbf{g}$ and $\boldsymbol{\gamma}$ as the vectors containing discrete filter coefficients,
$\mathbf{g}\in\mathbb{R}^{L_T\times 1}$, 
$\boldsymbol{\gamma}\in\mathbb{R}^{L_R\times 1}$, and
$L_T$ and $L_R$ denoting the required filter length for transmit and receive prototype filter, respectively. Using the above discrete expressions, 
the discrete counterpart of \eqref{eq:SINR_continuous} is 
\begin{equation}\label{eq:SINR_discrete}
\text{SINR}^\mathtt{D}_{\mathbf{g},\boldsymbol\gamma}\left(\mathbb{H}\right)=\frac{\boldsymbol{\gamma}^H\mathbf{G}_{0,0}\mathbf{C}_\mathbb{H}\mathbf{G}_{0,0}^{H}\boldsymbol{\gamma}}{\boldsymbol{\gamma}^H\left(\sum_{\left(m,n\right)\ne\left(0,0\right)}\mathbf{G}_{m,n}\mathbf{C}_\mathbb{H}\mathbf{G}_{m,n}^{H}\right)\boldsymbol{\gamma}+\sigma_\mathrm{n}^2},
\end{equation}
where $(\cdot)^{H}$ denotes the Hermitian operation.
\textcolor{black}{The $p$th column of $\mathbf{G}_{m,n}\in\mathbb{C}^{L_R\times P}$ contains the $p$th path propagated version of $\mathbf{g}_{m,n}$ (discrete delays/Doppler shifts in samples depend on the sampling rate). The notation  
$\mathbf{g}_{m,n}$ denotes the time-frequency shifted version of $\mathbf{g}$, analogous to its continuous counterpart 
(\ref{eq:gmn}).}
If the noise variance part $\sigma_\mathrm{n}^2$ is omitted in (\ref{eq:SINR_discrete}), it reduces to SIR metric. 
\end{itemize}
After formulating the SIR/SINR metric in both models, pulse shapes $g(t)$ and $\gamma(t)$ are jointly optimized to maximize this metric. Since SIR/SINR optimization requires channel statistics (e.g. scattering function) which may be unknown, some default scattering function (e.g., brick-shaped function \cite{MatzTWC2007}) could be assumed. 
\subsection{Design Methods}
Taking above-mentioned design criteria into consideration, the ultimate goal of pulse design is to have short pulses with maximal spectral efficiency, optimal time frequency localization, minimized interference and best SINR performance for arbitrary channels. Nevertheless, not all the requirements can be fulfilled simultaneously in reality, either due to contradictory conditions or practical constraints. Alternatively, in this section, we propose two approaches to design the transceiver pulse shapes for practical pulse shaped OFDM systems, where both can respect the arbitrary given length constraint. 
\begin{enumerate}
	\item Orthogonal design w/o channel statistics: For the case that the system has no reliable channel statistics for pulse optimization, \textcolor{black}{we seek to apply (almost) orthogonal transceiver pulse pair with good time frequency localization.}
	\item \textcolor{black}{Bi-orthogonal design} with channel statistics: For the case that the system has reliable channel statistics (e.g. scattering function) for pulse optimization, transceiver pulses are designed to achieve optimal link level performance (w.r.t. SIR/SINR) given such channels. 
\end{enumerate}
\subsubsection{Orthogonal Pulse Design w/o Channel Statistics}
As introduced before, orthogonal pulse design employs matched filtering at the transceiver in order to achieve the maximum SNR for AWGN channels. In the absence of channel statistics, we suggest to use such orthogonal pulse design with good TFL characteristic. The TFL property 
is desirable since the ISI/ICI over doubly dispersive channels is
related by the TFL of pulses. Note that $TF > 1$ is assumed here to increase the freedom for pulse design. In the following, a universal approach for producing orthogonal
pulses with constrained length as well as good
TFL has been proposed. 

Before detailing the proposed method, we first review the orthogonal pulse generation in the literature, which provides a basis for our proposal. The classical approach in \cite{MatzTWC2007,StrohmerTCOM2003} consists of the following steps.
\begin{itemize}
\item Select an initial well-localized pulse, e.g. a Gaussian pulse with a decaying factor $\alpha$.
$$g_\mathtt{gauss}^{\left(\alpha\right)}\left(t\right)=\left(2\alpha\right)^{1/4} e^{-\pi \alpha t^2}$$
\item Construct an orthogonal system $\left(g_\perp^{\left(\alpha\right)}, T, F\right)$ based on $g_\mathtt{gauss}^{\left(\alpha\right)}$:
\begin{equation}
g_\perp^{\left(\alpha\right)}=\mathrm{orth}\{g_\mathtt{gauss}^{\left(\alpha\right)},T,F\}
\end{equation}
\end{itemize}
Orthgonalization can be constructed according to \cite{StrohmerTCOM2003} or efficient numerical solution for orthgonalization can be obtained by matrix factorization methods \cite{Feichtinger1998, Sondergaard2009}.

It is proved in \cite{StrohmerTCOM2003} that by appropriately dilating or shrinking $g^{\left(\alpha\right)}_\mathtt{gauss}$, i.e., adjusting $\alpha$, one can trivially generate the optimal TFL pulses to match different channel dispersive properties. 

\textcolor{black}{The resulting orthogonal pulse $g^{\left(\alpha\right)}_\perp$ is sometimes treated as time-unconstrained}. In practice, a time-constrained short filter is desired. In order to generate such a pulse from $g^{\left(\alpha\right)}_\perp$ given the required filter duration $D_\mathrm{req}=KT$ with $K \geq 1$, the simplest approach is to directly truncate $g^{\left(\alpha\right)}_\perp$. However, it may simultaneously lead to non-orthogonality and worse TFL. 

For producing the orthogonal prototype filters with fixed length equal to symbol duration, (i.e. $K=1$), Pinchon et al. have derived two explicit expressions to compute the filter coefficients for two different optimization criteria: minimizing OOB energy and TFL \cite{Pinchon2011}. 
Using the discretization illustrated by Fig. \ref{fig:lattQAM}, 
the derivation requires the condition $N_0=M_0+1$ where $N_0=N/\mathrm{gcd}\left(N,M\right)$ and $M_0=M/\mathrm{gcd}\left(N,M\right)$. Such constraint renders the extension to more general cases is not straightforward.

\begin{figure} [!htp]
	\centering
	\includegraphics[scale=0.3]{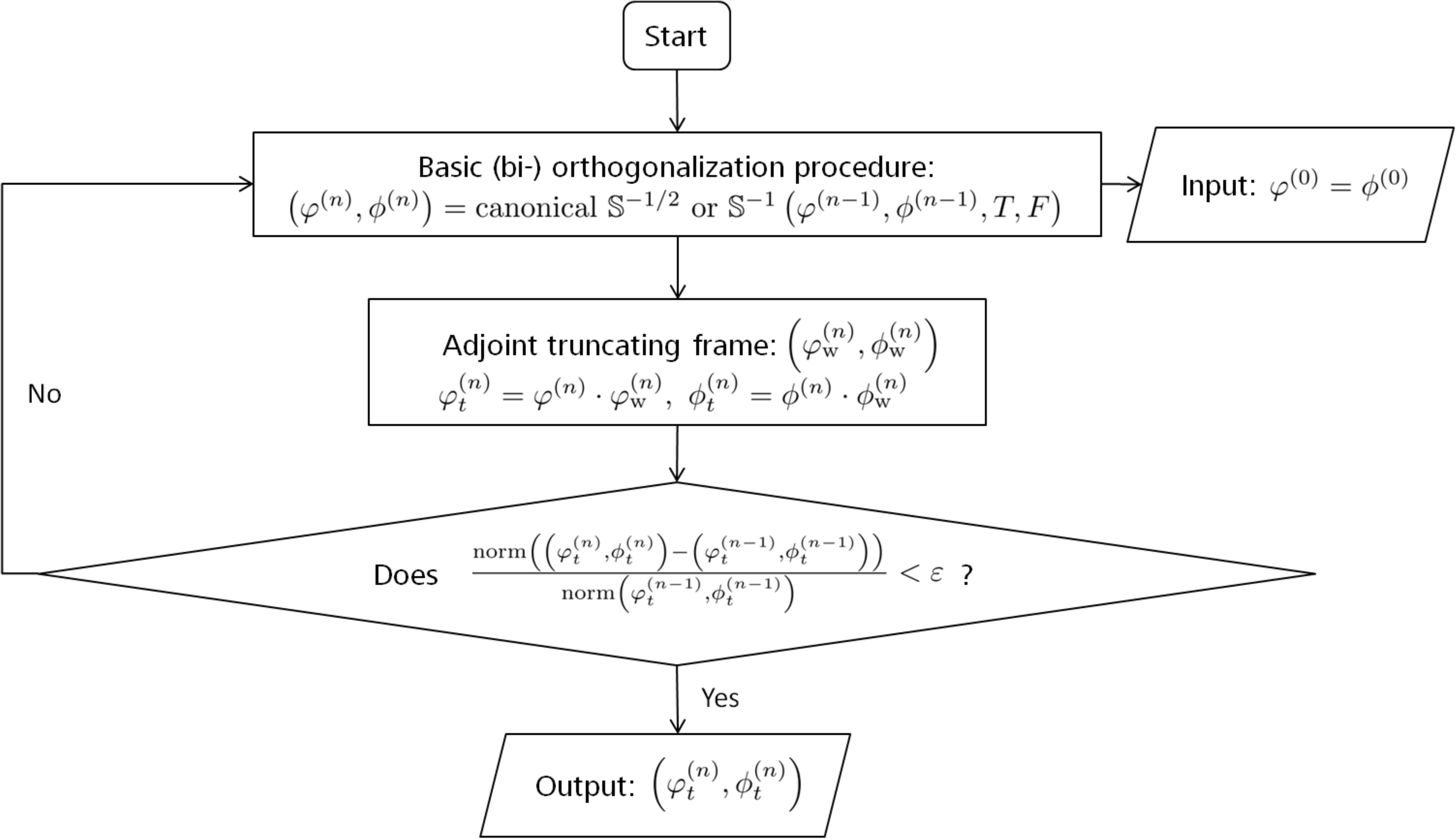}
	\caption{Proposed algorithm for constructing OFDM-based pulse with arbitrary length.} 
	\label{fig:proc}
	\vspace{-0.5cm}
\end{figure}

\begin{figure}[!htbp]
	\centering
	\includegraphics[scale=0.55]{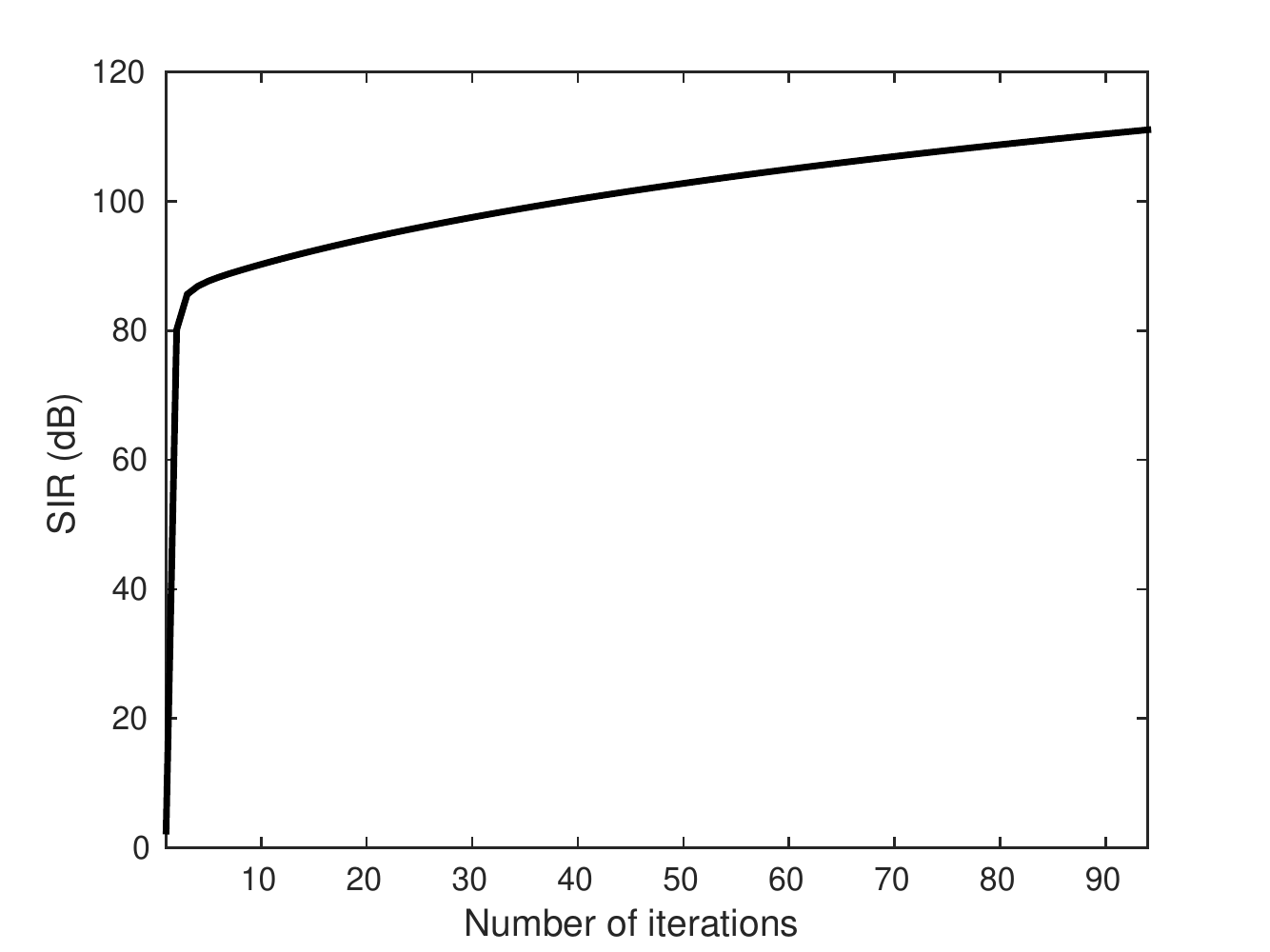}
	\caption{SIR vs. number of iterations.}
	\label{fig:SIR_per_iter}
		\vspace{-0.5cm}
\end{figure}

\begin{figure}[!htbp]
	\centering
	\includegraphics[scale=0.55]{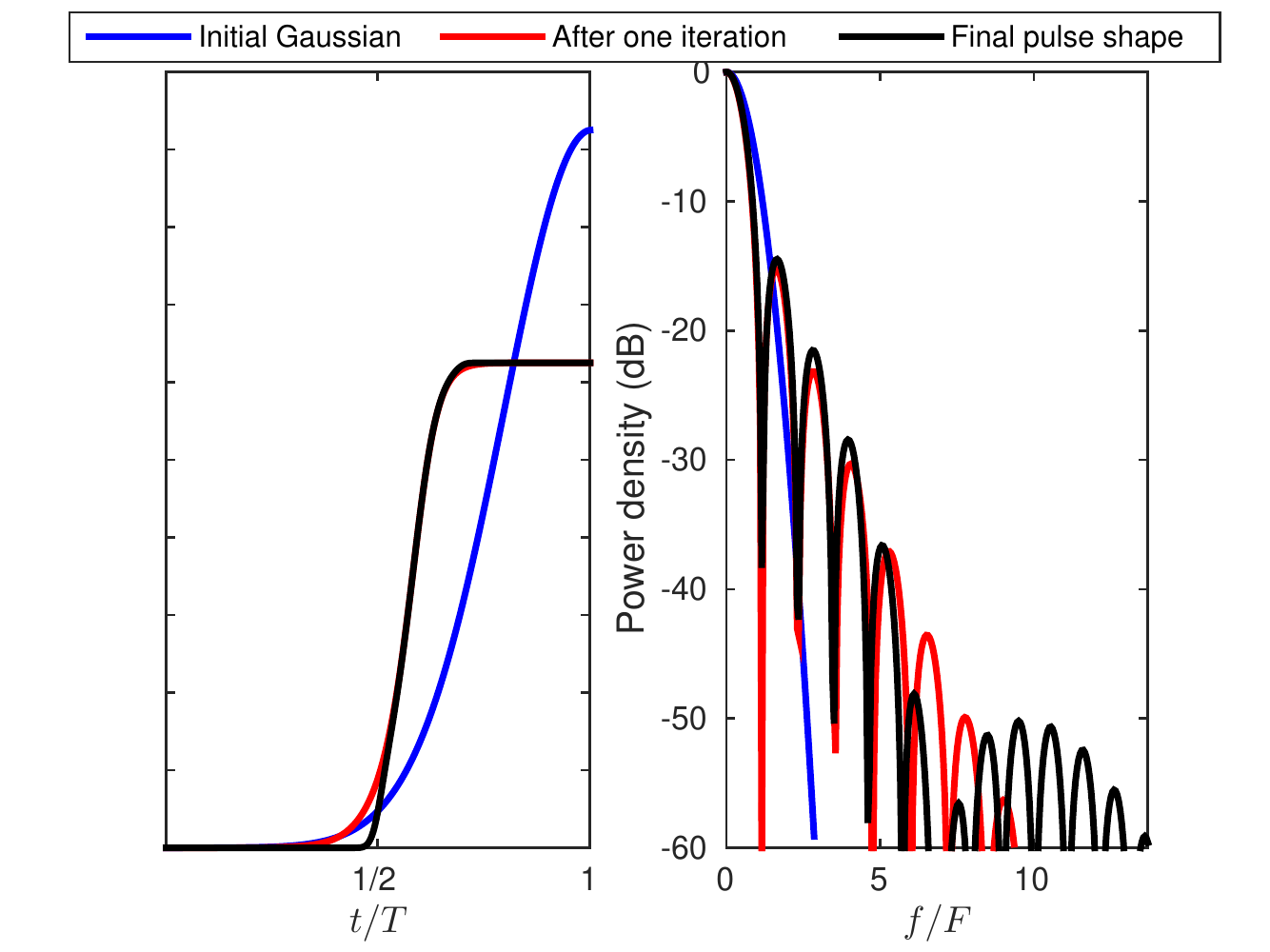}
	\caption{Time and frequency impulse responses for the initial Gaussian pulse, pulse after the first iteration, and the final pulse.}
	\label{fig:imp_freq_per_iter}
	\vspace{-0.8cm}
\end{figure}

We propose a method that aims at generating orthogonal pulses with arbitrary time constraint and maintaining good
TFL property \cite{GuoVTC2016}. Given an initial well-localized pulse,
by repeatedly performing orthogonalization and truncation,
the overall process will converge under a given convergence
criterion. A design example is described below.
Details of the algorithm are described in Algorithm \ref{algo_rx_wo_channel} and Fig. \ref{fig:proc}, which involves several essential variables:
\begin{algorithm}[!htbp]
	\caption{Iterative algorithm for constructing OFDM-based pulses with arbitrary length}\label{algo_rx_wo_channel}
	\small{\begin{algorithmic}[1]
			\STATE \textbf{Initialization}: Given $\varepsilon$ and $\alpha$. Let $n=0$ and $g^{\left(0\right)}=g^{\left(\alpha\right)}_\mathtt{gauss}$.\\
			\STATE \textbf{repeat} \{Main Loop\}\\
			\STATE\quad Compute $g^{\left(n\right)}=\left(\mathrm{orth}\{g^{\left(n-1\right)},T,F\}\right)\cdot g_\mathrm{w}^{\left(n\right)}$.\\
			\STATE\quad Let $n=n+1$.\\
			\STATE\textbf{until} $\frac{\left\|g^{\left(n\right)}-g^{\left(n-1\right)}\right\|}{\left\|g^{\left(n-1\right)}\right\|}\le\varepsilon$.\\
			\STATE \textbf{return} $g^{\left(n\right)}$. Truncate it to obtain $g$.\\
		\end{algorithmic}}
\end{algorithm}
\begin{itemize}
	\item  Initialize the pulse $g^{\left(0\right)}$: We choose a Gaussian pulse $g_\mathtt{gauss}^{(\alpha)}(t) = (2\alpha)^{1/4}e^{-\pi\alpha t^2}$ 
	as the initial pulse $g^{(0)}$ due to its optimal TFL~\cite{StrohmerTCOM2003}. This step is similar to the first step of above-mentioned standard method but described in the discrete manner. The
factor $\alpha$ determines the TFL of $g_\mathtt{gauss}^{(\alpha)}(t)$. In order to reduce ISI and ICI, it is suggested to choose $\alpha \approx \nu_{\text{max}}/\tau_{\text{max}}$~\cite{StrohmerTCOM2003}. 
In general, $\alpha$ can be adjusted to match different channel conditions.
	\item Orthogonalize $g^{(n-1)}[l]$ using the standard method, namely, by computing
	\begin{align}
	g^{(n)} = \textnormal{orth}\{ g^{(n-1)}, N, M \}.
	\end{align}
	\item Truncation is applied using a truncation window $g_{\textnormal{W}}$. The width of the window $L_{\textnormal{W}}$ corresponds to the desired pulse length. Common windows include rectangular $\left(\mathtt{RECT}\right)$, raised-cosine $\left(\mathtt{RC}\left(\beta\right)\right)$, and root raised-cosine $\left(\mathtt{RRC}\left(\beta\right)\right)$ windows, where $\beta$ is the roll-off factor. For $\beta\rightarrow 0$, $\mathtt{RC}\left(\beta\right)$ and $\mathtt{RRC}\left(\beta\right)$ converge to $\mathtt{RECT}$.
	\item Orthogonalization and truncation are iteratively applied by
	\begin{align}
	g^{(n)} = \big(\textnormal{orth}\{g^{(n-1)}, N, M\} \big)\cdot g_{\textnormal{W}}
	\end{align}
	until $\frac{\|g^{(n)} - g^{(n-1)}\|}{\|g^{(n-1)}\|} \leq \varepsilon$. The coefficient $\varepsilon$ can be interpreted as a tradeoff between orthogonality and TFL. Small $\varepsilon$ leads to higher number of iterations and improved orthogonality; large $\varepsilon$ leads to pulses with better TFL. Here, $\varepsilon$ is set to $10^{-4}$.
\end{itemize}
Both a fixed window $g_\mathrm{w}$ or an iteration-varying window can be used in the algorithm.
To illustrate the algorithm procedure, we first discuss the relationship between orthogonality and the number of iterations under a specific example, in which the orthogonality is measured by SIR. Essential parameter settings are listed in Table \ref{tab:example}. 
\begin{table}[!htbp]
	\caption{Parameter settings for one specific example.}
	\centering
	\label{tab:example}
	\renewcommand{\arraystretch}{1.2}
	\begin{tabular}{|p{.06\linewidth}|p{.06\linewidth}|p{.06\linewidth}|p{.06\linewidth}|p{.06\linewidth}|p{.06\linewidth}|}
		\hline $N$ & $M$ & $K$ & $\varepsilon$ & $g_\mathrm{w}$ & $\beta$ \\ 
		\hline $320$ & $256$ & $2$ & $10^{-4}$ & $\mathtt{RC}$ & $0.25$\\
		\hline
	\end{tabular}
		\vspace{-0.5cm}
\end{table}
As depicted in Fig. \ref{fig:SIR_per_iter}, it is obvious that by increasing the number of iterations, i.e., setting a small $\varepsilon$, the orthogonality of $g$ can be improved. Moreover, if taking the convergence time into consideration, confining the number of iterations to less than ten is reasonable as well, as the SIR is already more than 80dB after the first few iterations.
Fig. \ref{fig:imp_freq_per_iter} presents the time and frequency impulse responses for the initial Gaussian pulse, optimized pulse after the first iteration and the final result, which indicates how the number of iterations influences the time and frequency localization properties for the obtained pulse shapes.
\subsubsection{Bi-orthogonal Design with Channel Statistics}
Bi-orthogonal pulse design allows using different pulses at the transceiver to maximize the link performance. It employs mis-matched filtering to balance the robustness against ISI/ICI in doubly dispersive channels with the noise enhancement. In general, bi-orthogonal design 
capitalizes on more degree of freedom than orthogonal design, leading to better performance in practice, especially in self-interference-limited scenario.

Provided channel statistics are available, there are
two common approaches for bi-orthogonal design: first, fixing the transmit filter and design the optimal receiver filter; second, joint transmit and receive pulse design. SINR optimization is applied for such design. Since the resulting transmit and receive pulses are used and the orthogonality is slightly relaxed as in (\ref{eq:pr}), we still term it as "bi-orthogonal" design.
\begin{algorithm}[!htbp]
	\caption{Optimal receive filter design with channel statistics}\label{algo_rx}
	\small{\begin{algorithmic}[1]
			\STATE Given filter length $L$, transmit pulse $\mathbf{g}$, and channel statistics $\mathbf{C}_\mathbb{H}$.\\
			\STATE Compute $\mathbf{A}$ and $\mathbf{B}$ with \eqref{eq:A_and_B}. \\
			\STATE Perform the generalized eigendecomposition on $\mathbf{A}$ and $\mathbf{B}$. \\
			\STATE Find the generalized eigenvector $\boldsymbol{\gamma}_\mathrm{max}$ corresponding to the maximum generalized eigenvalue $\zeta_\mathrm{max}$.\\
			\STATE Return optimal receive filter $\boldsymbol{\gamma}_\mathrm{max}$ and achievable SINR $\zeta_\mathrm{max}$.\\
		\end{algorithmic}}
\end{algorithm}
\begin{itemize}
\item Optimized Receiver Filter Design: 
With regard to the pre-determined transmit pulse $\mathbf{g}$, we now derive the optimized receive pulse $\boldsymbol{\gamma}$ to maximize link performance (taking SINR as an example). $\text{SINR}^\mathtt{D}_{\mathbf{g},\boldsymbol\gamma}$ in \eqref{eq:SINR_discrete} can be further reformulated as
\begin{equation}\label{eq:SINR_derived}
\zeta = \text{SINR}^\mathtt{D}_{\mathbf g,\boldsymbol\gamma}=\frac{\boldsymbol{\gamma}^H\mathbf{A}\boldsymbol{\gamma}}{\boldsymbol{\gamma}^H\mathbf{B}\boldsymbol{\gamma}},
\end{equation}
where $\mathbf{A}$ and $\mathbf{B}$ are Hermitian matrices given respectively by 
\begin{align}\label{eq:A_and_B}
\mathbf{A}=\mathbf{G}_{0,0}\mathbf{C}_\mathbb{H}\mathbf{G}_{0,0}^H,\quad\mathbf{B}=\sum_{\left(m,n\right)\ne\left(0,0\right)}\mathbf{G}_{m,n}\mathbf{C}_\mathbb{H}\mathbf{G}_{m,n}^H+\sigma_\mathrm{n}^2\mathbf{I}.
\end{align}
Note that \eqref{eq:SINR_derived} is defined as a generalized Rayleigh quotient, which is associated with a generalized eigenvalue problem $\mathbf{A}\boldsymbol{\gamma}=\zeta\mathbf{B}\boldsymbol{\gamma}$ \cite{JungTCOM2007,Jung2007}. The maximum SINR target $\zeta_\mathrm{max}$ corresponds to the maximum generalized eigenvalue of $\mathbf{A}$ and $\mathbf{B}$, when the receive filter $\boldsymbol{\gamma}$ is chosen as $\boldsymbol{\gamma}_\mathrm{max}$ the corresponding generalized eigenvector, i.e.,
$\mathbf{A}\boldsymbol{\gamma}_\mathrm{max}=\zeta_\mathrm{max}\mathbf{B}\boldsymbol{\gamma}_\mathrm{max}.
$ 
Detailed implementation is stated in Algorithm \ref{algo_rx}. . 

\item Joint Transmitter and Receiver Design:
\begin{algorithm}[!t]
	\caption{Joint transmit and receive filters optimization with channel statistics}\label{algo}
	\small{\begin{algorithmic}[1]
			\STATE \textbf{Initialization}: Given convergence coefficient $\varepsilon$, initial transmit pulse $\mathbf{g}^{\left(0\right)}$, filter length $L$, and channel statistics $\mathbf{C}_\mathbb{H}$. Let iteration index $n=0$.\\
			\STATE \textbf{repeat} \{Main Loop\}\\
			\STATE\quad Given $\mathbf{g}^{\left(n-1\right)}$, compute $\boldsymbol{\gamma}^{\left(n\right)}$ by performing Algorithm \ref{algo_rx}.\\
			\STATE\quad Compute $\mathbf{g}^{\left(n\right)}$ based on $\boldsymbol{\gamma}^{\left(n\right)}$ following a similar manner.\\
			\STATE\quad Let $n=n+1$.\\
			\STATE\textbf{until} $\frac{\left\|\mathbf{g}^{\left(n\right)}-\mathbf{g}^{\left(n-1\right)}\right\|}{\left\|\mathbf{g}^{\left(n-1\right)}\right\|}\le\varepsilon$ and $\frac{\left\|\boldsymbol{\gamma}^{\left(n\right)}-\boldsymbol{\gamma}^{\left(n-1\right)}\right\|}{\left\|\boldsymbol{\gamma}^{\left(n-1\right)}\right\|}\le\varepsilon$.\\
			\STATE \textbf{return} $\mathbf{g}^{\left(n\right)}$ and $\boldsymbol{\gamma}^{\left(n\right)}$.\\
		\end{algorithmic}}
\end{algorithm}
Considering the joint optimization of the transmit and receive filters w.r.t. the provided channel statistics
for wide-sense stationary uncorrelated scattering (WSSUS) channels, 
\cite{JungTCOM2007} showed that the primal problem is a nonconvex problem. An efficient alternating algorithm has been proposed to achieve local optimum. Its detail implementation is listed in Algorithm \ref{algo}.
In general, this algorithm calculates the transmit and receive pulse alternatively until the overall process converges.
\end{itemize}

\section{Evaluation of Designed Pulse Shapes}
\label{sec:pulseshapeexample}
Section \ref{sec:pulseshapedesign} introduced two exemplary methods for transceiver pulse design. In practical systems, given the transmit pulse provided by the orthogonal or bi-orthogonal design, the receiver may fix or further adapt the receive pulse according to the available channel knowledge. Considering this issue into account, we propose in the section two solutions to design the receiver, and evaluate the link performance using SINR contour.  
\subsection{Evaluation Metric: SINR Contour}
For any doubly dispersive channel with predefined scattering function $C_\mathbb{H}\left(\tau,\nu\right)$ or $\mathbf{C}_\mathbb{H}$, the achievable SINR of given transceiver pulse pair can be computed by \eqref{eq:SINR_continuous} or \eqref{eq:SINR_discrete}. Therefore, we can draw a SINR contour w.r.t. \textcolor{black}{time delay spread $\tau$ and Doppler spread $\nu$} for WSSUS channels. Such contour is important to visualize the link performance. \textcolor{black}{Basically, the point $SINR(\tau,\nu)$ on the contour indicates the self-interference plus noise level when the signal modulated with a pulse pair is undergoing a TF dispersion with delay  region $[-\lvert\tau\rvert, \lvert\tau\rvert]$ and Doppler region  $[-\lvert\nu\rvert, \lvert\nu\rvert]$.}
To compute SINR via \eqref{eq:SINR_discrete} for SINR contour plot, the channel scattering function need to be a-priori known. In practice, however, accurate channel statistic is not available but only channel characteristics such as maximum delay $\tau_\mathrm{max}$ and maximum Doppler frequency $\nu_\mathrm{max}$. Without further specification, in this section, we assume that the "default" support region of the underspread WSSUS channel is an origin-centered rectangle-shaped \cite{Matz2013}, whose side lengths equal to $2\tau_\mathrm{max}$ and $2\nu_\mathrm{max}$, respectively. The diagonal entries of channel scattering function $\mathbf{C}_\mathbb{H}$ are set to be equal.
\subsection{SINR Evaluation based on Receiver Realizations}
Given a transmit pulse optimized according to the orthogonal or bi-orthogonal methods, two receive pulse design is considered here: so-called naive receiver which is designed without channel information, or max-SINR receiver which takes channel information into the design procedure. 
\subsubsection{Naive Receiver without Channel Knowledge}
\begin{itemize}
\item Transmit Pulse based on Orthogonal Design:
Provided the transmit pulse optimized by the orthogonal method, naive receiver refers to the receive pulse which adopts symmetric shape of the transmit pulse generated by Algorithm \ref{algo_rx_wo_channel}, i.e., $\gamma\left(t\right)=g\left(t\right)$. Herein, we provide several design examples in this section. 

A necessary condition for generating orthogonal pulses is to fulfill $TF>1$. On the other hand, larger $TF$ leads to smaller spectral efficiency. As a compromise, $TF$ is set to be slightly larger than 1. We choose $TF=1.07$ and $TF=1.25$ (same as normal/extended CP overhead in LTE) and \textcolor{black}{$\alpha = 1$}. Table \ref{tab:parameters} lists the key parameters in Algorithm \ref{algo_rx_wo_channel}.
\begin{table}[!htbp]
	\caption{Parameter settings for deriving short pulses with $K=1, 2$.}
	\centering
	\label{tab:parameters}
	\renewcommand{\arraystretch}{1.2}
	\begin{tabular}{|p{.25\linewidth}|p{.06\linewidth}|p{.06\linewidth}|p{.06\linewidth}|p{.06\linewidth}|p{.06\linewidth}|p{.06\linewidth}|p{.06\linewidth}|}
		\hline  & $N$ & $M$ & $TF$ & $K$ & $\varepsilon$ & $g_\mathrm{w}$ & $\beta$ \\ 
		\hline $g^{K=2}_1$ in Fig. \ref{fig:go_gausst_107_2} & $282$ & $256$ & $1.07$ & $2$ & $10^{-4}$ & $\mathtt{RECT}$ & $0$\\
		\hline $g^{K=2}_2$ in Fig. \ref{fig:go_gausst_125_2} & $320$ & $256$ & $1.25$ & $2$ & $10^{-4}$ & $\mathtt{RECT}$ & $0$\\
		\hline $g^{K=1}_1$ in Fig. \ref{fig:go_gausst_107_k1} & $282$ & $256$ & $1.07$ & $1$ & $10^{-4}$ & $\mathtt{RECT}$ & $0$\\
		\hline $g^{K=1}_2$ in Fig. \ref{fig:go_gausst_125_k1} & $320$ & $256$ & $1.25$ & $1$ & $10^{-4}$ & $\mathtt{RECT}$ & $0$\\
		\hline
	\end{tabular}
\end{table}

Fig. \ref{fig:go_gausst_2} illustrates the pulse shapes for overlapping factor set to $K=2$. Solid line and dashed line indicate the optimized pulse in this paper and \cite{Pinchon2011}, respectively. Both results are close to the pulse shapes used in windowed-OFDM. For the case of $TF=1.25$ (Fig. \ref{fig:go_gausst_125_2}), the optimized pulse in this paper converges to the analytically derived pulse shape with the optimal TFL in \cite{Pinchon2011}. Given the transmit pulse with $K=1$, the proposed pulse shapes are depicted in Fig \ref{fig:go_gausst_k1}. For $TF=1.25$, $g^{K=1}_2\left(t\right)$ coincides with the pulse proposed in  \cite{Pinchon2011}, which aim at minimizing the OOB leakage.
\begin{figure}[!htbp]
	\centering
	\begin{subfigure}[b]{0.4\textwidth}
		\centering
		\includegraphics[scale=0.7]{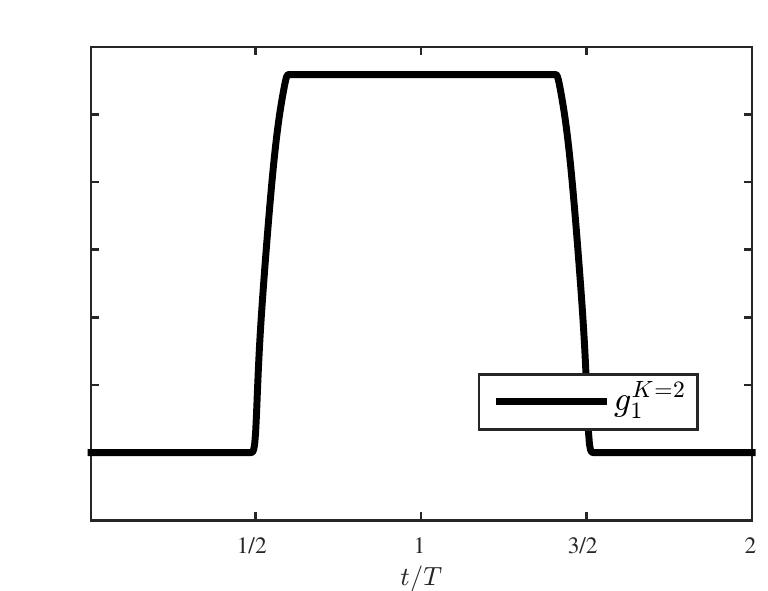}
		\caption{$g^{K=2}_1$ for $TF=1.07$} 
		\label{fig:go_gausst_107_2}
	\end{subfigure}
	\begin{subfigure}[b]{0.4\textwidth}
		\centering
		\includegraphics[scale=0.7]{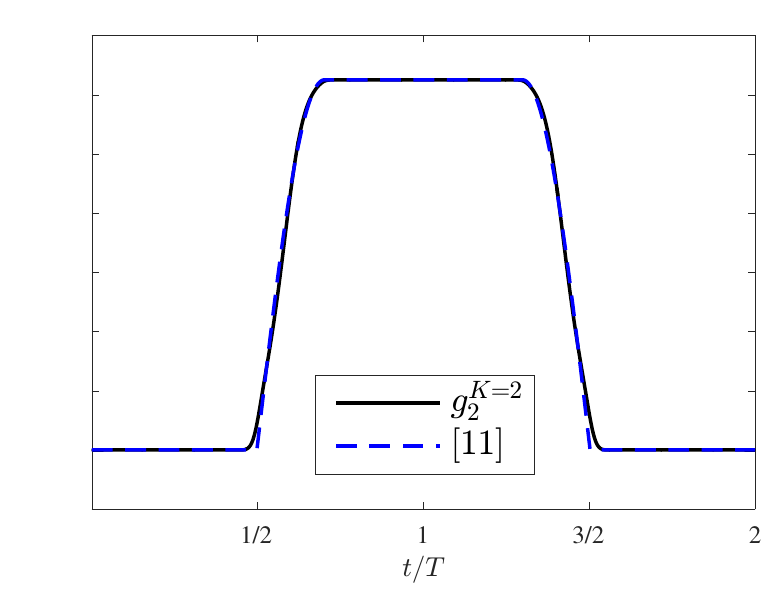}
		\caption{$g^{K=2}_2$ for $TF=1.25$} 
		\label{fig:go_gausst_125_2}
	\end{subfigure}
	\caption{Orthogonal pulse shape design for $K=2$.}
	\label{fig:go_gausst_2}
		\vspace{-0.5cm}
\end{figure}

This transceiver pulse design method is suitable to the scenario requiring good frequency localization, i.e., one symbol per transmission time interval (TTI) transmission, which is the extreme case of TDD transmission requiring the lowest round-trip time (RTT). In such scenario, owing to the guard periods inserted between the uplink and downlink, there is no ISI, and hence pulses that are well-localized in the frequency domain are favored. From the design perspective, we select $g_\mathtt{gauss}^{\left(\alpha\right)}$ with $\alpha\ll1$ as the initial pulse, as in this case we only need to consider suppressing ICI when designing the short pulses \cite{WangVTC2015}. 

\begin{figure}[!ht]
	\centering
	\begin{subfigure}[b]{.4\textwidth}
		\centering
		\includegraphics[scale=0.7]{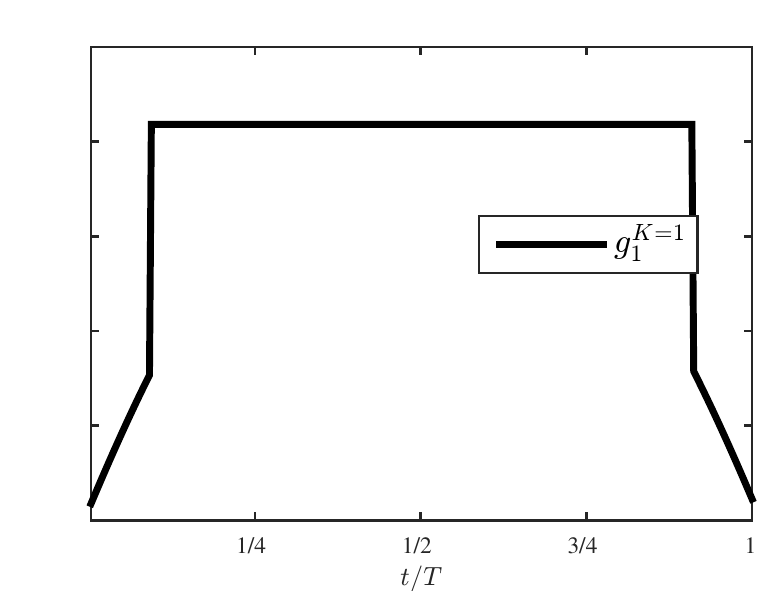}
		\caption{$g^{K=1}_1$ for $TF=1.07$} 
		\label{fig:go_gausst_107_k1}
	\end{subfigure}
	\begin{subfigure}[b]{.4\textwidth}
		\centering
		\includegraphics[scale=0.7]{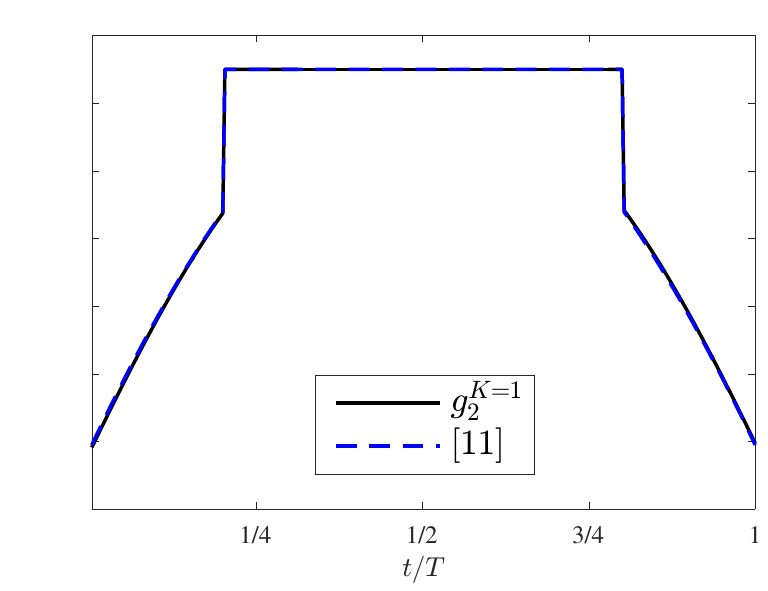}
		\caption{$g^{K=1}_2$ for $TF=1.25$} 
		\label{fig:go_gausst_125_k1}
	\end{subfigure}
	\caption{Orhogonal pulse shape design for $K=1$.}
	\label{fig:go_gausst_k1}
		\vspace{-0.5cm}
\end{figure}

Allowing long pulse, the exemplary orthogonal design results with $K=4$ are given in Fig. \ref{fig:go_gausst_107_k4} and \ref{fig:go_gausst_125_k4}. 
To compare with CP-OFDM, the SIR contour with pulse pair $g^{K=4}_1/\gamma^{K=4}_1$ (dashed) and $g_\text{rect}/g_\text{rect}$ (solid), 
as well as 
$g^{K=4}_2/\gamma^{K=4}_2$ (dashed) and $g_\text{rect}/g_\text{rect}$ (solid)
are depicted in Fig. \ref{fig:sirgo1.07} and Fig. \ref{fig:sirgo1.25}, respectively.
The number on contour line indicates the lowest achievable SINR level that a pulse pair could support within the closed region. In particular, compared with CP-OFDM, the proposed design possesses the strong robustness against time synchronization errors while maintaining similar support in F-domain, which could potentially enable timing advance (TA)-free transmission in uplink, or support downlink multi-point transmission with large coverage. Especially. the proposed design for $TF=1.25$ case support a similar T-F contour region for high order modulation (e.g. 64QAM) while achieving overall larger T-F contour support for lower modulation (e.g. QPSK and 16QAM). Thus, the $TF=1.25$ multi-carrier waveform is more robust in challenging dispersive scenarios, such as high speed vehicular transmission, to achieve high reliability.

\begin{figure}
\centering
\begin{subfigure}[b]{0.4\textwidth}
\centering
\includegraphics[scale=1]{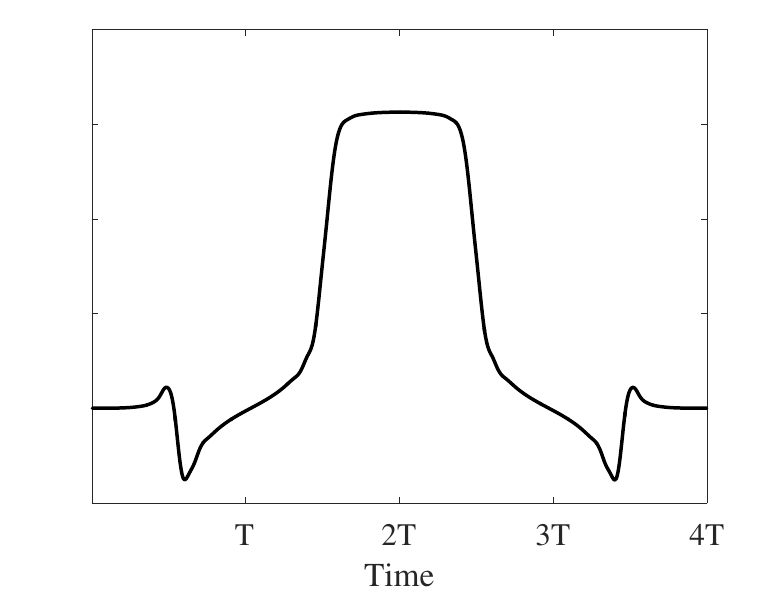}
\caption{$g^{K=4}_1$ for $TF=1.07$} 
\label{fig:go_gausst_107_k4}
\end{subfigure}
\hfill
\begin{subfigure}[b]{0.4\textwidth}
\centering
\includegraphics[scale=1]{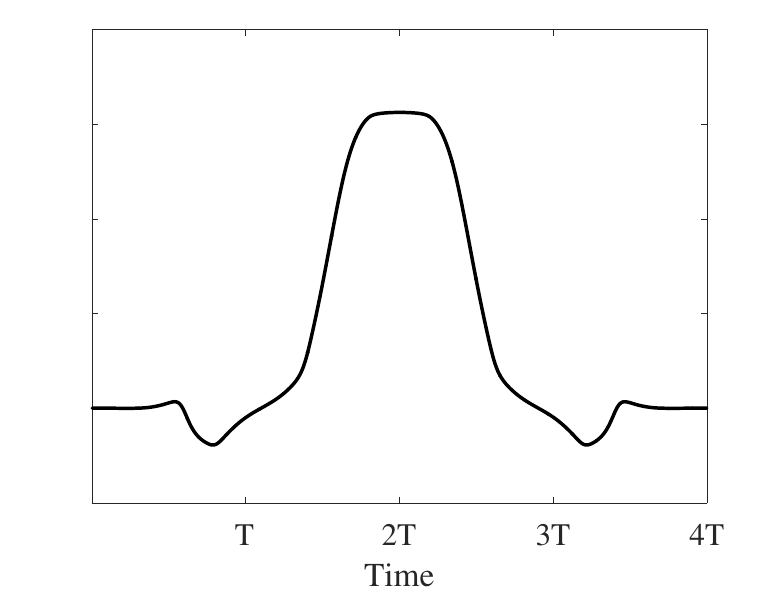}
\caption{$g^{K=4}_2$ for $TF=1.25$} 
\label{fig:go_gausst_125_k4}
\end{subfigure}
\hfill 
\caption{Orthogonal pulse shape design for $K=4$.}
\label{fig:go_gausst_k4}
\vspace{-0.3cm}
\end{figure}

\begin{figure}
\centering
\begin{subfigure}[b]{0.45\textwidth}
\centering
\includegraphics[height=6.5cm, width=7.5cm]{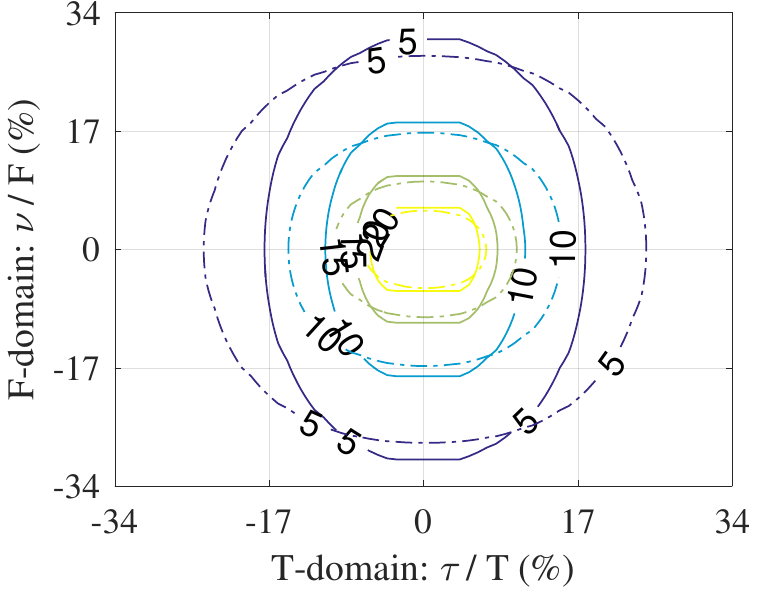}
\caption{$TF=1.07$} 
\label{fig:sirgo1.07}
\end{subfigure}
\hfill
\begin{subfigure}[b]{0.45\textwidth}
	\centering
	\includegraphics[height=6.5cm, width=7.5cm]{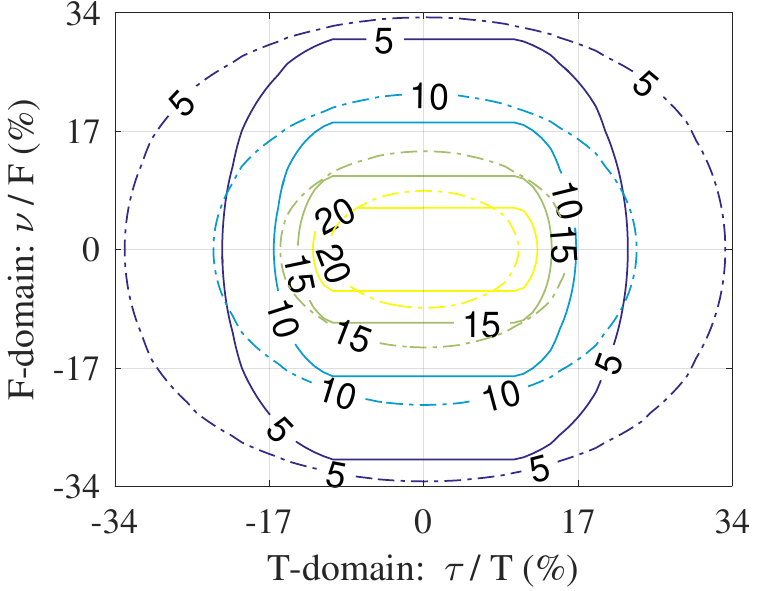}
	\caption{$TF = 1.25$} 
	\label{fig:sirgo1.25}
\end{subfigure}
\hfill 
\caption{SIR contour for orthogonal pulse design (dashed) and CP-OFDM (solid).}\label{fig:sir_go}
\vspace{-0.8cm}
\end{figure}

\item Transmit Pulse based on Bi-orthogonal Design:
Naive receiver for bi-orthogonal design indicates adopting mis-matched pulse shape of the transmit one without exploiting channel knowledge. To exemplify the receiver realization in this case, transmit pulse is fixed as two options: conventional rectangular pulse $\mathbf{g}_\text{RECT}$, and the raised-cosine (RC) shaped pulse $\mathbf{g}_\text{RC}$, which is commonly used in W-OFDM systems. We remark that other transmit pulse obtained from bi-orthogonal design is applicable.

For performance evaluation, $\mathbf{g}_\text{RC}$ is generated by the convolution with a window $\boldsymbol{w}$ with length $N_0$ and a rectangular window with length $N$. According to \cite{Farhang-Boroujeny2011}, any pulse shape satisfying $\sum_{i=0}^{N_0-1}w_i=1$ can be selected as a window. Without further specification, we choose
$\mathbf h$ as Hanning windowing and set $N_0={N_\text{CP}}/{2}$ with $N_{\text{CP}}=N-M$. All the essential parameters are listed in Table \ref{tab:parameter}. 
\begin{table}[!htbp]
	\caption{Parameters for bi-orthogonal based transmit pulse adopting the naive receiver.}
	\centering
	\label{tab:parameter}
	\renewcommand{\arraystretch}{1.2}
	\begin{tabular}{|p{.22\linewidth}|p{.22\linewidth}|}
		\hline \textit{Parameters} & \textit{Values}\\
		\hline Number of subcarriers $M$ & $256$ \\
		\hline \textcolor{black}{Samples per symbol} $N$ & $282$ \\
		\hline CP length $N_\mathrm{cp}$ & $26$\\
		\hline Seed window type & Hanning \\
		\hline Seed window length $N_0$ & $N_\mathrm{cp}/2$ \\
		\hline Filter length & \textcolor{black}{$310$} \\
		\hline Noise power $\sigma_\mathrm{n}^2$ & $-31, -25, -22, -19\mathrm{dB}$  \\
		\hline
	\end{tabular}
		\vspace{-0.5cm}
\end{table}

\begin{figure}[!htbp]
	\centering
	\begin{subfigure}[b]{.4\textwidth}
		\centering
		\includegraphics[scale=0.5]{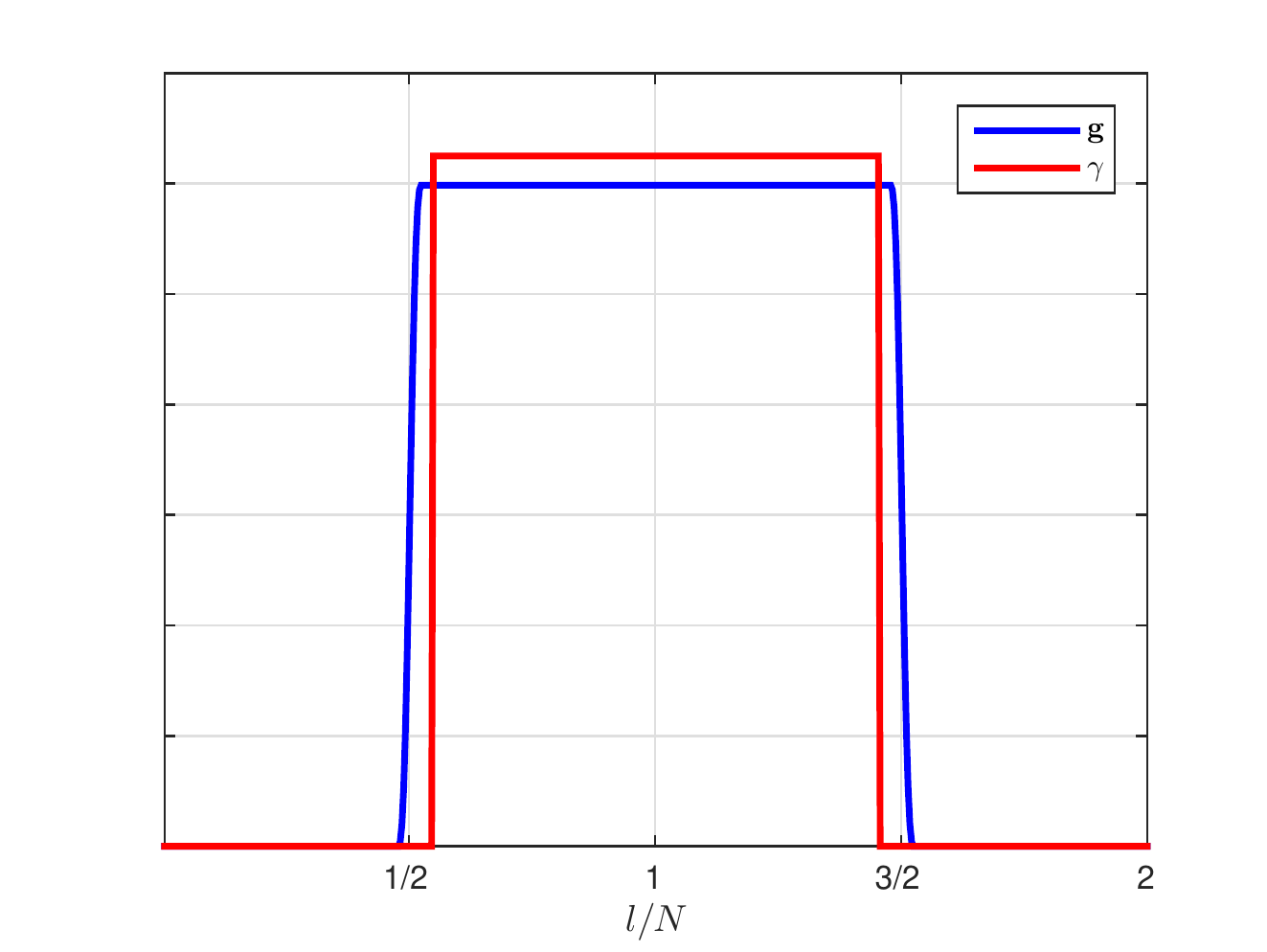}
		\caption{Rectangular $\boldsymbol{\gamma}$}		\label{fig:imp_resp_biorthogonal_naiive_rect}
	\end{subfigure}
	\begin{subfigure}[b]{.4\textwidth}
		\centering
		\includegraphics[scale=0.5]{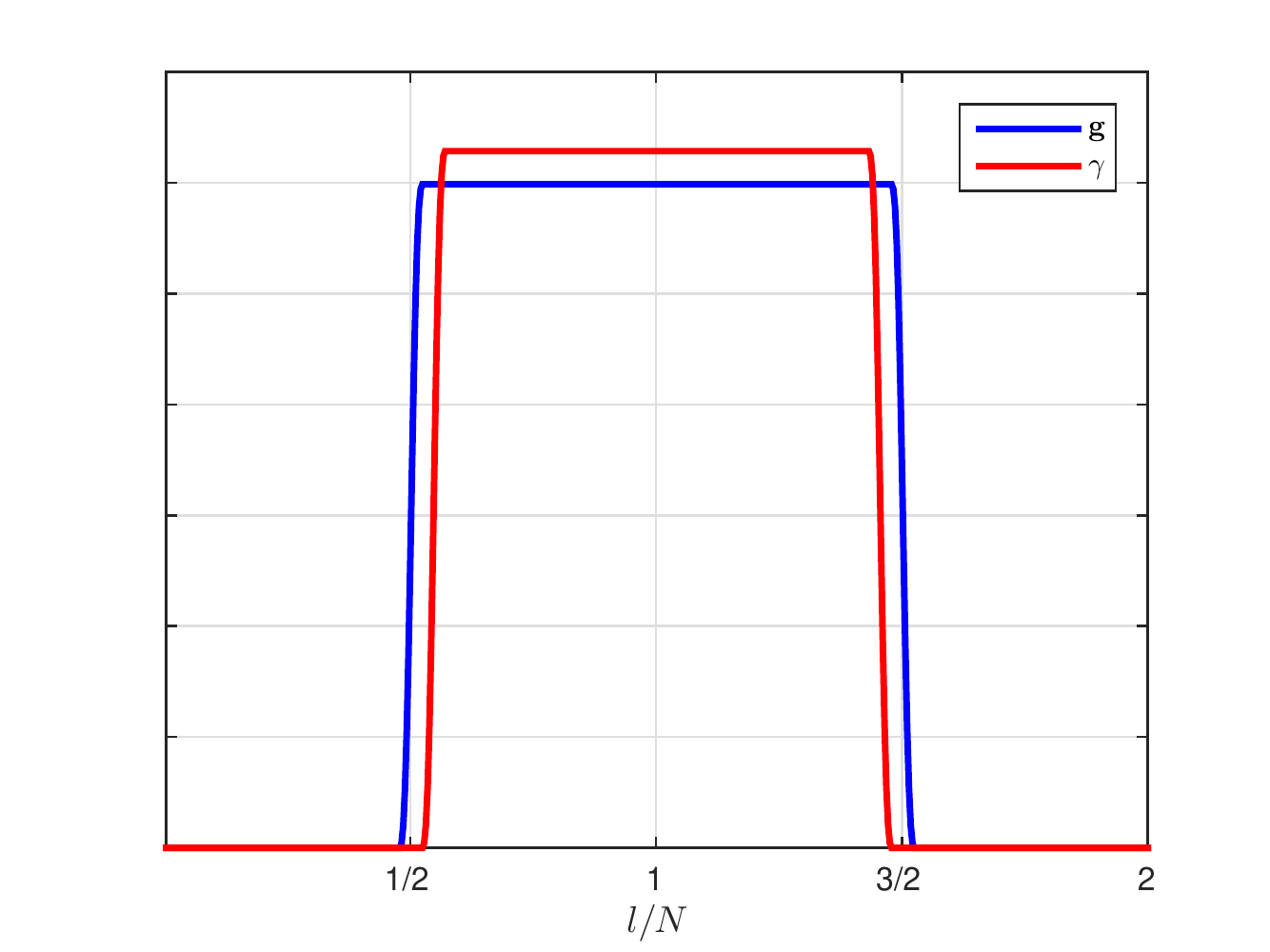}
		\caption{Raised-cosine $\boldsymbol{\gamma}$}
		\label{fig:imp_resp_biorthogonal_naiive_rc}
	\end{subfigure}
	\caption{Impulse response of bi-orthogonal based $\mathbf{g}_\text{RC}$ and naive $\boldsymbol{\gamma}$}
	\label{fig:imp_resp_biorthogonal_naiive}
	\vspace{-0.5cm}
\end{figure}
Fig. \ref{fig:imp_resp_biorthogonal_naiive_rect} and \ref{fig:imp_resp_biorthogonal_naiive_rc} show the RC transmit pulse combined with the rectangular receive pulse $\boldsymbol\gamma=\boldsymbol\gamma_\text{RECT}$ and raised-cosine receive pulse $\boldsymbol\gamma=\boldsymbol\gamma_\text{RC}$, respectively. The ratio $N/M$ is set to 1.1. The SINR contour with the pair $\mathbf{g}_\text{RECT}$/$\boldsymbol{\gamma}_\text{RECT}$ (solid) and $\mathbf{g}_\text{RC}$/$\boldsymbol{\gamma}_\text{RECT}$ (dashed) are depicted in Fig. \ref{fig:naiive_rx}. The x-axis denotes the delay $\tau$ normalized to symbol period $T$ in the time domain, while the y-axis represents the Doppler $\nu$ normalized to subcarrier spacing $F$ in the frequency domain. 
It can be observed that in noise-limited scenario, given rectangular receive pulse, $\mathbf{g}_\text{RC}$ achieves stronger robustness to asynchronization in the time domain and meanwhile supports similar dispersion $\mathbf{g}_\text{RECT}$ in the frequency domain. For the interference-limited scenario, i.e., noise variance equal to -31dB in Fig.  \ref{fig:contour_T0=0dot5_nonopt_31dB}, $\mathbf{g}_\text{RECT}/\boldsymbol{\gamma}_\text{RECT}$ and $\mathbf{g}_\text{RC}/\boldsymbol{\gamma}_\text{RECT}$ for 28dB SINR level have similar regions in contour plot, e.g., to support 256QAM on PDSCH in LTE.
\begin{figure}[!htbp]
	\centering
	\begin{subfigure}[b]{0.4\textwidth}
		\centering
		\includegraphics[scale=.5]{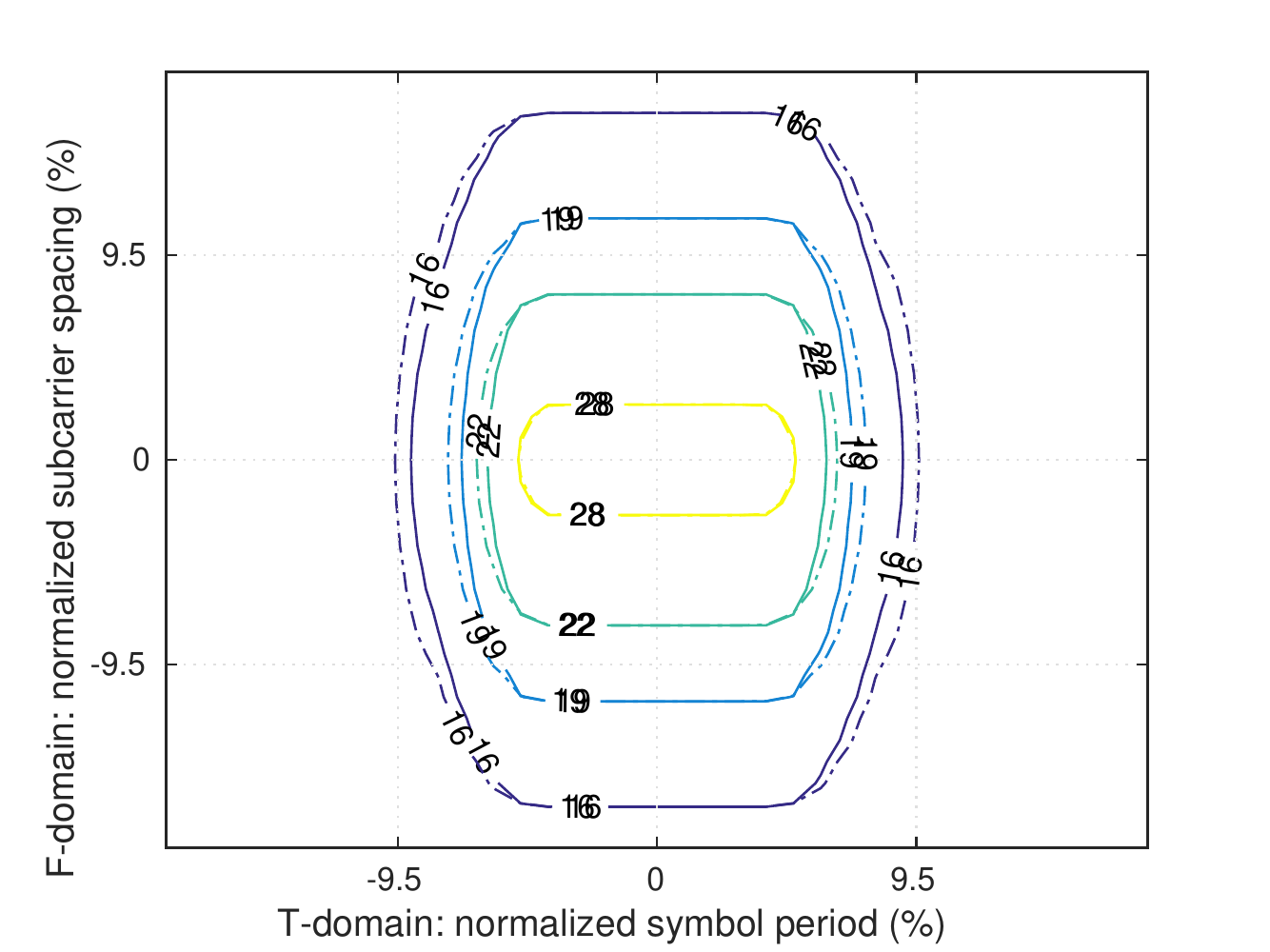}
		\caption{$\sigma_\mathrm{n}^2=-31\mathrm{dB}$}
		\label{fig:contour_T0=0dot5_nonopt_31dB}
	\end{subfigure}%
	~ 
	\begin{subfigure}[b]{0.4\textwidth}
		\centering
		\includegraphics[scale=.5]{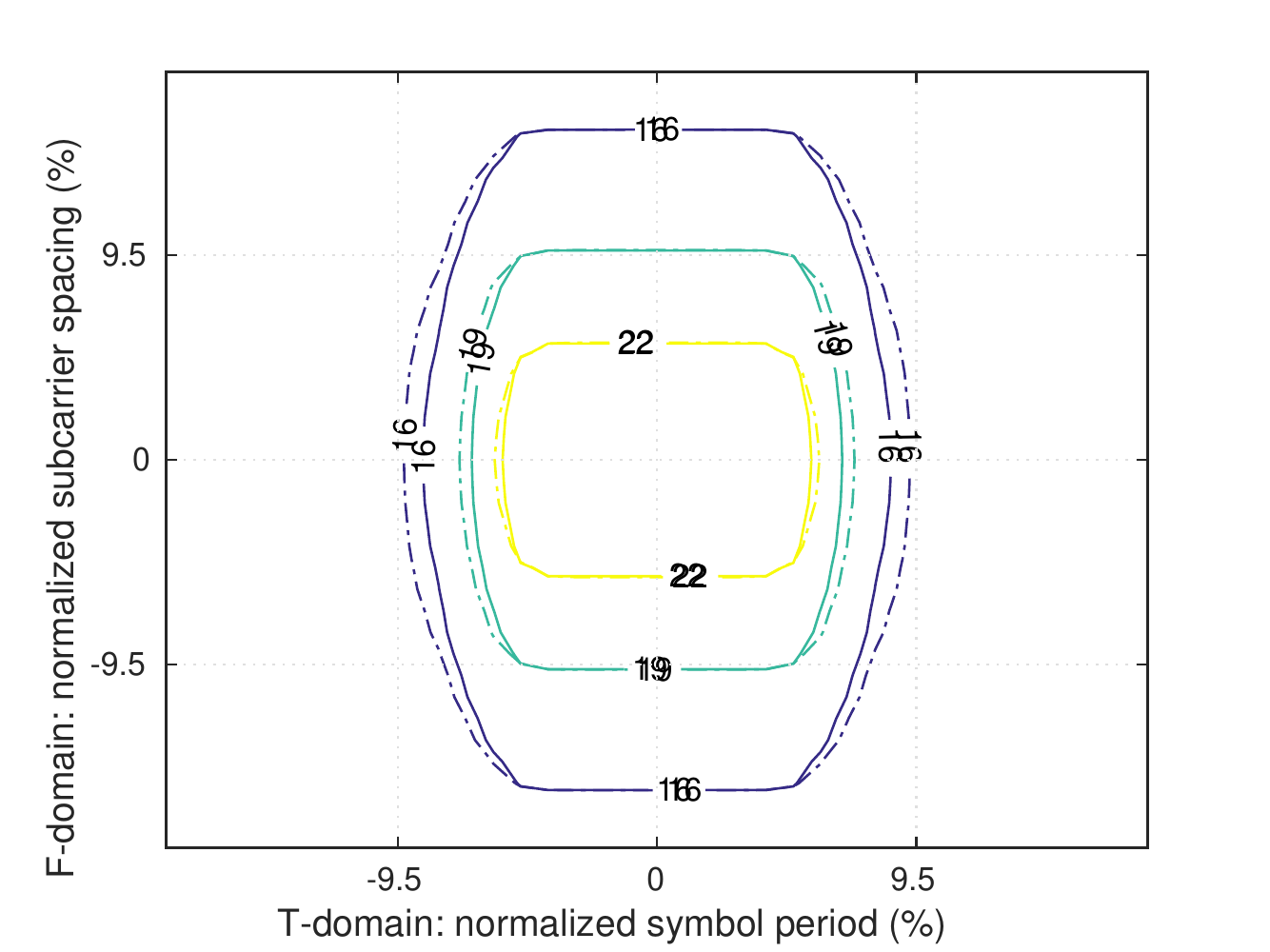}
		\caption{$\sigma_\mathrm{n}^2=-25\mathrm{dB}$}
		\label{fig:contour_T0=0dot5_nonopt_25dB}
	\end{subfigure}\par\medskip
	\begin{subfigure}[b]{0.4\textwidth}
		\centering
		\includegraphics[scale=.5]{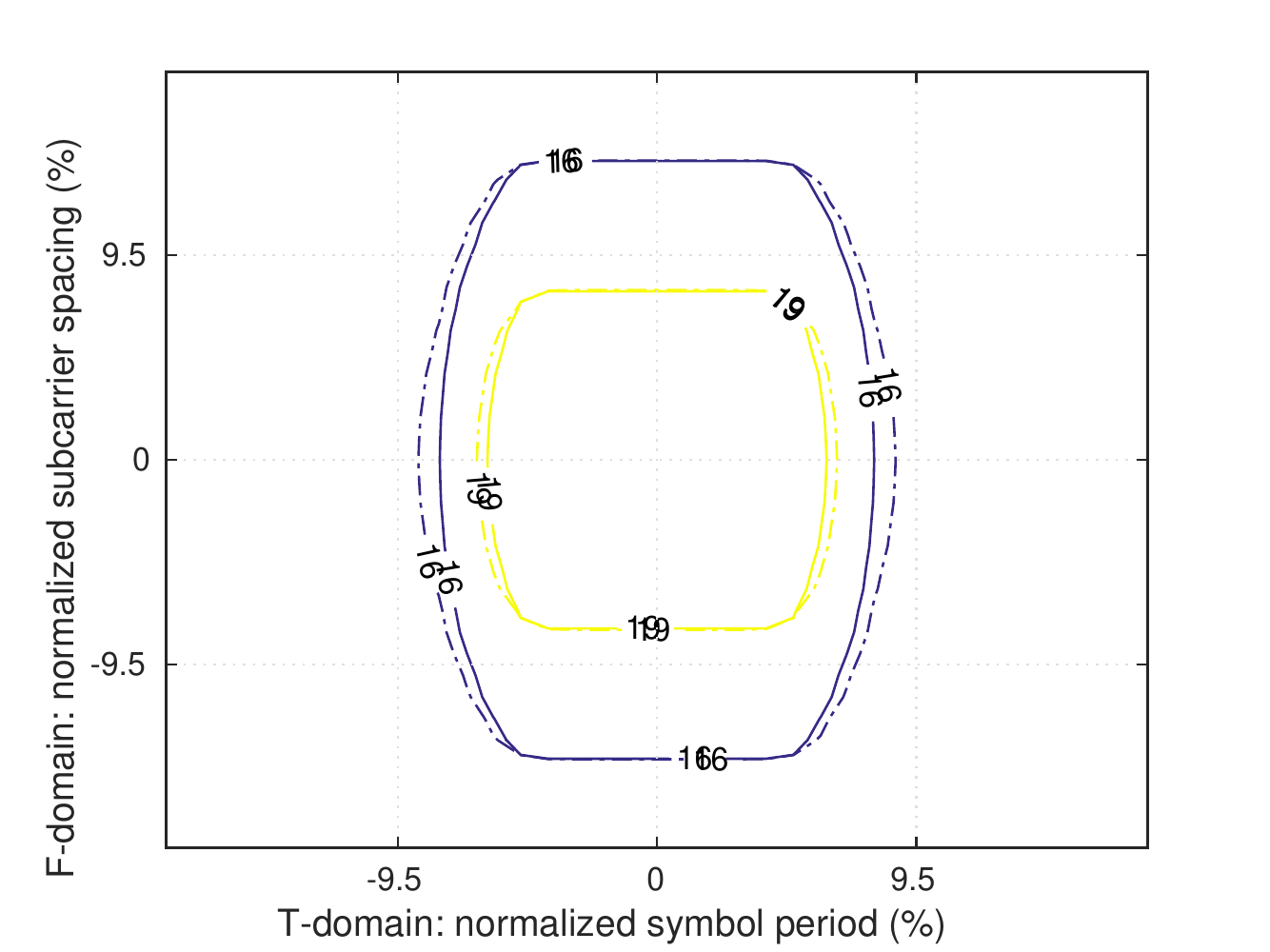}
		\caption{$\sigma_\mathrm{n}^2=-22\mathrm{dB}$}
		\label{fig:contour_T0=0dot5_nonopt_22dB}
	\end{subfigure}%
	~ 
	\begin{subfigure}[b]{0.4\textwidth}
		\centering
		\includegraphics[scale=.5]{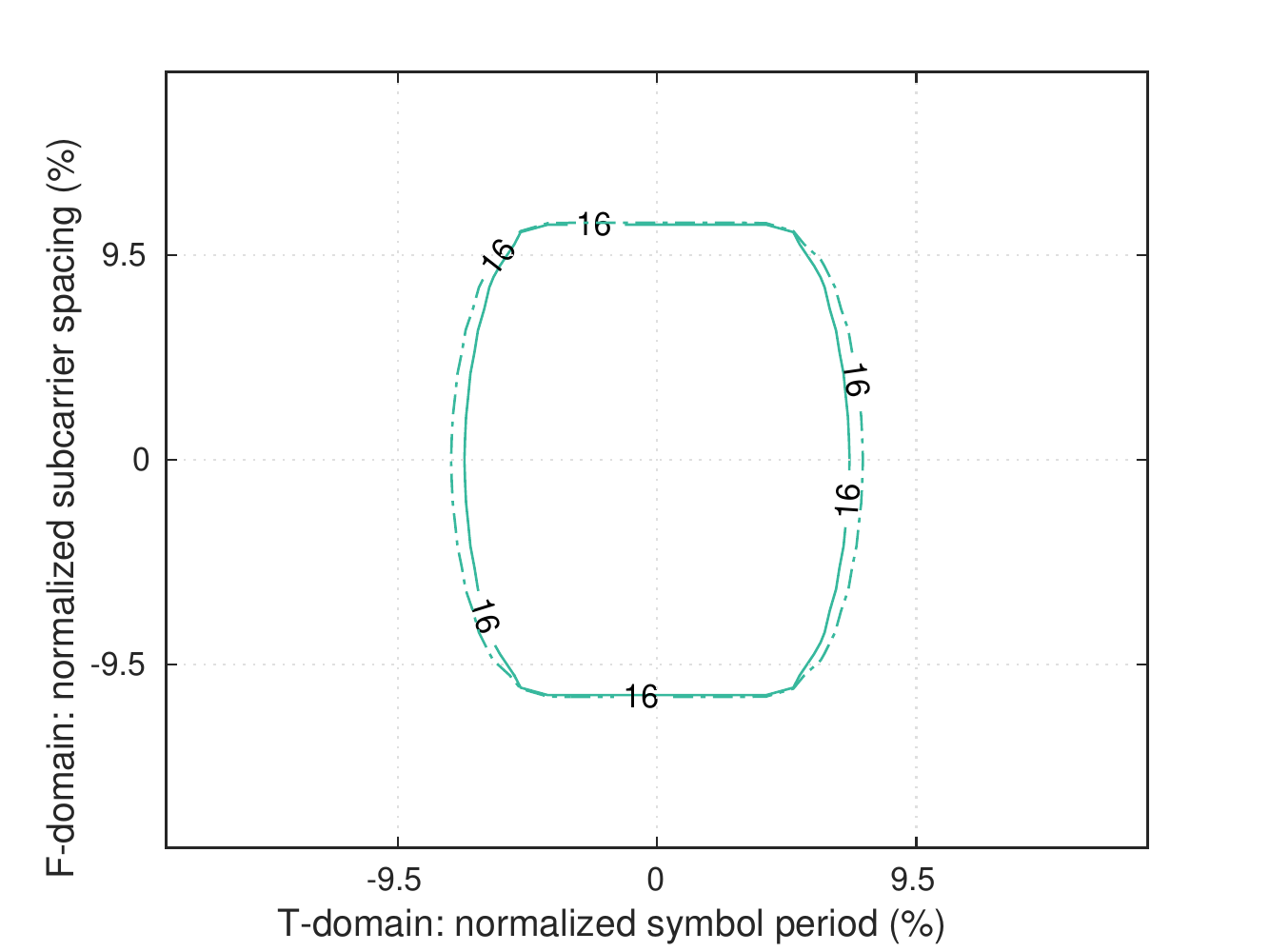}
		\caption{$\sigma_\mathrm{n}^2=-19\mathrm{dB}$}
		\label{fig:contour_T0=0dot5_nonopt_19dB}
	\end{subfigure}
	\caption{{\textcolor{black}{SINR contour of $\mathbf{g}_\text{RECT}$ (solid) and $\mathbf{g}_\text{RC}$ (dashed) w.r.t. naive receive $\boldsymbol{\gamma}_\text{RECT}$.} }}\label{fig:naiive_rx}
	\vspace{-0.8cm}
\end{figure}
\end{itemize}
\subsubsection{Max-SINR Receiver with Channel Statistical Knowledge}
With the assumption of a rectangular-shaped channel scattering function, we evaluate the performance with transmit pulse from both orthogonal and bi-orthogonal design, and its corresponding
max-SINR receive pulse.
\begin{itemize}
\item Transmit Pulse based on Orthogonal Design:
Choosing the transmit pulse for $g_1^{K=2}$ shown in Fig. \ref{fig:go_gausst_107_2}, we evaluate its SINR operational range w.r.t. double dispersion and make a comparison to $g_\text{cpofdm}$. The receive pulse is chosen calculated by Algorithm \ref{algo_rx}. Main simulation parameters have the same setting as in Table \ref{tab:parameters}.\\
As observed in Fig. \ref{fig:contour_Pinchon_TFL_opt}, compared with $g_\text{cpofdm}$, $g_1^{K=2}$ and its respective max-SINR receive pulse are more robust to time dispersion in high-noise-power regions, i.e., noise variance equal to -25, -22 and -19dB. For the case when $\sigma_\mathrm{n}^2$ is -31dB, the performance of $g_1^{K=2}$ on the level of 28dB is worse than $g_\text{cpofdm}$, thus making it an undesirable choice for enabling 256QAM in such case.
\begin{figure}[!htbp]
	\centering
	\begin{subfigure}[b]{0.4\textwidth}
		\centering
		\includegraphics[scale=.5]{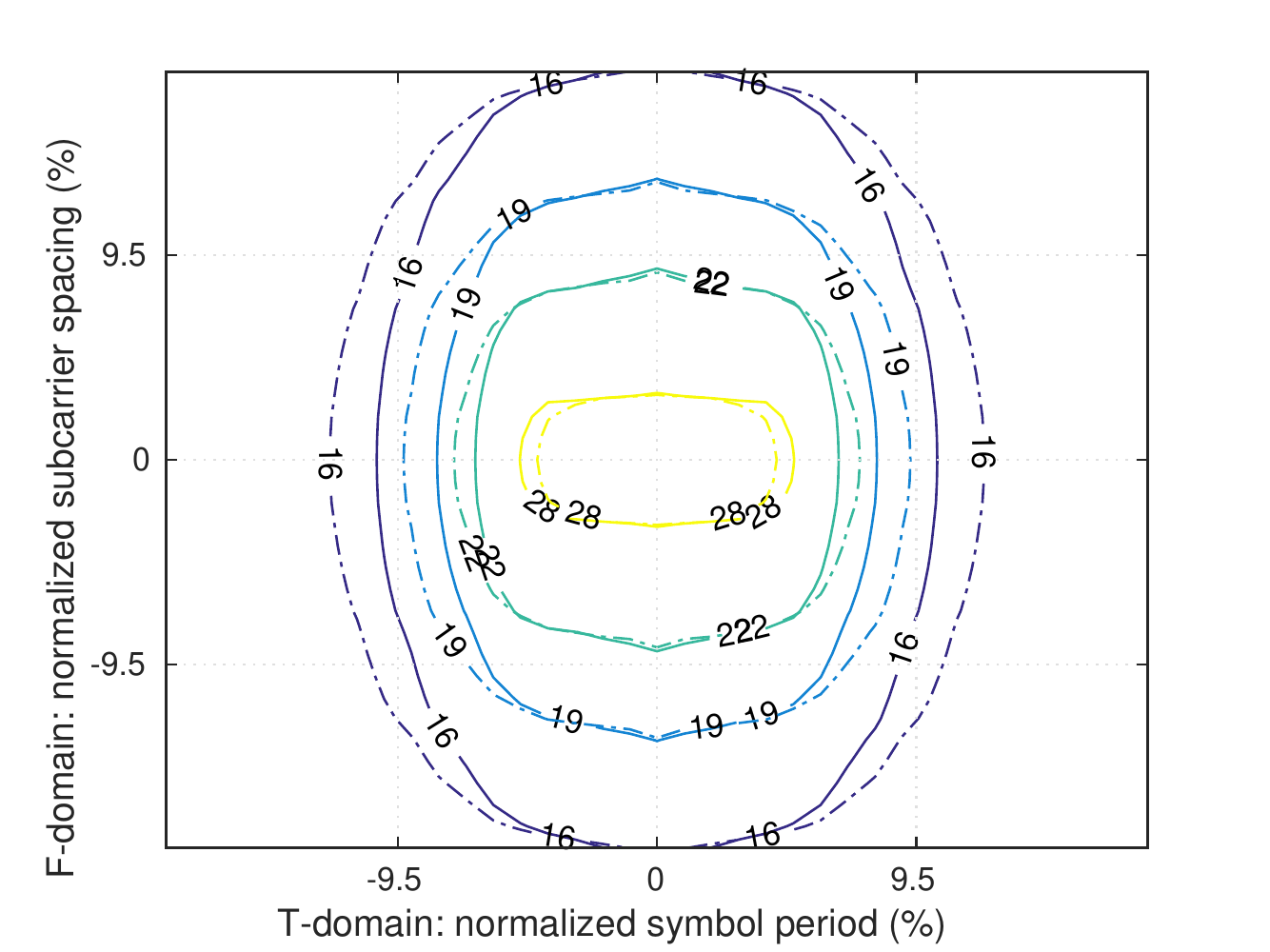}
		\caption{{$\sigma_\mathrm{n}^2=-31\mathrm{dB}$}}
		\label{fig:contour_Pinchon_TFL_opt_31dB}
	\end{subfigure}%
	~ 
	\begin{subfigure}[b]{0.4\textwidth}
		\centering
		\includegraphics[scale=.5]{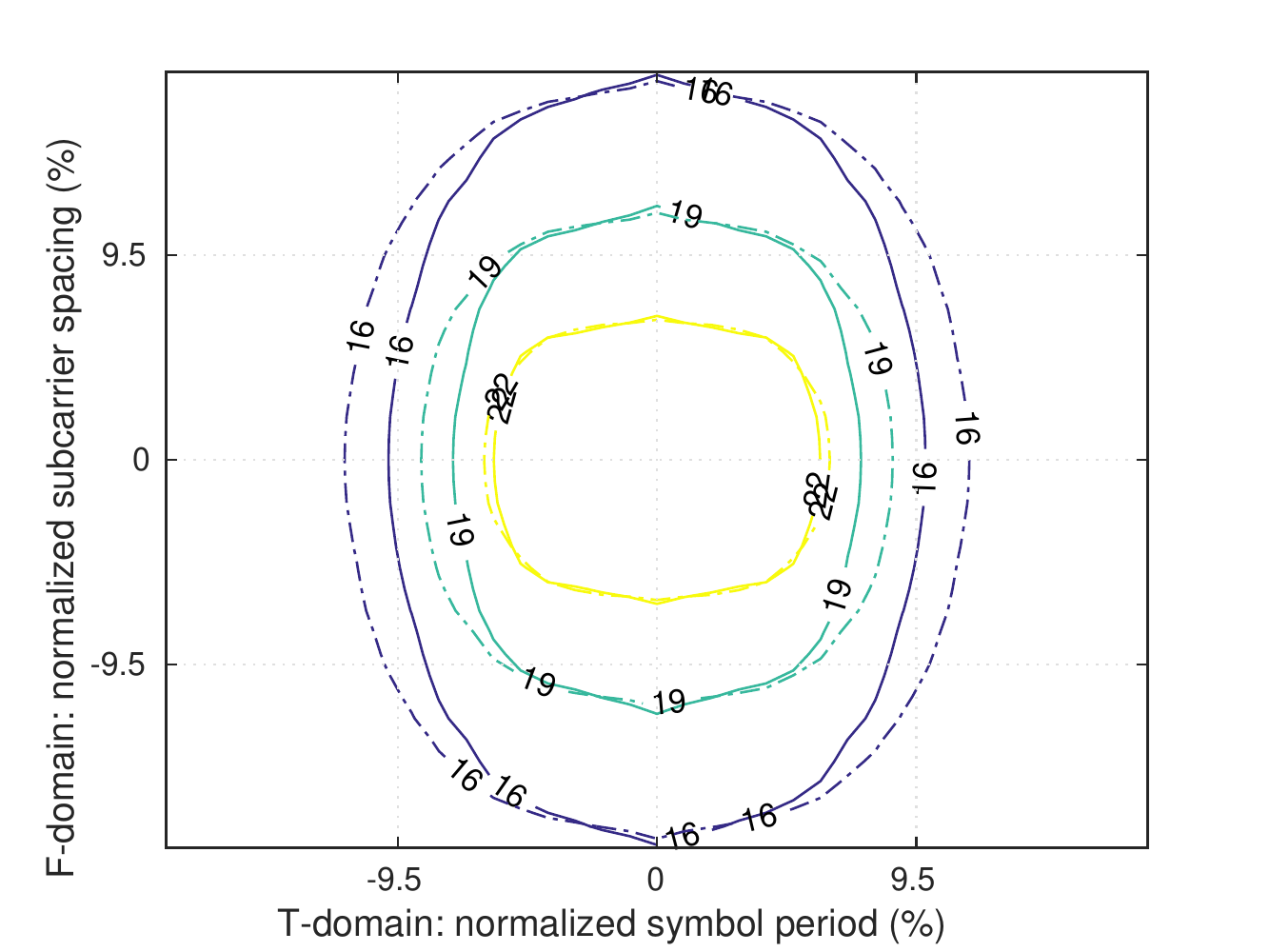}
		\caption{$\sigma_\mathrm{n}^2=-25\mathrm{dB}$}
		\label{fig:contour_Pinchon_TFL_opt_25dB}
	\end{subfigure}\par\medskip
	\begin{subfigure}[b]{0.4\textwidth}
		\centering
		\includegraphics[scale=.5]{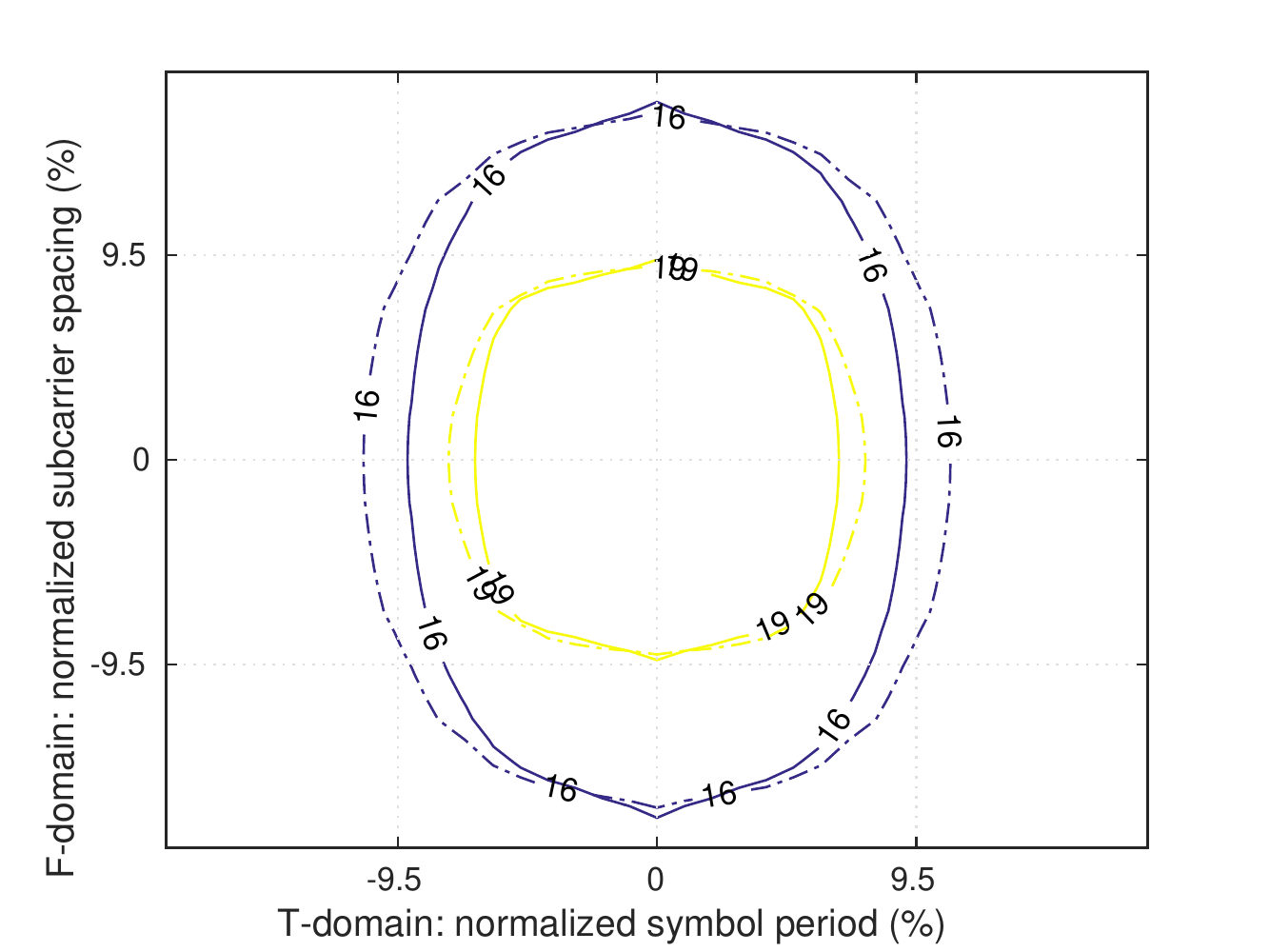}
		\caption{$\sigma_\mathrm{n}^2=-22\mathrm{dB}$}
		\label{fig:contour_Pinchon_TFL_opt_22dB}
	\end{subfigure}%
	~ 
	\begin{subfigure}[b]{0.4\textwidth}
		\centering
		\includegraphics[scale=.5]{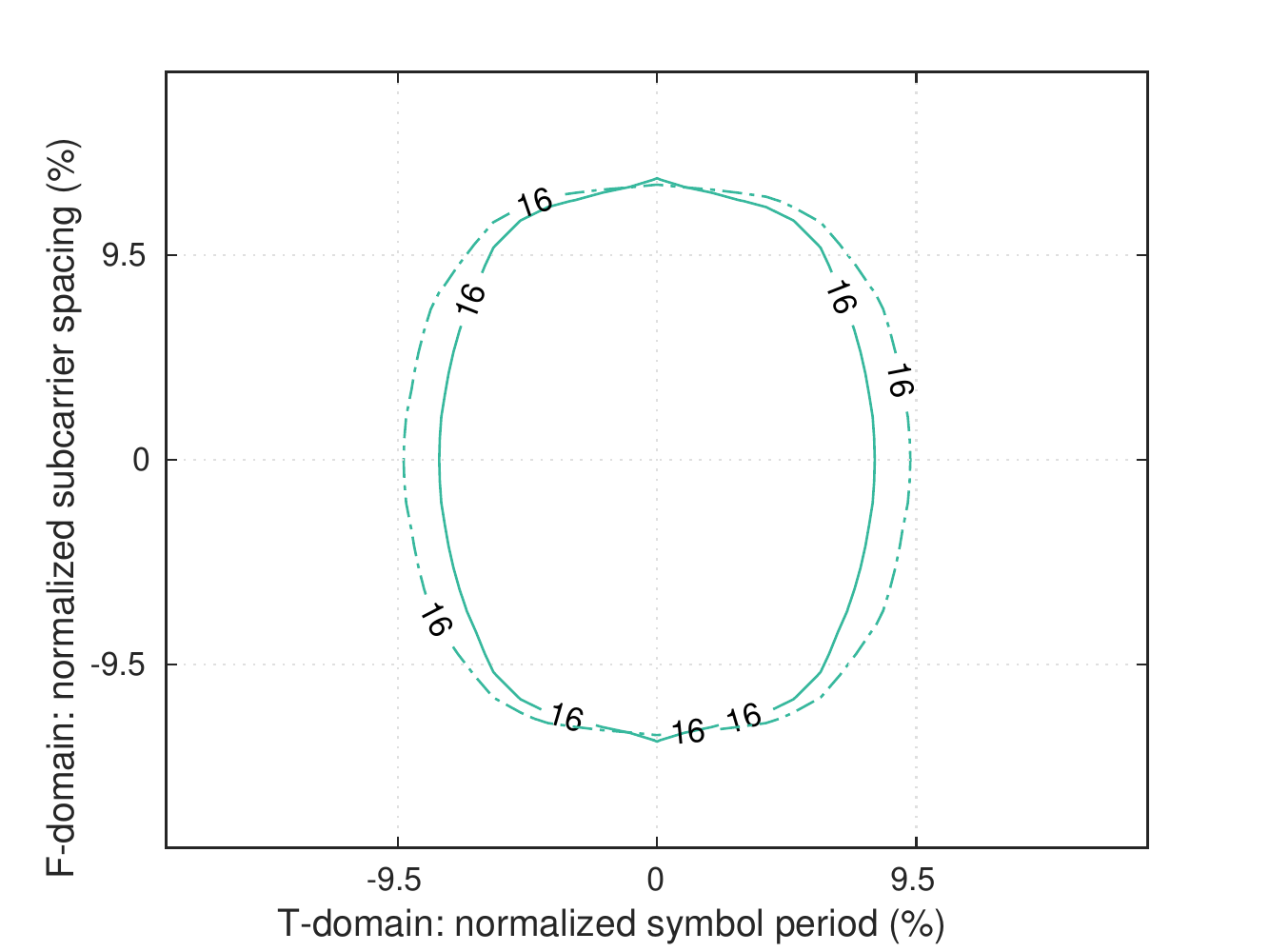}
		\caption{$\sigma_\mathrm{n}^2=-19\mathrm{dB}$}
		\label{fig:contour_Pinchon_TFL_opt_19dB}
	\end{subfigure}
  	\caption{\textcolor{black}{SINR contour of $g_\text{cpofdm}$ (solid) and $g_1^{K=2}$ (dashed) w.r.t. max-SINR receiver.}}\label{fig:contour_Pinchon_TFL_opt}
	\vspace{-0.8cm}
\end{figure}
\item Transmit Pulse based on Bi-orthogonal Design
We analyze in this section the SINR contours of $\mathbf{g}_\text{RECT}$ and $\mathbf{g}_\text{RC}$ \textcolor{black}{with its corresponding receive pulse calculated by Algorithm \ref{algo_rx} according to channel statistics}. Noise power level is set as the same in Fig. \ref{fig:contour_Pinchon_TFL_opt}  and parameter settings are given in Table \ref{tab:parameters}.

As depicted in Fig. \ref{fig:optimized_rx}, given max-SINR receiver, $\mathbf{g}_\text{RC}$
outperforms $\mathbf{g}_\text{RECT}$ w.r.t. robustness to timing misalignment, while maintaining comparable robustness to frequency dispersion. Moreover, comparing Fig. \ref{fig:optimized_rx} and Fig. \ref{fig:naiive_rx}, the optimized receiver is more robust against the frequency misalignment than the naive one, especially when the time shift close to zero.
\begin{figure}[!htbp]
	\centering
	\begin{subfigure}[b]{0.4\textwidth}
		\centering
		\includegraphics[scale=.5]{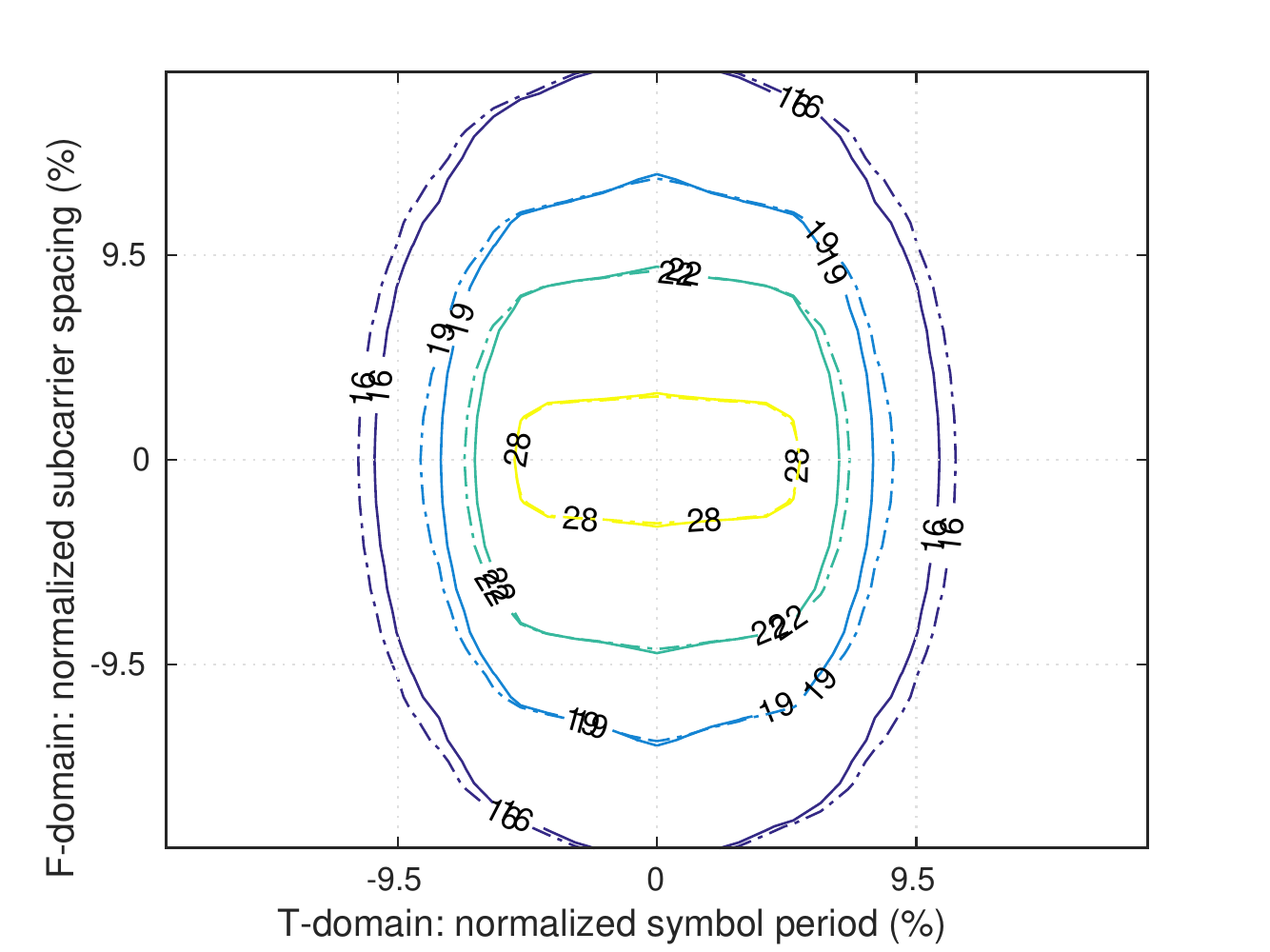}
		\caption{$\sigma_\mathrm{n}^2=-31\mathrm{dB}$}
		\label{fig:contour_T0=0dot5_opt_31dB}
	\end{subfigure}%
	~ 
	\begin{subfigure}[b]{0.4\textwidth}
		\centering
		\includegraphics[scale=.5]{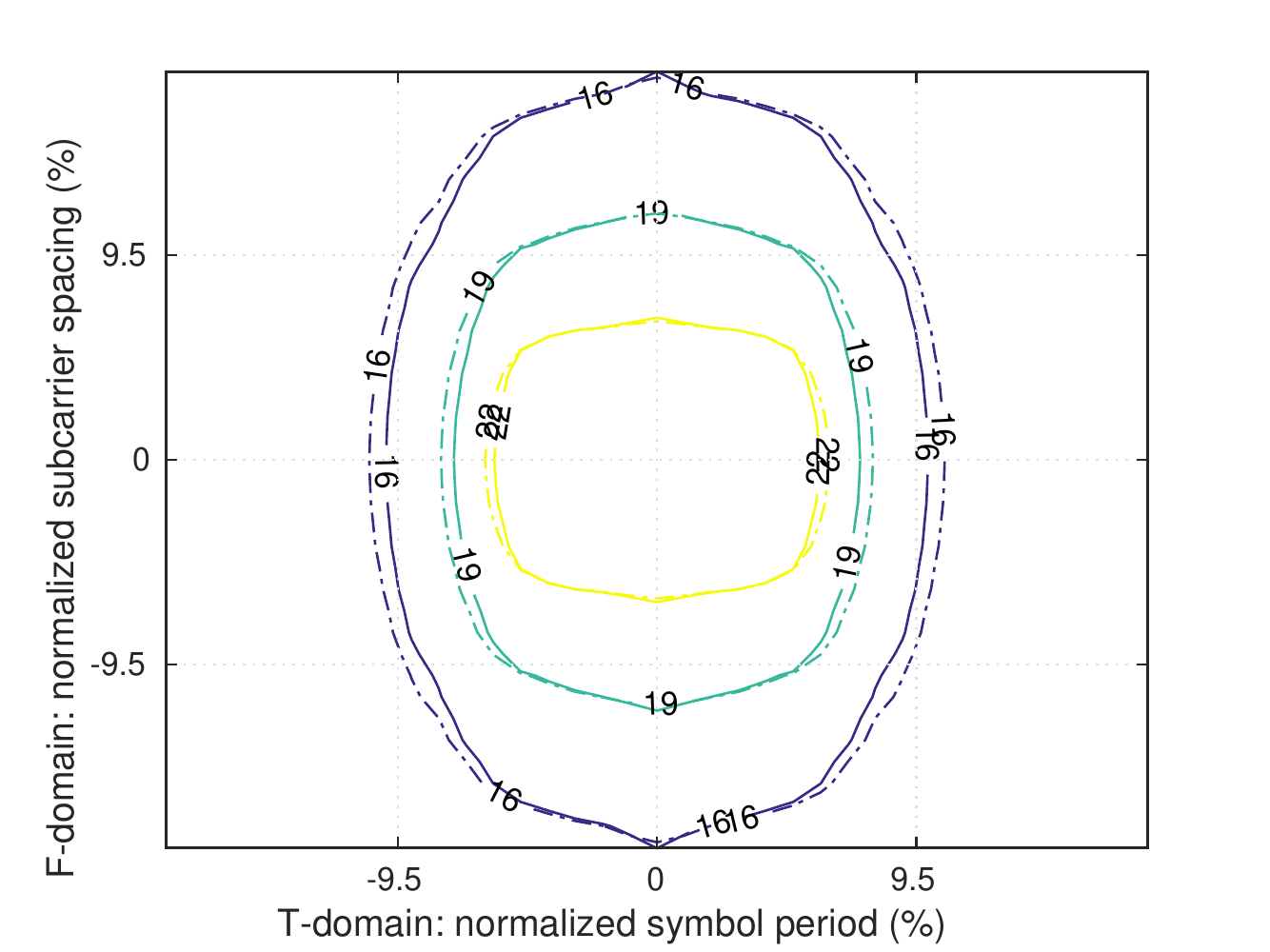}
		\caption{$\sigma_\mathrm{n}^2=-25\mathrm{dB}$}
		\label{fig:contour_T0=0dot5_opt_25dB}
	\end{subfigure}\par\medskip
	\begin{subfigure}[b]{0.4\textwidth}
		\centering
		\includegraphics[scale=.5]{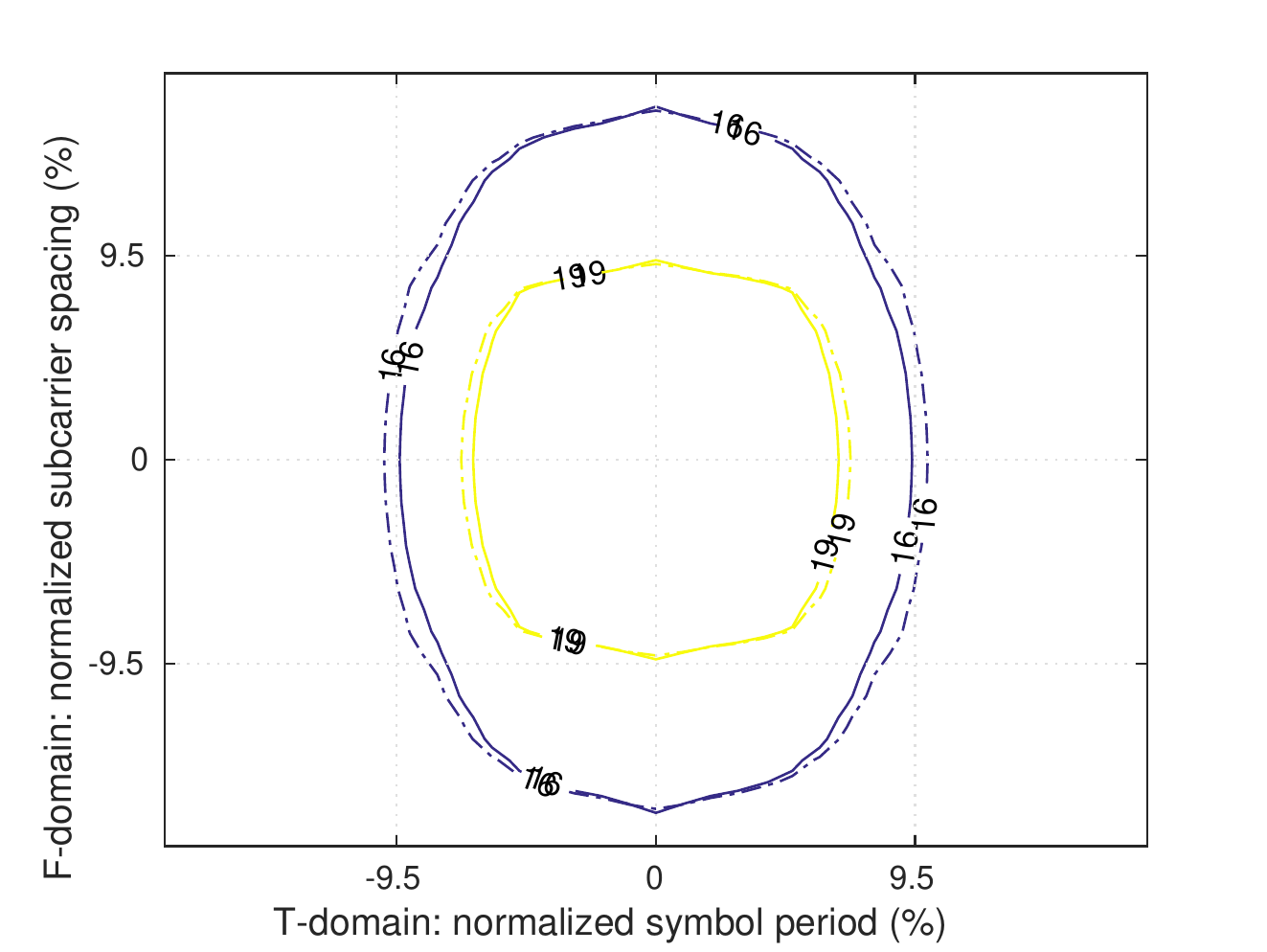}
		\caption{$\sigma_\mathrm{n}^2=-22\mathrm{dB}$}
		\label{fig:contour_T0=0dot5_opt_22dB}
	\end{subfigure}%
	~ 
	\begin{subfigure}[b]{0.4\textwidth}
		\centering
		\includegraphics[scale=.5]{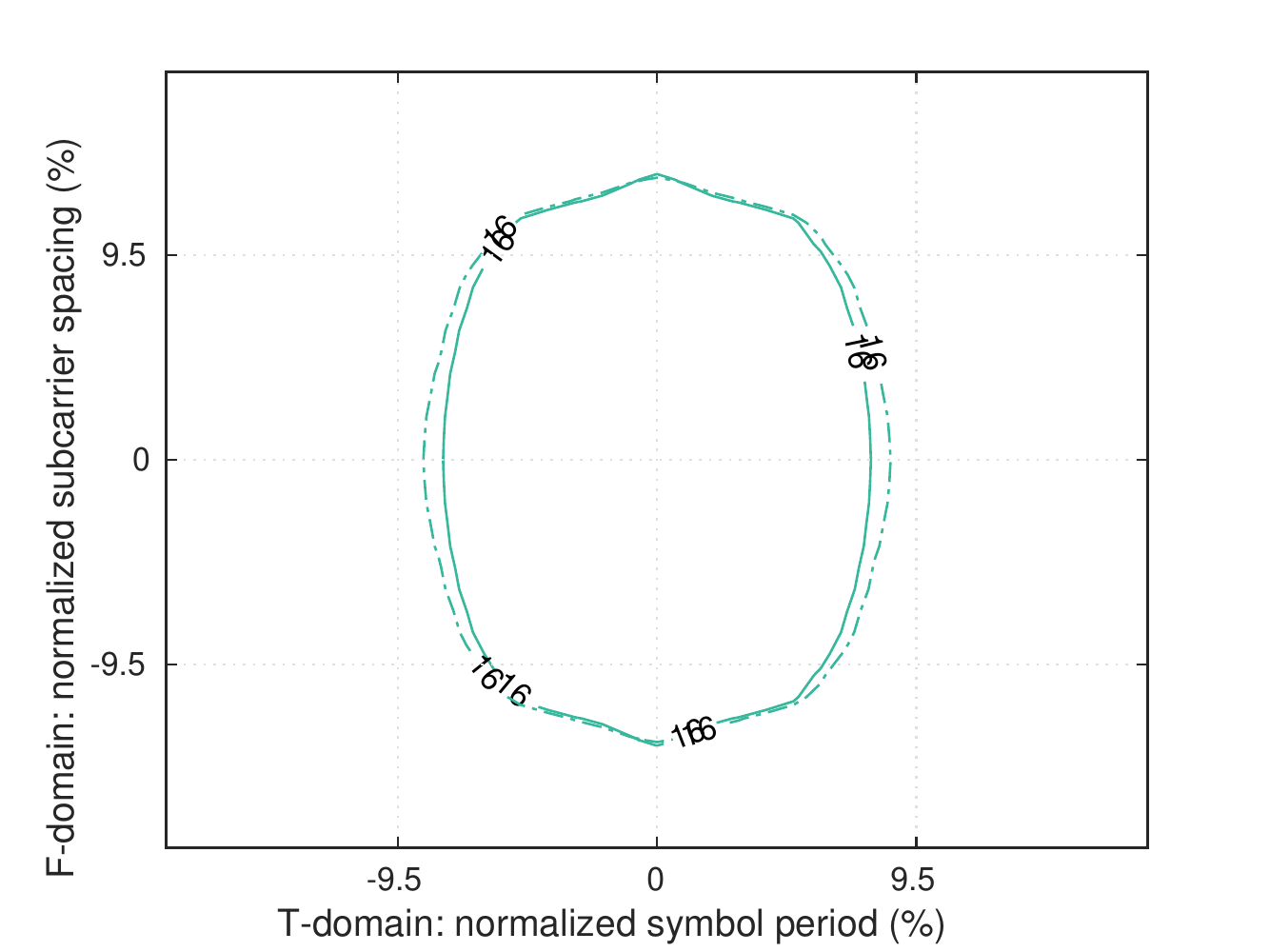}
		\caption{$\sigma_\mathrm{n}^2=-19\mathrm{dB}$}
		\label{fig:contour_T0=0dot5_opt_19dB}
	\end{subfigure}
	\caption{\textcolor{black}{SINR contour of $\mathbf{g}_\text{RECT}$ (solid) and $\mathbf{g}_\text{RC}$ (dashed) w.r.t. max-SINR receiver}}\label{fig:optimized_rx}
	\vspace{-0.8cm}
\end{figure}
\end{itemize}
\begin{table}[!htbp]
	\caption{Simulation parameters for extreme cases}
	\centering
	\label{tab:parameter_extreme}
	\renewcommand{\arraystretch}{1.2}
	\begin{tabular}{|p{.4\linewidth}|p{.1\linewidth}|}
		\hline \textit{Parameters} & \textit{Values}\\
		\hline Number of subcarriers & $128$ \\
		\hline Samples per symbol & $160$ \\
		\hline Filter length & $320$ \\
		\hline Convergence coefficient & $10^{-4}$ \\
		\hline Noise power (low) & 0  \\
		\hline Noise power (high) & $-1\mathrm{dB}$  \\
		\hline Normalized maximum time delay in Fig. \ref{fig:invariant_channel} & $10\%$ \\
		\hline Normalized maximum Doppler shift in Fig. \ref{fig:invariant_channel} & $0$ \\
		\hline Normalized maximum time delay in Fig. \ref{fig:variant_channel} & $5\%$ \\
		\hline Normalized maximum Doppler shift in Fig. \ref{fig:variant_channel} & $\approx1.6\%$ \\
		\hline
	\end{tabular}
\end{table} 
\subsubsection{Joint Transmit and Receive Design}
In this section, we provide several transceiver pulse pairs optimized according to Algorithm \ref{algo}, both for time-invariant and time-varying channels. Detailed simulation parameter setting is presented in Table \ref{tab:parameter_extreme}, in which two extreme noise power levels are selected.

\begin{figure}[!htbp]
	\centering
	\begin{subfigure}[b]{0.45\textwidth}
		\centering
		\includegraphics[scale=0.95]{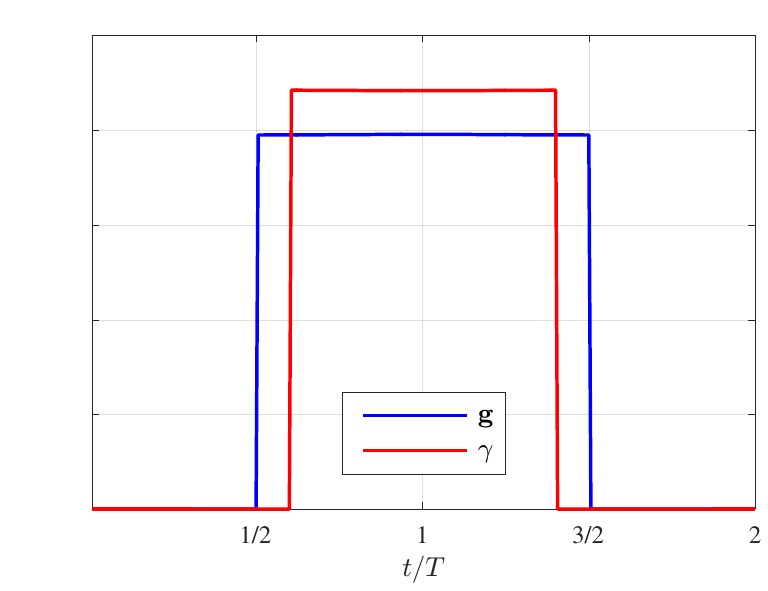}
		\caption{Low noise power} 
		\label{fig:invariant_channel_high_SNR}
	\end{subfigure}
	\begin{subfigure}[b]{0.45\textwidth}
		\centering
		\includegraphics[scale=0.95]{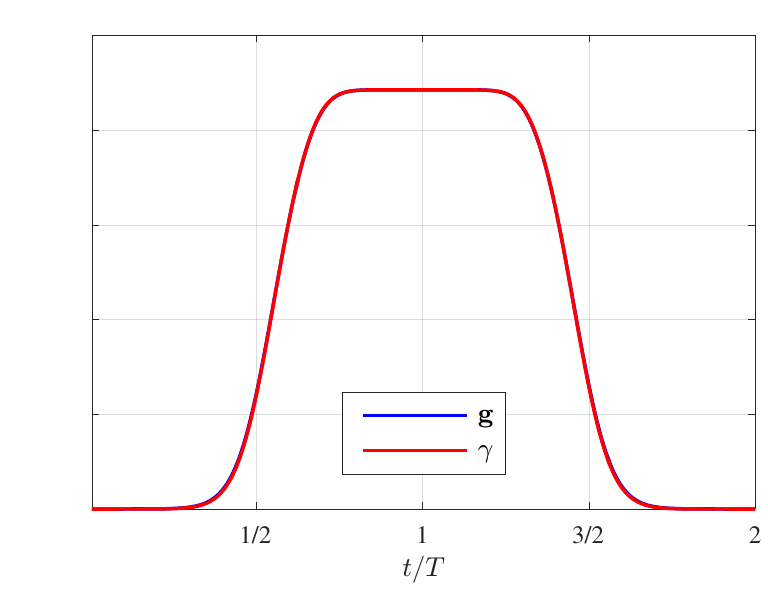}
		\caption{High noise power} 
		\label{fig:invariant_channel_low_SNR}
	\end{subfigure}
	\caption{Pulse shapes designed for time-invariant channel}
	\label{fig:invariant_channel}
	\vspace{-0.8cm}
\end{figure}

Fig. \ref{fig:invariant_channel_high_SNR} and \ref{fig:invariant_channel_low_SNR} depict the computed pulse shapes respectively for low and high noise power levels in time-invariant channels, where the normalized maximum frequency shift is $\nu_\mathrm{max}/F=0$  and the normalized maximum time delay is $\tau_\mathrm{max}/T=10\%$.
An interesting observation is that for the case of low noise power level, the proposed pulses converge to the pulses used in conventional CP-OFDM. This result makes sense since CP-OFDM is known to be optimal in the high SNR scenario with low Doppler spreads. For the case of high noise power, Fig. \ref{fig:invariant_channel_low_SNR} shows the transceiver pulses are close to a matched pulse pair. Intuitive interpretation of this result is that since the SNR loss due to transceiver mismatching becomes dominating in such noise-limited region, matched filtering is desirable.
\begin{figure}[!htbp]
	\centering
	\begin{subfigure}[b]{0.45\textwidth}
		\centering
		\includegraphics[scale=0.95]{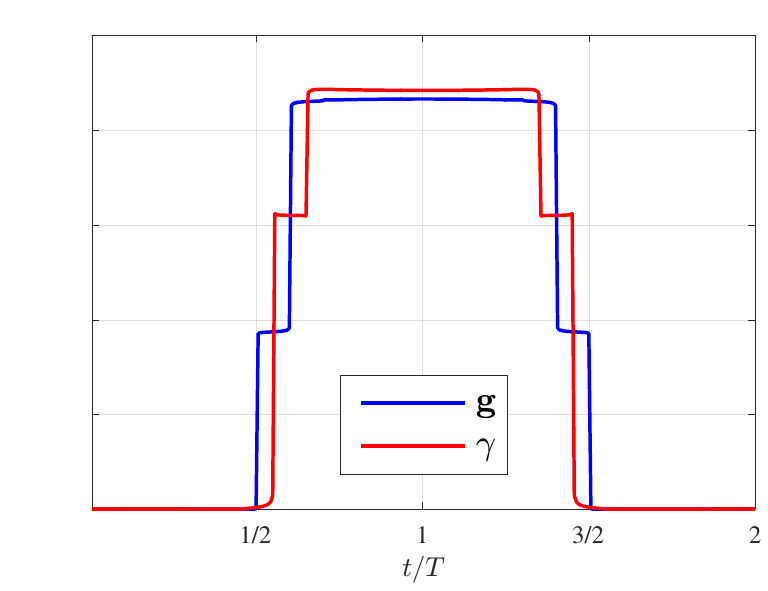}
		\caption{Low noise power} 
		\label{fig:variant_channel_high_SNR}
	\end{subfigure}
	\begin{subfigure}[b]{0.45\textwidth}
		\centering
		\includegraphics[scale=0.95]{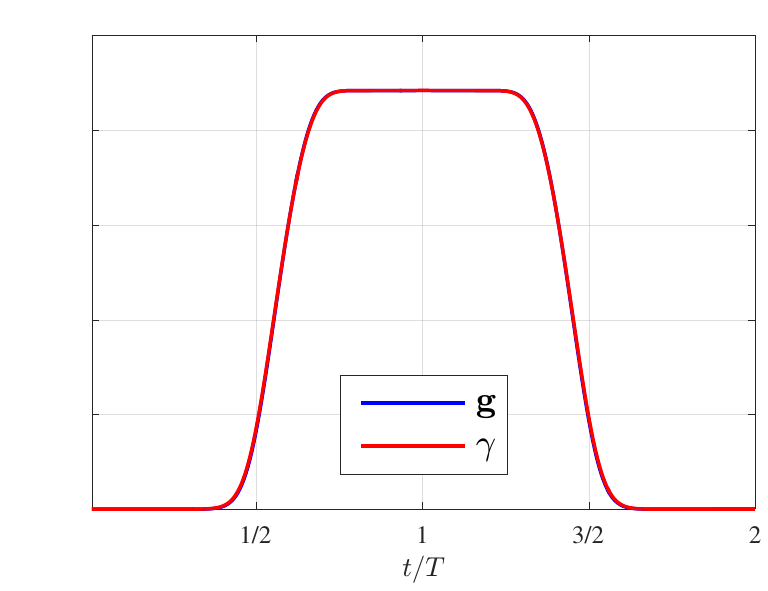}
		\caption{High noise power} 
		\label{fig:variant_channel_low_SNR}
	\end{subfigure}
	\caption{Pulse shapes designed for time-variant channel.}
	\label{fig:variant_channel}
	\vspace{-0.8cm}
\end{figure}

Propagation channels are commonly time-variant in practical communication systems. To evaluate the performance in this case, we select $\tau_\mathrm{max}/T=5\%$ and $\nu_\mathrm{max}\approx1.6\%$ by assuming that an object moves at a relatively high velocity in a medium delay spread environment, e.g., extended vehicular A (EVA) model \cite{361042015}.
Fig. \ref{fig:variant_channel} illustrates the derived pulse shapes for both low and high noise power levels. Similar to Fig. \ref{fig:invariant_channel_high_SNR}, the optimized link-adapted pulse pair for a doubly dispersive channel in high SNR region is also close to rectangular-shaped. However, due to the frequency shifts, both $\mathbf{g}$ and $\boldsymbol{\gamma}$ have ``steps" at the filter head and tail. For the channel with high noise power level, Fig. \ref{fig:variant_channel_low_SNR} shows that $\mathbf{g}$ and $\boldsymbol{\gamma}$ are nearly matched, as can be explained analogous to Fig.\ref{fig:invariant_channel_low_SNR}.

Ideally, pulse shape optimization aims at fulfilling the orthogonal condition, achieve good TFL and SIR/SINR performance. In reality, pulse shapes need to be properly designed according to the system requirements and available resources and channel information. Several exemplary design methods have been addressed in detail in this section. 

\section{Numerology Design and Implementation}
\label{sec:implementation}
\subsection{Numerology Design}
Following the introduction in Section \ref{sec:numerdesign}, the principle of numerology design for pulse shaped OFDM is briefly discussed in this section. Since future mobile communication systems are expected to accommodate multiple services, numerology and pulse shapes may need to be carefully adapted to diverse requirements facing in different services.

Considering the service MCC, short pulse is potentially desirable due to enabling low-latency transmission and fast switching between uplink and downlink. Under such circumstance, pulse shaped OFDM is reduced to W-OFDM or pulse shaped OFDM with small oversampling factor. Consequently, numerology design for pulse shaped OFDM can be hightly based on that of OFDM/W-OFDM, followed by a slight tuning with consideration of designed pulses. Alternatively, for the MTC service, good TFL property is desirable to combat interference introduced by sporadic or random moving of the objects. Therefore, pulse shaped OFDM with relatively long pulse and comparable large $TF$ product is recommended for MTC transmission, enabling more degree of freedom in pulse design to minimize the effects of doubly TF dispersion. Hence, the numerology of MTC needs to be designed according to service requirements and channel characteristics, followed by further amendment with the applied pulses. 
\subsection{Implementation and Complexity}
Using the specification in Fig. \ref{fig:lattQAM} for symbol period $T= N T_\text{s}$ and subcarrier spacing $F= {1}/{M T_\text{s}}$, the transmit and receive signal can be synthesized and analyzed with PPN implementation efficiently (e.g. Fig. \ref{fig:transceiver}). For detailed realization of PPN structure please refer to \cite{VaidyanathanProcIEEE1990}. 
Recalling the definition of oversampling factor $K$, the implementation of the state-of-the-art single and multi-carrier waveforms can be unified with PPN structure, as shown in Table \ref{tb:wfframework}. We remark that, alternatively, a system featuring with multi-rate multi-pulse shaping synthesis/analysis could also benefit from the implementation with frequency oversampling based filter banks \cite{Bellanger2012, Renfors2014}. 

\begin{table} \centering \caption{Summary of SoTA Waveform Complexity by PPN Implementation Framework.}
	\begin{tabular} {|c |p{1cm} |p{2.2cm} | p{3.8cm}| p{3.5cm}|}
		\hline 
		Waveform & $TF$ Density & Overlapping Factor $K$ & Pulse Shape & PPN Implementation \\ \hline
		General pulse shaped OFDM & $\frac{N}{M}$&  Arbitrary $K$ & Arbitrary $g_\text{tx}(t)$& General PPN Framework \\ \hline
		CP-OFDM & $\frac{N}{M}$ & $K=1$ & Rectangular & Downgrades to "add CPP" \\ \hline
		W-OFDM & $\frac{N}{M}$ &  \textcolor{black}{$2>K>1$} & Hamming, RC, RRC, etc. & Downgrades to scalar multiplication \\ \hline
		TF Localized OFDM & $\frac{N}{M}$ & Typically $K\geq 4$ 
		& Orthogonal Gaussian-based function & General PPN Framework \\ \hline
		FBMC/OQAM & 2 & Typically $K\geq 4$ & PHYDYAS, IOTA, etc. & With offset signaling $\frac{TF}{2}$ \\ \hline
	\end{tabular} \label{tb:wfframework} 
	\vspace{-0.5cm}
\end{table}

Furthermore, we exemplify the complexity comparison as follows. Assuming symmetric transceiver pulse design, namely, $g(t)=\gamma(t)$,
and $M=2048, TF=1.07$, the number of operations including complex multiplication and summation for implementing different waveforms are summarized in the Table \ref{tb:complex2048}. 
As seem from the table, the overall complexity overhead introduced by the PPN-based implementation for pulse shaped OFDM is minor compared to CP-OFDM. Taking the whole PHY-layer baseband processing into account, where multi-rate sampling and conversion, MIMO processing, coding and decoding is considered, the complexity overhead  for modulator and demodulator part due to PPN implementation is rather marginal.

\begin{table} [!hbp] \centering \caption{Number of Multiplications for P-OFDM (FFT size $M=2048$, $TF = 1.07$)}
\begin{tabular} {|p{2.8cm} | p{3cm} | p{3cm} |p{2cm} |p{2cm}| }
\hline 
& \multicolumn{4}{c |}{Transmitter} \\ \hline
& IDFT/DFT ($M log_2 M$) & Pulse Shaping ($KN$) & Total & Complexity  \\ \hline
CP-OFDM & 22528 & 0 & 22528 & 100\% \\ \hline
P-OFDM ($K=1$) & 22528 & 288 & 22816 & 101\%  \\ \hline
P-OFDM ($K=2$) & 22528 & 4384 & 26912 & 119\% \\ \hline
P-OFDM ($K=4$) & 22528 & 8768 & 31296 & 139\%  \\ \hline 
\end{tabular} \label{tb:complex2048} 
\vspace{-0.5cm}
\end{table}

\section{\textcolor{black}{Spectrum Containment and Coexistence}}
CP-OFDM systems suffers from poor OOB leakage in its power spectral density (PSD) due to the slow frequency decay property of the rectangular pulse. \textcolor{black}{One intuitive solution to suppress the sidelobes of CP-OFDM is to adopt a subband-wise low pass filtering to shape and fit the subband signal constituted of several adjacent subcarriers into the spectral mask \cite{Schaich2014, Zhangglobecom2015}. However, this filtering method needs to be carefully designed since it usually yields an in-band ISI and thus an EVM loss. Moreover, subband-wise filtering needs to adapt the filter coefficients to the particular width of the subband, rendering it less flexible to adapt to any requirement of a particular application or service.}
 
\begin{figure}[!htbp]
	\centering
	\begin{subfigure}[b]{0.45\textwidth}
		\centering
		\includegraphics[height=6.5cm, width=7cm]{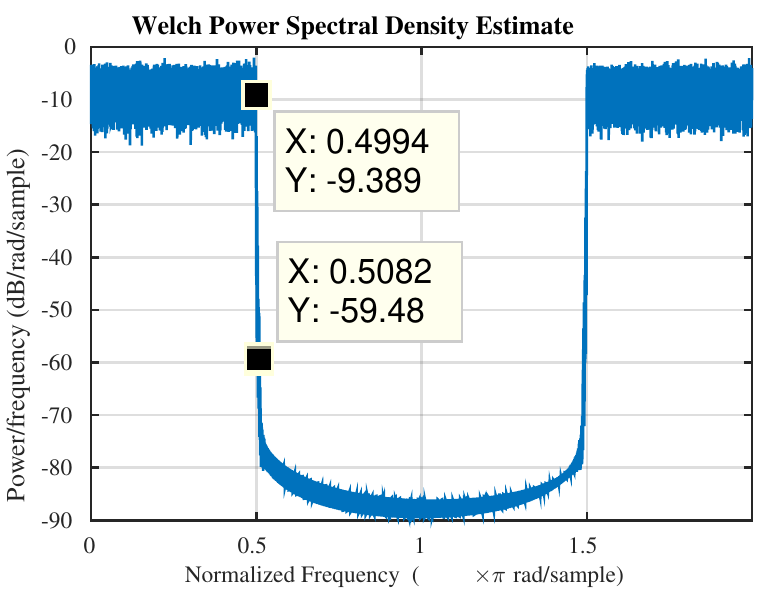}
		\caption{$K=4, TF=1.07$} 
		\label{fig:psd_filter107_K=4_M=2048}
	\end{subfigure}
	\hfill
	\begin{subfigure}[b]{0.45\textwidth}
		\centering
		\includegraphics[height=6.5cm, width=7cm]{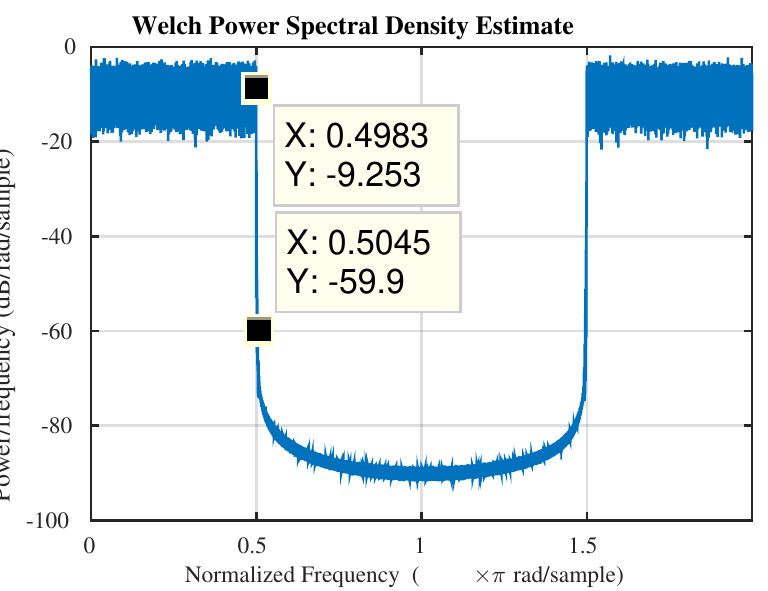}
		\caption{$K=4, TF=1.25$} 
		\label{fig:psd_filter125_K=4_M=2048}
	\end{subfigure}
	\par\medskip
	\begin{subfigure}[b]{0.5\textwidth}
		\centering
		\includegraphics[height=6.5cm, width=7cm]{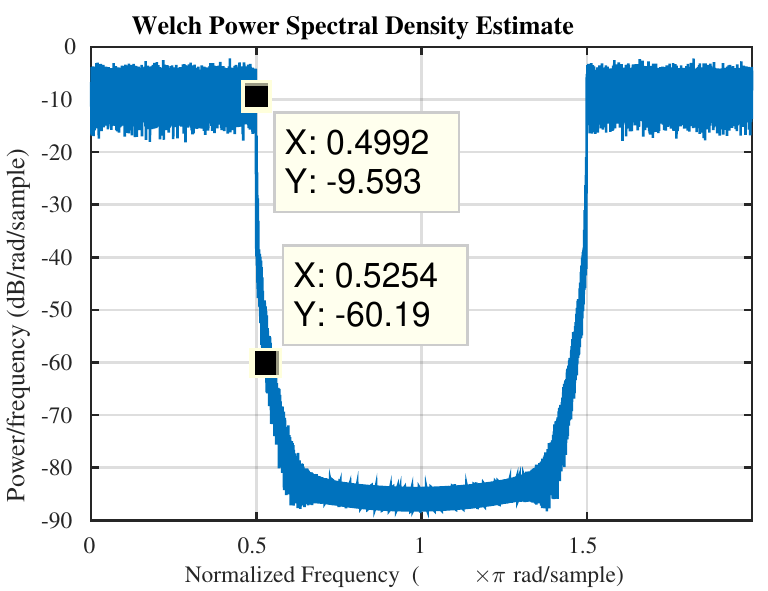}
		\caption{$K=1.07, TF=1.07$} 
		\label{fig:psd_filter107_K=2_M=2048}
	\end{subfigure}
	\hfill
	\begin{subfigure}[b]{0.45\textwidth}
		\centering
		\includegraphics[height=6.5cm, width=7cm]{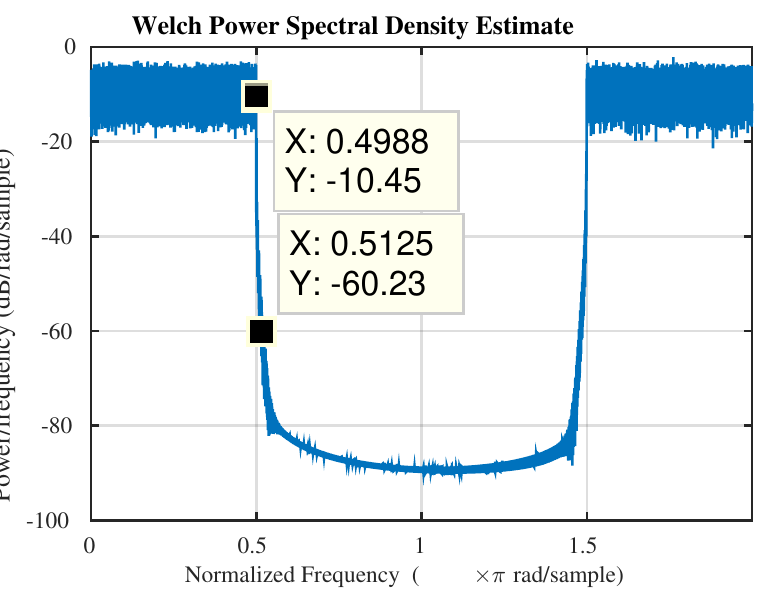}
		\caption{$K=1.07, TF=1.25$} 
		\label{fig:psd_filter125_K=2_M=2048}
	\end{subfigure}
	\caption{\textcolor{black}{PSD analysis of pulse shaped OFDM prototype filters.}}
	\vspace{-0.8cm}
\end{figure}

\textcolor{black}{Alternatively, subcarrier-based filtering as in pulse shaped OFDM can also improve the OOB emission properties of OFDM systems without link performance degradations.}
%
For the case that $K=4$ (see in Fig. \ref{fig:psd_filter107_K=4_M=2048} and \ref{fig:psd_filter125_K=4_M=2048}), where large degree of freedom for constructing the localized filtering is available, the resulting PSD of pulse shaped OFDM is very satisfactory even without the spectral mask filtering, i.e., no EVM loss is incurred. 
For the case that $K=1.07 \approx 1$ (see in Fig. \ref{fig:psd_filter107_K=2_M=2048} and \ref{fig:psd_filter125_K=2_M=2048}), while relatively satisfactory PSD can be achieved so that the number of guard subcarriers for spectral coexistence can be kept small.

\begin{table} [!htbp] \centering \caption{Guard subcarrier requirement (single side) and EVM.}
	\begin{tabular} {|c| p{1.2cm}| p{1.4cm}| p{1.4cm}| p{1.4cm}| p{1.4cm}| p{1.4cm}| p{1.4cm}| p{1.4cm}|}
		\hline 
		& \multicolumn{4}{c|}{$TF=1.07$} & \multicolumn{4}{c|}{$TF=1.25$} \\ \hline
		& Guard Subc. & Overhead (comp. 20MHz) & EVM for Edge Subc. & EVM for Central Subc. & Guard Subc. & Overhead (comp. 20MHz) & EVM for Edge Subc. & EVM for Central Subc. \\ \hline
		$K=4$ & 9 & $0.7\%$ & $-48.9$ dB & $-48.9$ dB & 7 & $0.53\%$ & $-56.8$ dB & $-56.8$ dB \\ \hline
		$K=1.07$ & 27 & $2\%$ & $-57.2$ dB & $-57.3$ dB & 14 & $1.05\%$ & $-55.8$ dB & $-55.8$ dB \\
		\hline 
	\end{tabular} \label{tb:psdguard} 
\end{table}

The above-shown PSD figures shown are based on the LTE setting of 15 KHz subcarrier spacing for 20 MHz bandwidth. Assuming -50 dBc/Hz as the required spectral leakage, the required number of guard sucarriers is summarized in Table \ref{tb:psdguard}.

\begin{figure}[!htbp]
	\centering
	\begin{subfigure}[b]{0.49\textwidth}
		\centering
		\includegraphics[scale=1]{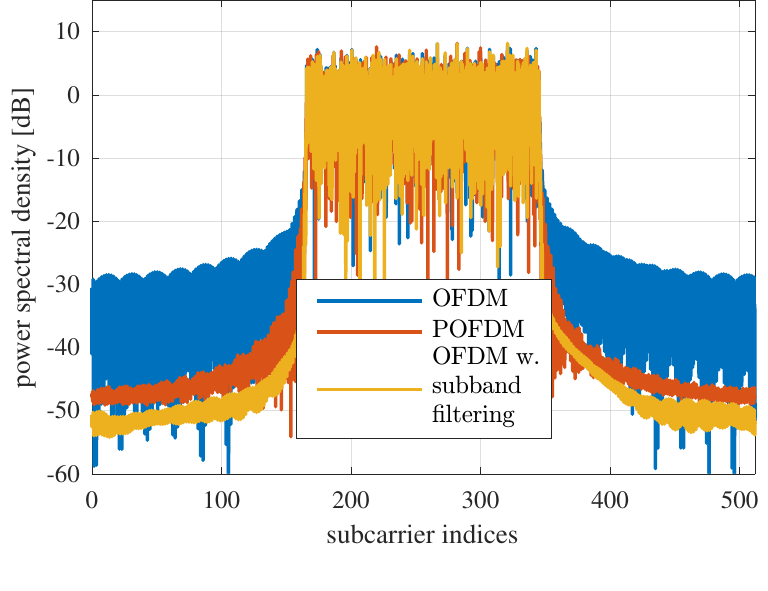}
		\caption{PSD before the PA} 
		\label{fig:psdbeforepa}
	\end{subfigure}
	\begin{subfigure}[b]{0.49\textwidth}
		\centering
		\includegraphics[scale=1]{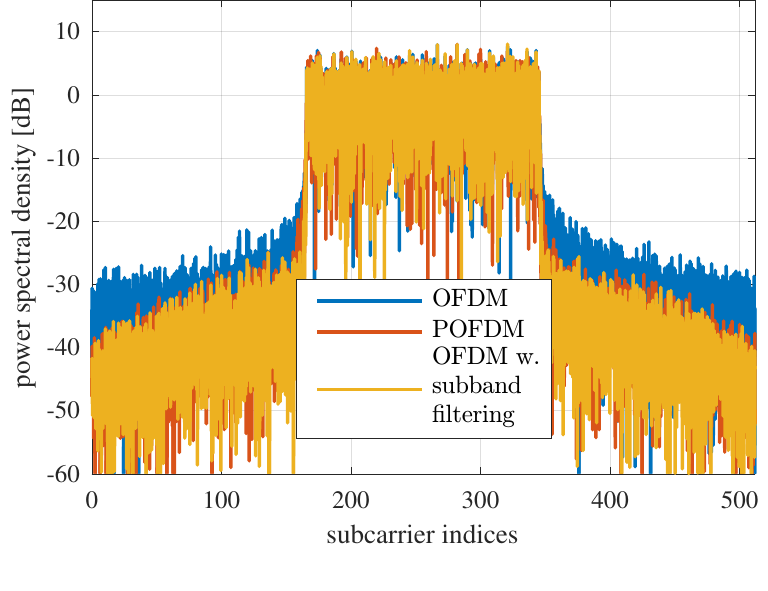}
		\caption{PSD after the PA} 
		\label{fig:psdafterpa}
	\end{subfigure}
	\caption{PSD of different waveforms before and after the nonlinear power amplifier.}
	\label{fig:psdpa}
	\vspace{-0.7cm}
\end{figure}

The linearity of RF unit should be considered for meeting the actual spectral mask requirement. Hence, the PSD under nonlinear power amplifier (PA) is vital to indicate the spectral confinement property for different waveforms. In Fig. \ref{fig:psdbeforepa} and Fig. \ref{fig:psdafterpa}, the PSD performance (before and after the PA) of OFDM, pulse shaped OFDM, and OFDM with subband-filtering are shown, respectively. The product $TF$ is set to 1.07 for the first two waveforms and $K=1$ is used for pulse-shaped OFDM, while OFDM with subband-filtering employs half-symbol length FIR filter.
We observe that  the OFDM system with subband-filtering shows better spectral containment compared to pulse shaped OFDM and OFDM. Nevertheless, 
this advantage in OOB emission becomes marginal if taking PA non-linear effects into account.

\textcolor{black}{Aiming at better spectral containment of the OFDM signals, pulse shaped OFDM or windowed-OFDM is an effective method, facilitating a suitable trade-off between improved OOB emission and link-level performance.} We remark that ror a more aggressive spectrum usage requiring minimum guard subcarrier overhead, additional subband-wise filtering can also be applied to pulse shaped OFDM signal. However, the trade-off between EVM, OOB leakage, and particularly the linearity for RF unit (cost and power efficient) at both BS/UE sides should be carefully reviewed.

\section{Application Examples}
In the section, we provide some applications of pulse shaped OFDM and evaluate the link performance in respective scenarios.
\label{sec:appexamples}
\subsection {Uplink TA-free Access}

\begin{figure}[!htbp] \centering 
	\centering	\includegraphics[scale=0.3]{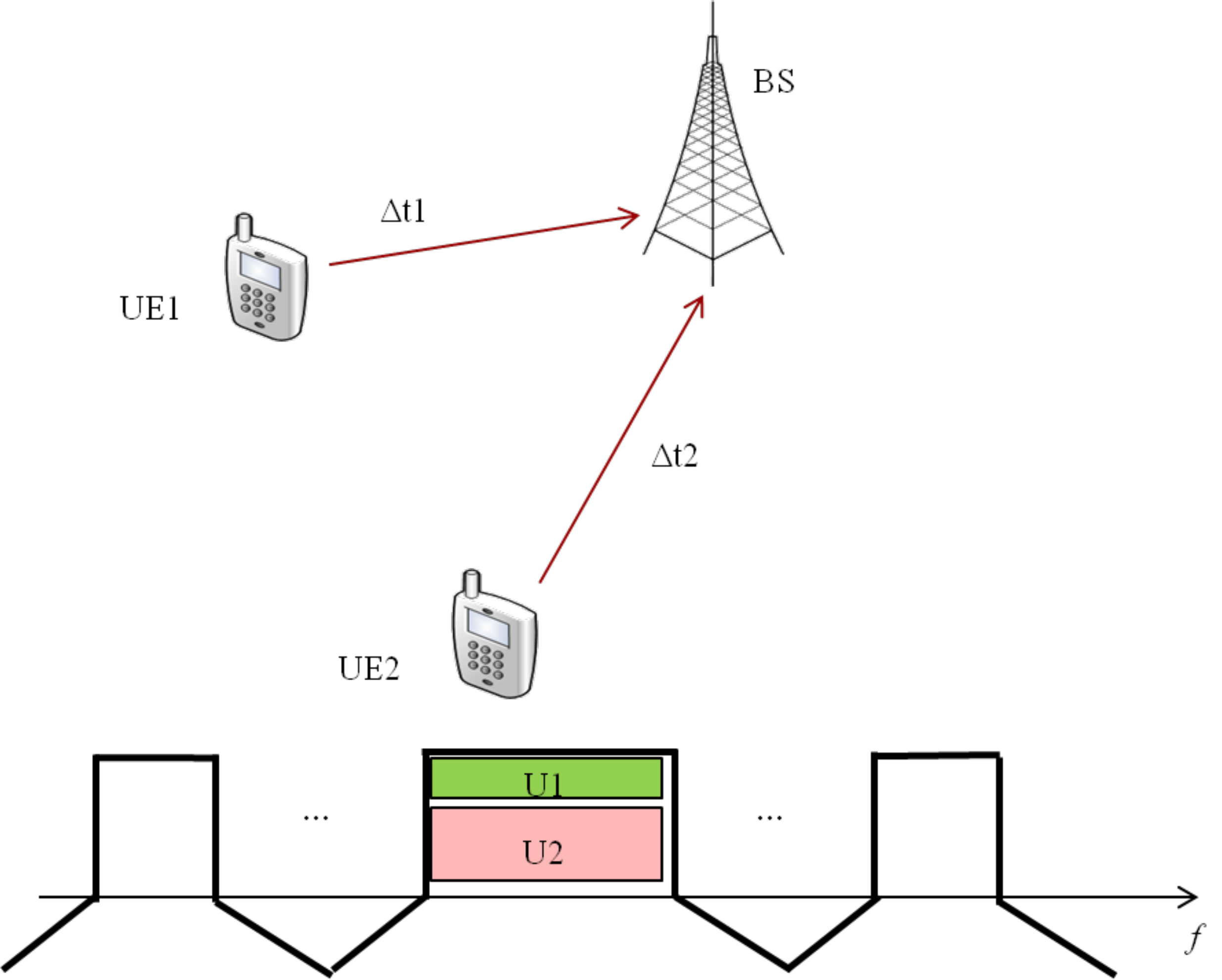}
	\caption{Scenario for uplink massive access without TA.} 
	\label{fig:taul}
	\vspace{-0.6cm}
\end{figure}

In current uplink transmission, due to radio propagation latency, the timing misalignment is present for the uplink signal arrival to the Base Station, unless an close-loop TA adjustment is performed. For example, if a cell radius is about $1732$m, TA misalignment could be in a range of $0\sim 13\mu$s.
In the circumstances of massive machine connections, each UE needs only to send a small data packet for a long period of time. The one-by-one TA adjustment procedure is becoming an overhead to the system, especially when the UE mobility is considered. 

\begin{figure} [!htbp]
	\centering
	\begin{subfigure}[b]{0.8\textwidth}
		\centering
		\includegraphics[scale=1]{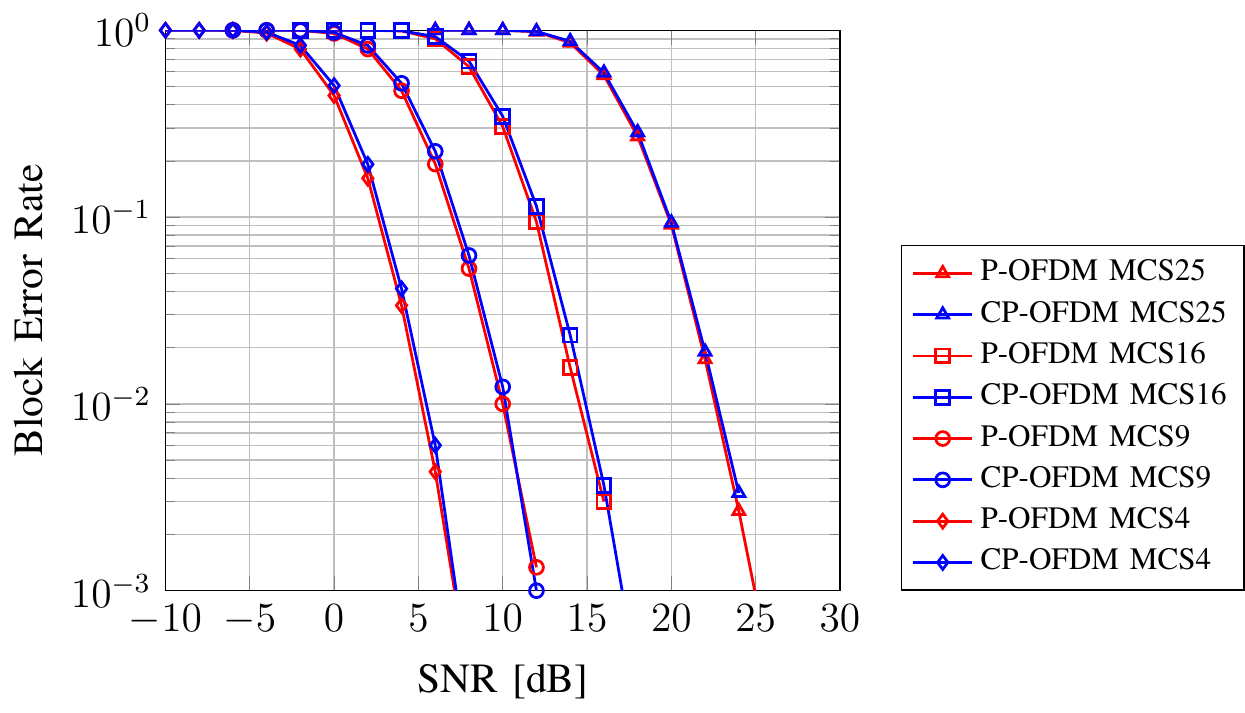}
		\caption{TA-sync cases: 1U2R ETU channel 3kmh.} 
		\label{fig:ta1t2rsync}
	\end{subfigure}
	\hfill
	\begin{subfigure}[b]{0.8\textwidth}
		\centering
		\includegraphics[scale=1]{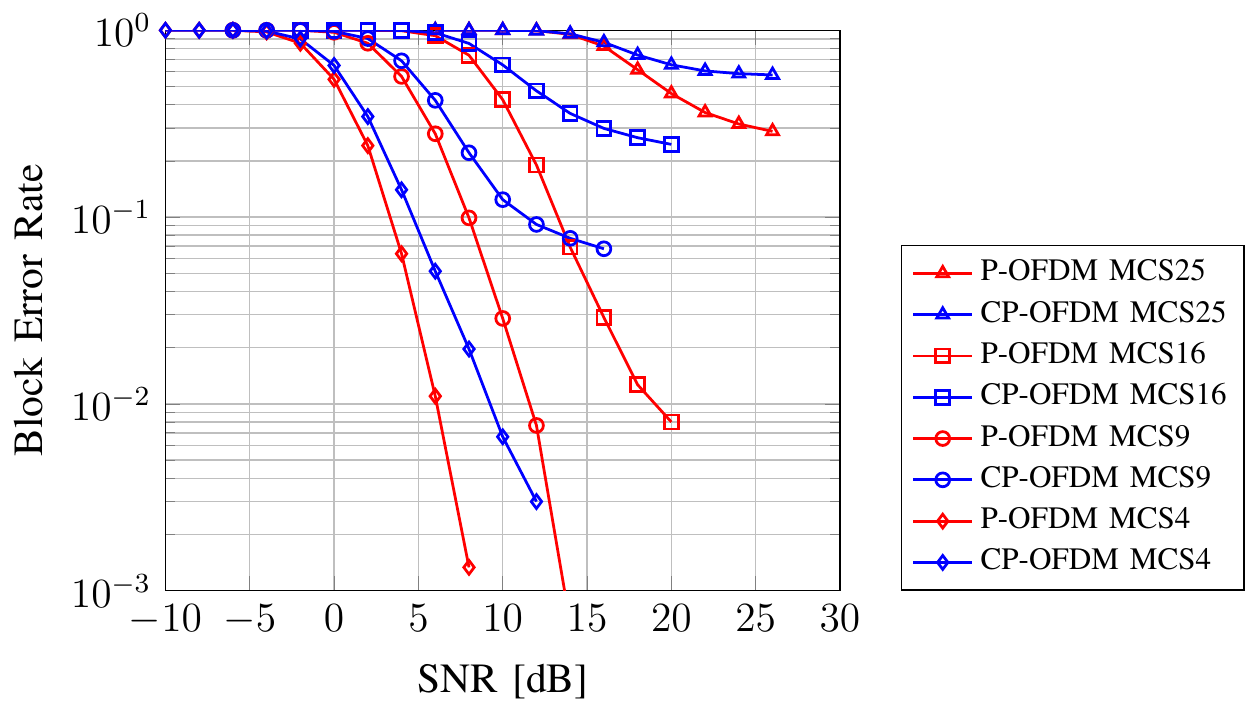}
		\caption{TA-unsync cases: 1U2R ETU channel 3kmh.} 
		\label{fig:ta1t2runsync}
	\end{subfigure}
	\caption{Comparison vs. CP-OFDM for 1U2R.}\label{fig:tafree1t2r}
		\vspace{-0.6cm}
\end{figure}

\begin{figure} [!htbp]
	\centering
	\begin{subfigure}[b]{0.8\textwidth}
		\centering
		\includegraphics[scale=1]{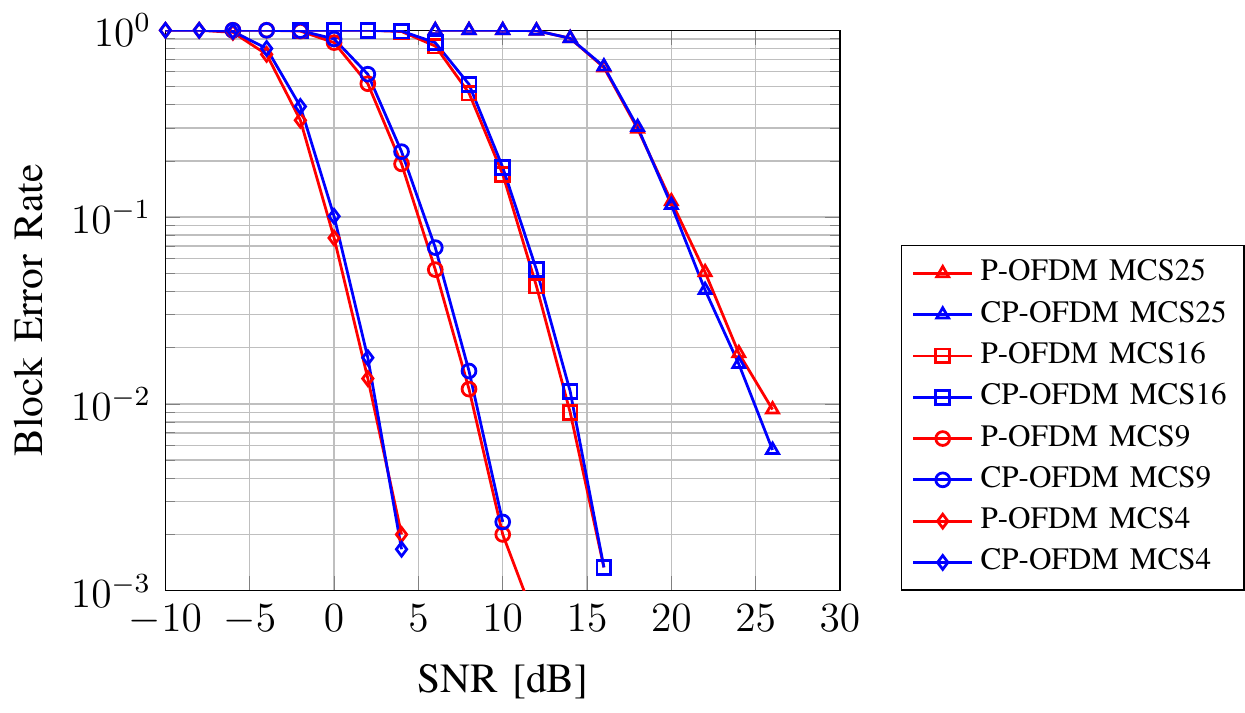}
		\caption{TA-sync cases: 2U4R ETU channel 3kmh.} 
		\label{fig:ta2t4rsync}
	\end{subfigure}
	\hfill
	\begin{subfigure}[b]{0.8\textwidth}
		\centering
		\includegraphics[scale=1]{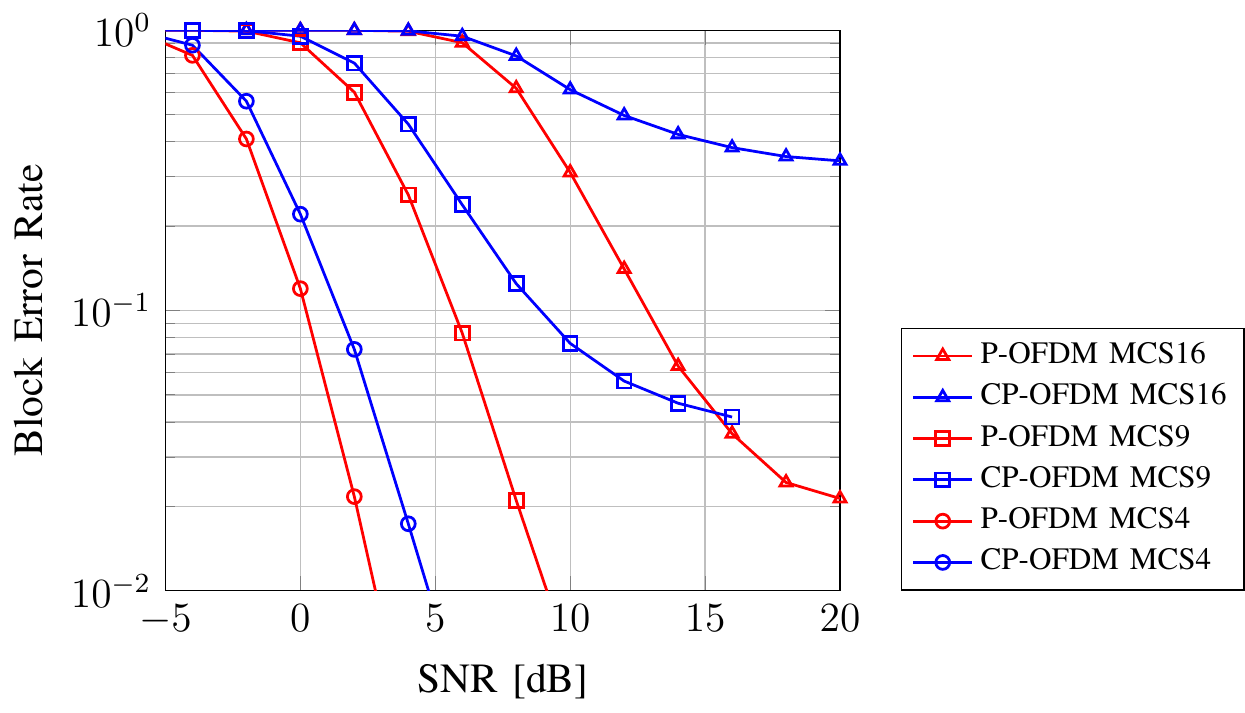}
		\caption{TA-unsync cases: 2U4R ETU channel 3kmh.} 
		\label{fig:ta2t4runsync}
	\end{subfigure}
	\caption{Comparison vs. CP-OFDM for 2U4R.}\label{fig:tafree2t4r}
	\vspace{-0.6cm}
\end{figure}

From Fig. \ref{fig:sir_go}, we observe that pulse-shaped OFDM with long pulse can be utilized to enable uplink TA-free or relaxed TA transmission. This is particularly useful when combining non-orthogonal multiple access (NOMA) with different users, where the base stations can hardly synchronize to each user at receiver side under reasonable complexity \cite{ZhaoAsilomar2015}. See 
Fig. \ref{fig:taul} for the scenario and 
Table \ref{tb:uplinksim} for simulation assumptions. The pulse shape is depicted in Fig. \ref{fig:go_gausst_107_k4} \textcolor{black}{with the overlapping factor $K=4$}.
The simulation results in Fig. \ref{fig:tafree1t2r} and \ref{fig:tafree2t4r} confirms the advantages for pulse shaped OFDM in the scenarios with substantial link performance gain $3\sim 5$ dB.

\begin{table} [!htbp] \centering \caption{Link level simulation for uplink TA-free access.}
	\begin{tabular} {|c |l |}
		\hline 
		System bandwidth & 10 MHz \\ \hline
		Duplex & FDD UL \\ \hline
		Cell Size & 1732 m \\ \hline
		TA error (open loop) & $0\sim 13.3 \mu$s \\ \hline
		Subcarrier Spacing & 15 KHz \\ \hline
		$TF$ & 1.07 \\ \hline
		\multirow{2}{*}{Antenna Configuration} & 1 Tx at UE \\ & 2 Rx at BS \\ \hline
		User Configuration & 1 or 2 UE \\ \hline
		PRB Allocation & 15 PRBs to one UE \\ \hline 
		MIMO Mode & SIMO-MU-MIMO \\ \hline
		Channel estimation & Real channel and noise estimation (LS based) \\ \hline
		MCS & LTE MCS 4,9,16 \\ \hline
		Channel Models & ETU 3km/h uncorrelated \\ \hline
		Hybrid ARQ & Not modelled \\ \hline
		Receiver & LMMSE or QRD-ML \\ \hline
		Reference signal & LTE R-8 DL CRS  \\ \hline
		\hline
	\end{tabular} \label{tb:uplinksim} 
\end{table}

\subsection {HST/V2V with High Mobility}
\begin{figure} [!htbp] \centering 
	\centering	\includegraphics[scale=0.3]{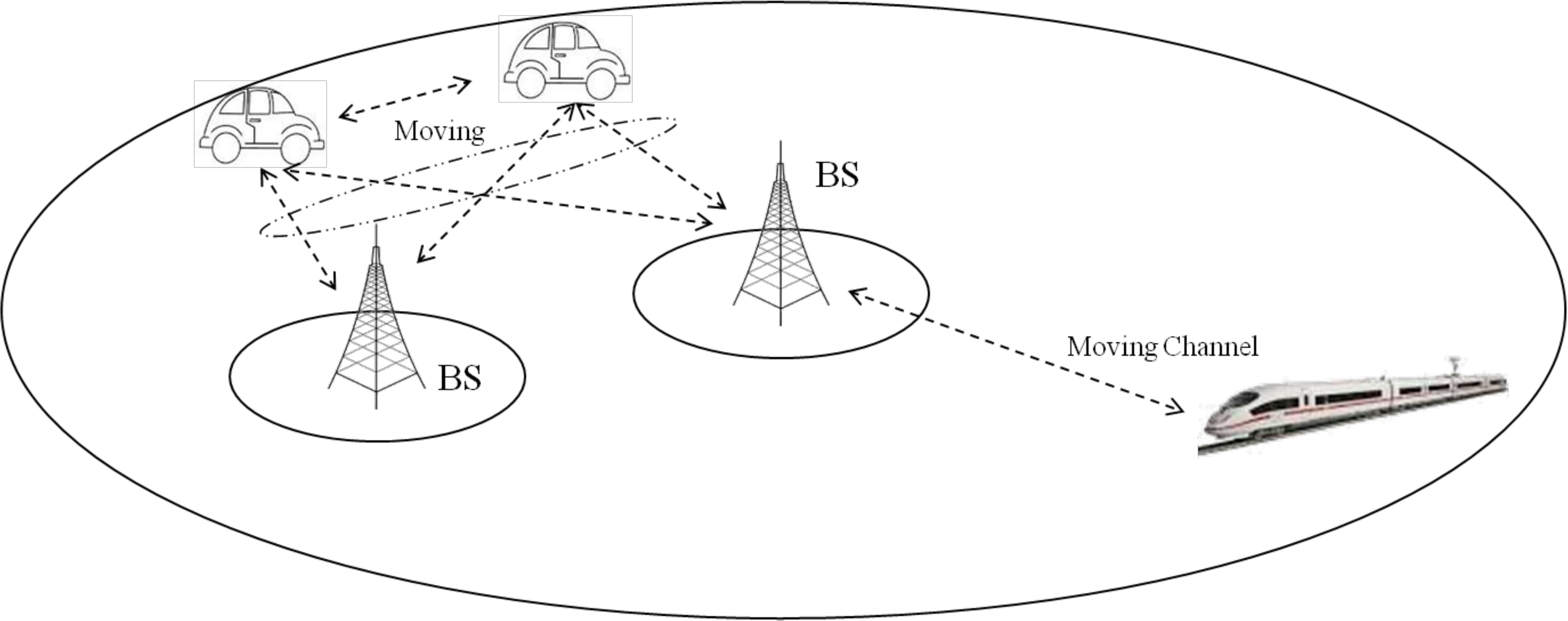}
	\caption{High Mobility scenario for HST/V2V.} 
	\label{fig:v2v}
	\vspace{-0.5cm}
\end{figure}

High mobility scenarios become of great importance for future wireless communications. The high speed train (HST) and vehicular-to-vehicular (V2V) scenarios are illustrated in Fig. \ref{fig:v2v}.

\begin{figure} [!htbp] 
	\centering
    \includegraphics[scale=1]{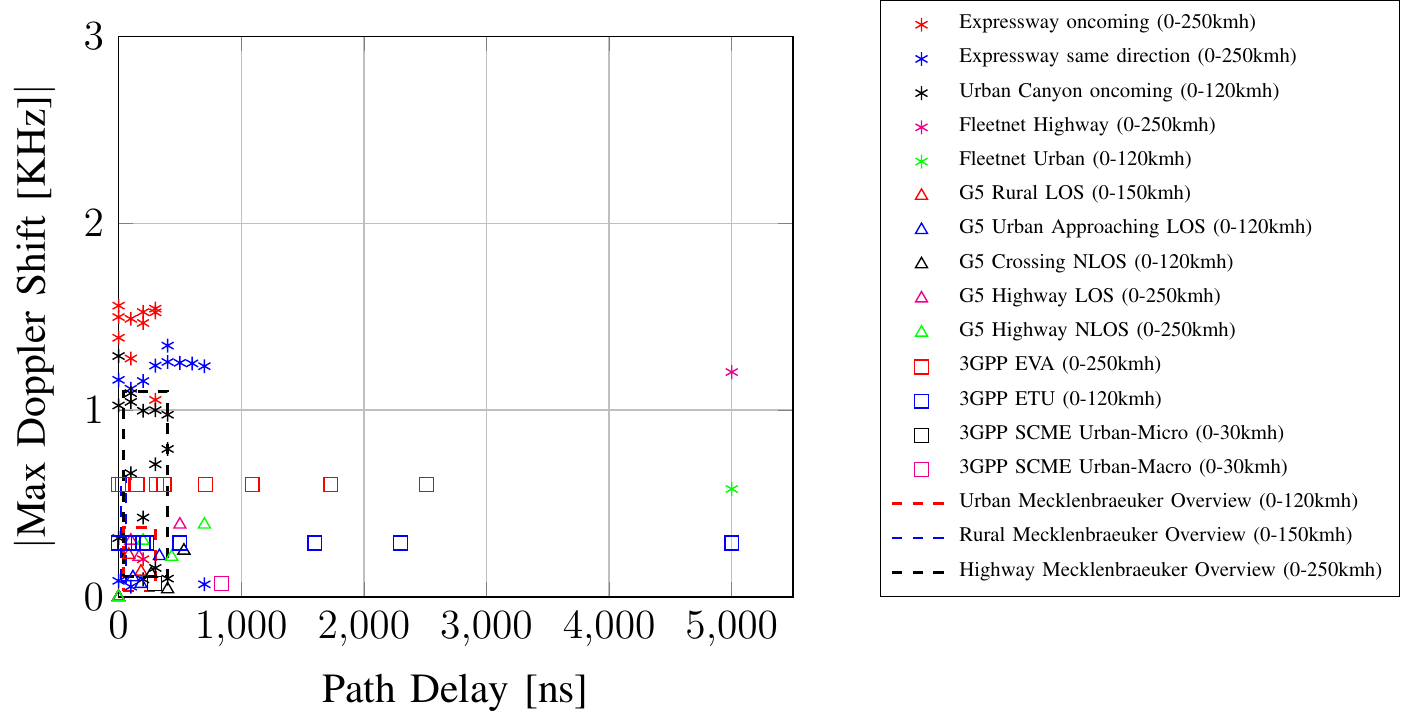}
	\caption{Doppler vs. Delay relationship of cellular and V2V channels.} 
	\label{fig:v2xchan}
		\vspace{-0.6cm}
\end{figure}

To derive a reasonable ($T, F$) setting and overhead for this scenario, it is necessary to understand the channel first. In this scenario, as high mobility of the objects is involved, the channels are often characterized as "doubly dispersive". Based on the modeling report \cite{Mecklenbrauker2011,Acosta-Marum2007,D.S.Baum2005}, the possible channel TF modeling under the fading effect can be summarized in the Fig. \ref{fig:v2xchan}.
From the channel modeling, we consider that the $(T, F)$ lattice should be adjusted best between 60 KHz and 75 KHz for an isotropic design with $TF=1.25$ for guaranteed performance in many sub-scenarios each extreme high velocity happens (LTE uses 15KHz with $TF=1.07$, 802.11p uses 156 KHz with $TF=1.25$). The pulse shape is depicted in Fig. \ref{fig:go_gausst_k4}. Refer to Table \ref{tb:hstsim} and Table \ref{tb:v2vsim} concerning the corresponding parameter setting for link performance comparison. 

\begin{table} [!htbp] \centering \caption{Link Level Simulation for High Speed Train}
	\begin{tabular} {|c |l |}
		\hline 
		System bandwidth & 10 MHz \\ \hline
		Duplex & FDD DL \\ \hline
		Subcarrier Spacing & 15 KHz \\ \hline
		$TF$ & 1.07 \\ \hline
		\multirow{2}{*}{Antenna Configuration} & 2 Tx at BS \\ &2 Rx at UE  \\ \hline
		PRB Allocation & 15 PRBs to one UE \\ \hline 
		MIMO Mode & Full Rank Open Loop-MIMO \\ \hline
		Channel estimation & Real channel and noise estimation \\ \hline
		MCS & LTE MCS 4, 9, 16 \\ \hline
		Channel Models & 3GPP EVA 500km/h low correlation \\ \hline
		Hybrid ARQ & Not modelled \\ \hline
		Receiver & LMMSE or QRD-ML \\ \hline
		Reference signal & LTE R-8 DL CRS  \\ \hline
		\hline
	\end{tabular} \label{tb:hstsim} 
		\vspace{-0.2cm}
\end{table}

\begin{table} [!htbp] \centering \caption{Link Level Simulation for Highway Vehicular to Vehicular}
	\begin{tabular} {|c |l |}
		\hline 
		System bandwidth & 10 MHz \\ \hline
		Duplex & TDD \\ \hline
		Subcarrier Spacing & 60 KHz \\ \hline
		$TF$ & 1.25 \\ \hline
		\multirow{2}{*}{Antenna Configuration} & 1 Tx at BS \\ &1 Rx at UE  \\ \hline
		PRB Allocation & 15 PRBs to one UE \\ \hline 
		MIMO Mode & Full Rank Open Loop-MIMO \\ \hline
		Channel estimation & Real channel and noise estimation \\ \hline
		MCS & LTE MCS 4, 9, 16, 25 \\ \hline
		Channel Models & 802.11p 250km/h Onway \\ \hline
		Hybrid ARQ & Not modelled \\ \hline
		Receiver & LMMSE or QRD-ML \\ \hline
		Reference signal & LTE R-8 DL CRS  \\ \hline
		\hline
	\end{tabular} \label{tb:v2vsim} 
		\vspace{-0.6cm}
\end{table}

\begin{figure}[!t]
	\centering
	    \includegraphics[scale=1]{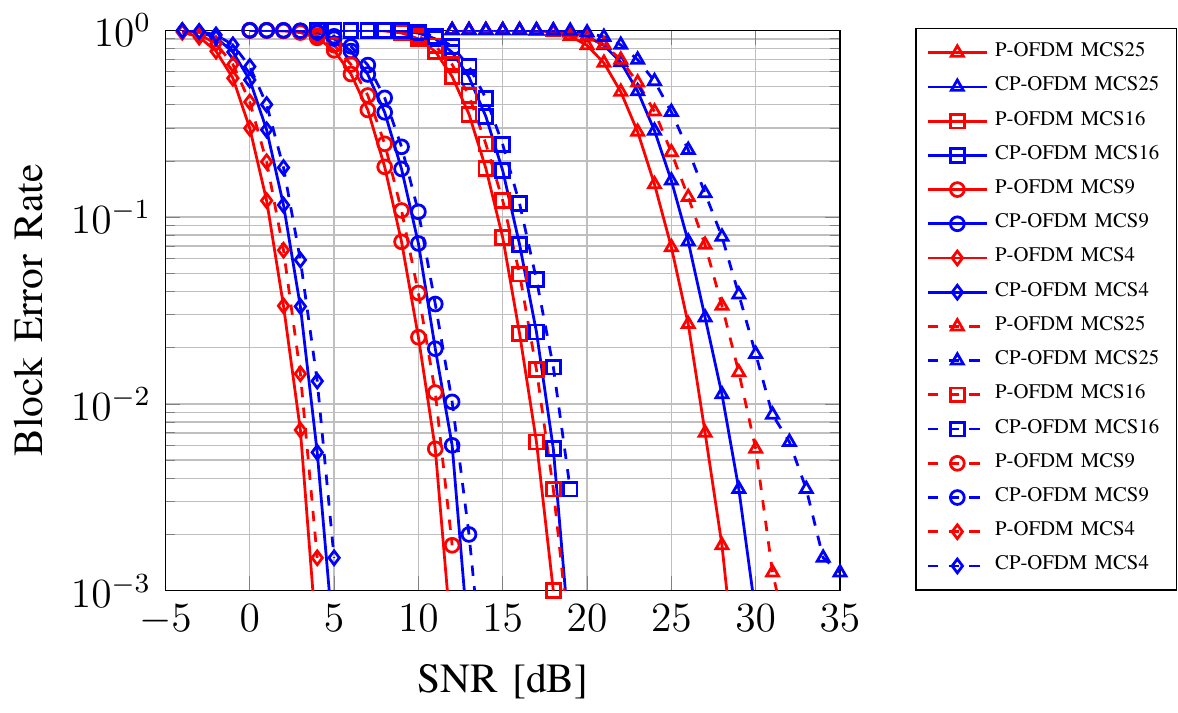}
	\caption{Link level performance for high speed train (EVA channel, 500Kmh, 60KHz subcarrier spacing).}
	\label{fig:bler1t1reva500kmh60khz}
		\vspace{-0.7cm}
\end{figure}

\begin{figure}[!t]
	\centering
		    \includegraphics[scale=1]{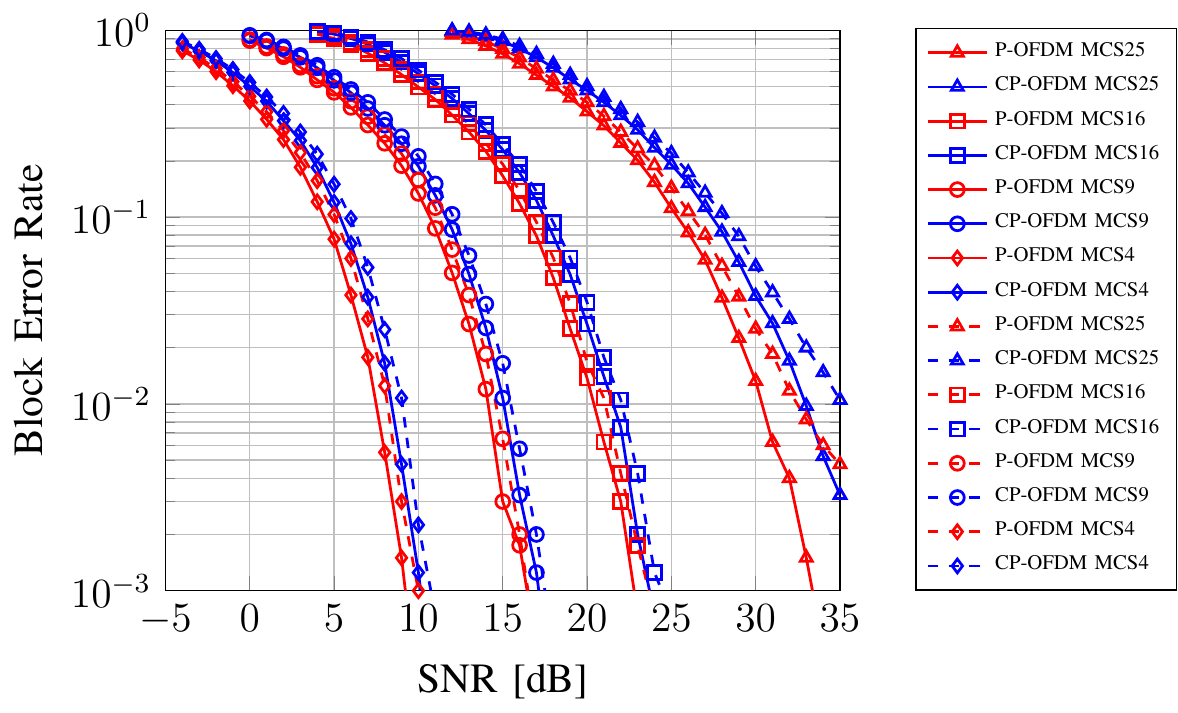}
	\caption{Link level performance for Highway V2V (250Kmh, 60KHz subcarrier spacing).}
	\label{fig:bler1t1ro2o500kmh60khz}
	\vspace{-0.7cm}
\end{figure}

From the BLER-SNR performance depicted in
Fig. \ref{fig:bler1t1reva500kmh60khz} and \ref{fig:bler1t1ro2o500kmh60khz} (solid- ideal channel estimation, dash- LS-based channel estimation), we see about 1-3 dB performance gain by pulse shaped OFDM due to TF well-localized pulse shape design.

\section{Conclusion}
\label{sec:conclusion}
\textcolor{black}{OFDM family contains the most promising waveform candidates for future 5G system due to its superior performance w.r.t. spectral efficiency, good MIMO compatibility, high flexibility, and efficient implementations. While maintaining the advantages of CP-OFDM and meliorating its drawbacks, such as poor OOB emission and sensitivity to frequency distortions and temporal distortions beyond CP region, pulse shaped OFDM with the proposed pulse design methods has been shown to be an effective solution.}

\textcolor{black}{Different from CP-OFDM where CP-overhead is used to combat temporal distortions like ISI and timing offsets, the idea of pulse shaped OFDM is to exploit pulse shaping to provide a trade-off for the robustness against both temporal and spectral distortions. Exemplary pulse shaping methods have been elaborated on, taking into consideration different design criteria and the practical receiver complexity. The common target is to achieve better time-frequency localization of the subcarrier signals, yielding better spectral containment of the OFDM signals and improved SINR performance. For example, taking a LTE system with 20 MHz bandwidth, only 100 resource blocks (RBs) are used (amounting to 18 MHz signal bandwidth). By using pulse shaped OFDM, up to 108 RB can be used to fit the spectrum mask without performance degradation, yielding ca. 8\% spectrum efficiency gain.}

Two major degrees of freedom can be exploited in the OFDM multicarrier waveform design, namely its numerology and its transceiver pulse shape. The numerology mainly aims at designing the time-frequency operational range, while the pulse shape is for further refining the time-frequency localization for matching particular system requirements. 
Table \ref{tb:wf2g5g} reviews the transmit waveform specified by 3GPP mobile system standards. If flexibility and forward-compatibility are considered as vital properties for future 5G waveforms, flexible OFDM based waveform with configurable sets of its numerology settings and pulse shapes
can be considered an ideal candidate.

\begin{table}[!ht] \centering \caption{Summary of  transmit signal model in the standardized digital mobile systems.} 
\begin{tabular} {|p{1.5cm} |p{1.6cm} | p{1.6cm} | p{1.6cm} | p{1.7cm}| p{5.5cm}|}
\hline 
Generation & Modulation & Frequency Spacing ($F$) & Symbol Period ($T$) & Pulse Shaping $g_\text{tx}(t)$& Signal Generation $s(t)$ (Simplified) \\ \hline
2G & GMSK & \multicolumn{2}{c |}{Single Carrier Only} & {Laurent pulses} & $s(t)\approx \sum_n j^{A} g_\text{tx}(t-nT) $ \\ \hline
3G & DSSS & \multicolumn{2}{c |}{Single Carrier Only} & RRC & $s(t)= \sum_n a_n \sum_c s(c) g_\text{tx}(t-n T/N)$	\\ \hline
\multirow{2}{*}{4G} &  {CP-OFDM (DL)} & $15$ KHz or $7.5$ KHz & $TF=1.07$ or $1.25$ & Rectangular & $s(t)= \sum_m \sum_n a_{m,n} g_\text{tx}(t-nT) e^{j2\pi  m F t}$\\ \cline{2-6}
& DFTs-OFDM (UL) & \multicolumn{2} {c |}{Single Carrier $TF=1.07$ or $1.25$ }  & Dirichlet & $s(t)= \sum_n a_n g_\text{tx}(t-nT)$ \\ \hline
5G &? & {Configurable set?} &  {Configurable set?} & {Configurable set? } & $s(t)= \sum_m \sum_n a_{m,n} g_\text{tx}(t-nT) e^{j2\pi m F t}$ \\ \hline
\end{tabular} \label{tb:wf2g5g} 
\end{table}

\appendices
\section{Overview of Different Candidate Waveform Proposals}
\label{sec:relationmc}
Pulse shaped OFDM generalizes several state-of-the-art OFDM-based waveform candidates for future mobile systems. Here we detail this relation as follows.
\begin{enumerate}
	\item \textbf{"CP-OFDM"} is a special case of pulse shaped OFDM, where $g(t)$ and $\gamma(t)$ are rectangular pulse shapes with the overlapping factor $K=1$.
	Specifically, prototype filters $g_\text{cpofdm}\left(t\right)$ and $\gamma_\text{cpofdm}\left(t\right)$ are given by\footnote{For simplifying the analysis, causality of the system is firstly ignored, and thus CP-OFDM is modeled as half-prefixed and half-suffixed OFDM.} 
	\begin{align}
	g_\text{cpofdm}\left(t\right)&=\begin{cases}\begin{array}{lcl}
	\frac{1}{\sqrt{T}}\qquad\quad & & \textrm{for}\ t\in\left[-\frac{T}{2},\frac{T}{2}\right] \\ \nonumber
	0 & & \textrm{otherwise} \end{array}\end{cases} \\
	\gamma_\text{cpofdm}\left(t\right)&=\begin{cases}\begin{array}{lcl}
	\frac{1}{\sqrt{T-T_\mathrm{cp}}} & & \textrm{for}\ t\in\left[-\frac{T-T_\mathrm{cp}}{2},\frac{T-T_\mathrm{cp}}{2}\right] \\ 
	0 & & \textrm{otherwise} \end{array}\end{cases}
	\end{align}
	and the spectral efficiency is proportional to $1/TF = {\left(T-T_\mathrm{cp}\right)}/T$.
	Note that applying the transmit and receive pulses $g_\text{cpofdm}\left(t\right)$ and $\gamma_\text{cpofdm}\left(t\right)$ are equivalent to the ``CP addition" and ``CP removal" operations in CP-OFDM technology.
	In an additive white Gaussian noise (AWGN) channel, the discrepancy of transmit and receive pulses, namely, $g_\text{cpofdm} \neq \gamma_\text{cpofdm}$, there is a mismatching SNR loss following Cauchy-Schwarz inequality.
	Using the common setting in LTE systems with $7 \%$ or $25 \%$ CP overhead, the mismatching SNR loss is about
	0.3dB for $TF = 1.07$, while about 1dB for $TF = 1.25$.
	\item \textbf{"ZP-OFDM"} is also a special case of pulse shaped OFDM, where the transmit pulse $g_\text{zpofdm}(t)$ is a rectangular pulse of length $L=T$ and the receive filter $\gamma_\text{zpofdm}(t)$ is also rectangular shaped with length $T+T_\mathrm{ZP}$. The overlapping
	factor is $K=1$.
	\item \textbf{"Windowed-OFDM"} can be also considered as a special case under pulse shaped OFDM framework, with overlapping factor $1<K<2$ (usually $K$ is slightly larger than 1). The pulse shape can be flexibly adjusted.
	\item \textcolor{black}{\textbf{Fitlered multitone (FMT)} is a pulse shaped OFDM system with frequency localized filters; the length of filter may thus be very long in time and may thus overlap with successively transmitted symbols. \cite{SahinSurvey2014}.}
	\item \textbf{"DFTs-OFDM"} is a special case of pulse shaped OFDM where a single carrier modulation is used ($M=1$). \textcolor{black}{The pulse shaping is carried out with a circular convolution, which corresponds to periodically time-varying filters.} The transmit pulse $g(t)$ can be considered as the Dirichlet sinc function. The $k$th DFT spreading block is upsampled with ${N_\text{IDFT}}/{N_\text{DFTs,k}}$ where ${N_\text{IDFT}}$ and $N_\text{DFTs,k}$ are number of subcarriers of IDFT block and the size of DFT spreading block, respectively. 
	\item \textbf{"ZT-OFDM"}\cite{Berardinelli2013} is an extended single carrier modulation ($M=1$) based on "DFTs-OFDM", where the transmit pulse $g(t)$ can be considered also as a $N_\text{zp}$-expanded Dirichlet sinc function. Similar to DFTs-OFDM, the upsampling ratio for $k$th DFT
	spreading block is ${N_\text{FFT}}/{N_\text{DFTs,k}+N_\text{zp}}$.
\end{enumerate} 

\section{EVM Requirements for Mobile Communications}
\label{sec:appendixevm}
In \cite{361042015}, error vector magnitude (EVM) indicates a measurement of the difference between the ideal and measured symbols after equalization. 
Following its definition, relationship between the required EVM and SINR without channel and noise effect (in linear scale) is given by
\begin{equation}
\text{EVM}^2=\frac{1}{\text{SINR}}.
\end{equation}

The limit of the EVM of each E-UTRA carrier for different modulation schemes on PDSCH \cite{361042015} along with the associated minimum SINR are summarized in the second and the third columns of Table \ref{tab:evmrequirement}, respectively.
\begin{table}[!htbp]
	\caption{EVM limit in LTE and corresponding minimum SINR requirement}
	\centering
	\label{tab:evmrequirement}
	\renewcommand{\arraystretch}{1.2}
	\begin{tabular}{|p{.25\linewidth}|p{.25\linewidth}|p{.25\linewidth}|}
		\hline Modulation scheme for PDSCH & Required EVM & Required minimum SINR (dB)\\ 
		\hline QPSK  & 17.5\% & 15.14 \\
		\hline 16QAM  & 12.5\% & 18.06 \\
		\hline 64QAM  & 8\% & 21.94 \\
		\hline 256QAM  & 3.5\% & 29.12 \\
		\hline
	\end{tabular}
\end{table}

\bibliographystyle{IEEEtran}
\bibliography{pofdm}

\end{document}